\newif\ifpdflatex    
\pdflatextrue           

\documentclass[fleqn,usenatbib,useAMS,twocolumn,numberedappendix]{aastex}

\usepackage{amsmath,esint}
\usepackage{url,aasmacros}
\usepackage{natbib}
\usepackage{soul, CJK}
\usepackage{multirow}
\usepackage{tablefootnote}
\usepackage{appendix}
\usepackage{bm}
\usepackage{xspace}
\usepackage{color}
\usepackage{enumitem}
\usepackage{booktabs}
\usepackage{lineno}

\usepackage{graphicx}	
\usepackage{amsmath}	
\usepackage{amssymb}	
\usepackage{bm}		
\usepackage{url}
\usepackage{amsbsy}
\usepackage{xspace}
\usepackage{hyperref}
\usepackage{longtable}
\usepackage[colorinlistoftodos]{todonotes}
\usepackage[flushleft]{threeparttable}

\usepackage[T1]{fontenc}
\usepackage{ae,aecompl}

\def\lesssim{\mathrel{\hbox{\rlap{\hbox{\lower5pt\hbox{$\sim$}}}\hbox{$<$}}}}
\def\gtrsim{\mathrel{\hbox{\rlap{\hbox{\lower5pt\hbox{$\sim$}}}\hbox{$>$}}}}

\def\Msun{{\rm M$_{\odot}$}\xspace}

\usepackage{upgreek}                                                  
\usepackage{xspace}                                                       
 
\def\apjl{ApJ}%
\def\apjs{ApJS}%
\def\aap{A\&A}%
\shorttitle{ZTF on S250206dm}
\shortauthors{}

\begin{document}
\title{LIGO/Virgo/KAGRA neutron star merger candidate S250206dm:\\ Zwicky Transient Facility observations}

\author[0000-0002-2184-6430]{Tomás Ahumada}
\affiliation{Cahill Center for Astrophysics, California Institute of Technology, MC 249-17, 1216 E California Boulevard, Pasadena, CA, 91125, USA}

\author[0000-0003-3768-7515]{Shreya Anand}
\altaffiliation{LSST-DA Catalyst Postdoctoral Fellow} \affiliation{Kavli Institute for Particle Astrophysics and Cosmology, Stanford University, 452 Lomita Mall, Stanford, CA 94305, USA} \affiliation{Department of Astronomy, University of California, Berkeley, CA 94720-3411, USA}

\author[0000-0002-8255-5127]{Mattia Bulla}
\affiliation{Department of Physics and Earth Science, University of Ferrara, via Saragat 1, I-44122 Ferrara, Italy} \affiliation{INFN, Sezione di Ferrara, via Saragat 1, I-44122 Ferrara, Italy} \affiliation{INAF, Osservatorio Astronomico d’Abruzzo, via Mentore Maggini snc, 64100 Teramo, Italy}

\author[0000-0002-7672-0480]{Vaidehi Gupta}
\affiliation{School of Physics and Astronomy, University of Minnesota, Minneapolis, Minnesota 55455, USA}

\author[0000-0002-5619-4938]{Mansi Kasliwal}
\affiliation{Cahill Center for Astrophysics, California Institute of Technology, MC 249-17, 1216 E California Boulevard, Pasadena, CA, 91125, USA}

\author[0000-0003-2434-0387]{Robert Stein}
\affiliation{Department of Astronomy, University of Maryland, College Park, MD 20742, USA} \affiliation{Joint Space-Science Institute, University of Maryland, College Park, MD 20742, USA} \affiliation{Astrophysics Science Division, NASA Goddard Space Flight Center, MC 661, Greenbelt, MD 20771, USA}

\author[0000-0003-2758-159X]{Viraj Karambelkar}
\affiliation{Cahill Center for Astrophysics, California Institute of Technology, MC 249-17, 1216 E California Boulevard, Pasadena, CA, 91125, USA}

\author[0000-0001-8018-5348]{Eric C. Bellm}
\affiliation{DIRAC Institute, Department of Astronomy, University of Washington, 3910 15th Avenue NE, Seattle, WA 98195, USA}

\author[0009-0003-6181-4526]{Theophile Jegou du Laz}
\affiliation{Cahill Center for Astrophysics, California Institute of Technology, MC 249-17, 1216 E California Boulevard, Pasadena, CA, 91125, USA}

\author[0000-0002-8262-2924]{Michael W. Coughlin}
\affiliation{School of Physics and Astronomy, University of Minnesota, Minneapolis, Minnesota 55455, USA}

\author[0000-0002-8977-1498]{Igor Andreoni}
\affiliation{Department of Physics and Astronomy, University of North Carolina, Chapel Hill, NC 27599, USA}

\author[0000-0001-6595-2238]{Smaranika Banerjee}
\affiliation{The Oskar Klein Centre, Department of Astronomy, Stockholm University, AlbaNova, SE-10691 Stockholm, Sweden}

\author[0009-0008-2714-2507]{Aleksandra Bochenek}
\affiliation{Astrophysics Research Institute, IC2, Liverpool Science Park, 146 Brownlow Hill, Liverpool L3 5RF, UK}

\author[0000-0002-0129-806X]{K-Ryan Hinds}
\affiliation{Astrophysics Research Institute, IC2, Liverpool Science Park, 146 Brownlow Hill, Liverpool L3 5RF, UK}

\author[0000-0001-7201-1938]{Lei Hu}
\affiliation{McWilliams Center for Cosmology and Astrophysics, Department of Physics, Carnegie Mellon University, 5000 Forbes Avenue, Pittsburgh, PA 15213}

\author[0000-0002-6011-0530]{Antonella Palmese}
\affiliation{McWilliams Center for Cosmology and Astrophysics, Department of Physics, Carnegie Mellon University, 5000 Forbes Avenue, Pittsburgh, PA 15213}

\author[0000-0001-8472-1996]{Daniel Perley}
\affiliation{Astrophysics Research Institute, IC2, Liverpool Science Park, 146 Brownlow Hill, Liverpool L3 5RF, UK}

\author[0009-0008-8062-445X]{Natalya Pletskova}
\affiliation{Department of Physics, Drexel University, Philadelphia, PA 19104, USA}

\author[0000-0003-3173-4691]{Anirudh Salgundi}
\affiliation{Department of Physics, IIT Bombay, Powai, Mumbai 400076, India}

\author[0000-0003-2091-622X]{Avinash Singh}
\affiliation{The Oskar Klein Centre, Department of Astronomy, Stockholm University, AlbaNova, SE-10691 Stockholm, Sweden}

\author[0000-0003-1546-6615]{Jesper Sollerman}
\affiliation{The Oskar Klein Centre, Department of Astronomy, Albanova, Stockholm University, SE-106 91 Stockholm, Sweden}

\author[0000-0002-7942-8477]{Vishwajeet Swain}
\affiliation{Department of Physics, IIT Bombay, Powai, Mumbai 400076, India}

\author[0000-0002-9998-6732]{Avery Wold}
\affiliation{IPAC, California Institute of Technology, 1216 E. California Blvd, Pasadena, CA 91125, USA}

\author[0000-0002-6112-7609]{Varun Bhalerao}
\affiliation{Department of Physics, Indian Institute of Technology Bombay, Powai, 400 076, India}

\author[0000-0003-1673-970X]{S. Bradley Cenko}
\affiliation{Astrophysics Science Division, NASA Goddard Space Flight Center, MC 661, Greenbelt, MD 20771, USA} \affiliation{Joint Space-Science Institute, University of Maryland, College Park, MD 20742, USA}

\author[0000-0002-6877-7655]{David O. Cook}
\affiliation{IPAC, California Institute of Technology, 1200 E. California Blvd, Pasadena, CA 91125, USA}

\author[0000-0001-7983-8698]{Chris Copperwheat}
\affiliation{Astrophysics Research Institute, IC2, Liverpool Science Park, 146 Brownlow Hill, Liverpool L3 5RF, UK}

\author[0000-0002-3168-0139]{Matthew Graham}
\affiliation{Cahill Center for Astrophysics, California Institute of Technology, MC 249-17, 1216 E California Boulevard, Pasadena, CA, 91125, USA}

\author[0000-0001-6295-2881]{David L. Kaplan}
\affiliation{Department of Physics, University of Wisconsin-Milwaukee, P.O. Box 413, Milwaukee, WI 53201, USA}

\author[0000-0001-9898-5597]{Leo P. Singer}
\affiliation{Astrophysics Science Division, NASA Goddard Space Flight Center, MC 661, Greenbelt, MD 20771, USA}

\author{Niharika Sravan}
\affiliation{Department of Physics, Drexel University, Philadelphia, PA 19104, USA}

\author[0009-0001-0574-2332]{Malte Busmann }
\affiliation{University Observatory, Faculty of Physics, Ludwig-Maximilians-Universität, Scheinerstr. 1, 81679 Munich, Germany}

\author[0009-0008-2754-1946]{Julius Gassert}
\affiliation{University Observatory, Faculty of Physics, Ludwig-Maximilians-Universität, Scheinerstr. 1, 81679 Munich, Germany}

\author[0000-0003-3270-7644]{Daniel Gruen}
\affiliation{University Observatory, Faculty of Physics, Ludwig-Maximilians-Universität, Scheinerstr. 1, 81679 Munich, Germany} \affiliation{Excellence Cluster ORIGINS, Boltzmannstr. 2, 85748 Garching, Germany}

\author[0000-0002-1154-8317]{Julian Sommer}
\affiliation{University Observatory, Faculty of Physics, Ludwig-Maximilians-Universität, Scheinerstr. 1, 81679 Munich, Germany}

\author[0000-0003-2976-8198]{Yajie Zhang}
\affiliation{University Observatory, Faculty of Physics, Ludwig-Maximilians-Universität, Scheinerstr. 1, 81679 Munich, Germany}

\author[0000-0003-3433-2698]{Ariel Amsellem}
\affiliation{McWilliams Center for Cosmology and Astrophysics, Department of Physics, Carnegie Mellon University, 5000 Forbes Avenue, Pittsburgh, PA 15213}

\author[0000-0002-1270-7666]{Tomás Cabrera}
\affiliation{McWilliams Center for Cosmology and Astrophysics, Department of Physics, Carnegie Mellon University, 5000 Forbes Avenue, Pittsburgh, PA 15213}

\author[0000-0002-9364-5419]{Xander J. Hall}
\affiliation{McWilliams Center for Cosmology and Astrophysics, Department of Physics, Carnegie Mellon University, 5000 Forbes Avenue, Pittsburgh, PA 15213}

\author[0009-0000-4830-1484]{Keerthi Kunnumkai}
\affiliation{McWilliams Center for Cosmology and Astrophysics, Department of Physics, Carnegie Mellon University, 5000 Forbes Avenue, Pittsburgh, PA 15213}

\author[0000-0002-9700-0036]{Brendan O'Connor}
\affiliation{McWilliams Center for Cosmology and Astrophysics, Department of Physics, Carnegie Mellon University, 5000 Forbes Avenue, Pittsburgh, PA 15213}

\author[0000-0002-4843-345X]{Tyler Barna}
\affiliation{School of Physics and Astronomy, University of Minnesota, Minneapolis, Minnesota 55455, USA}

\author[0000-0001-7129-1325]{Felipe Fontinele Nunes}
\affiliation{School of Physics and Astronomy, University of Minnesota, Minneapolis, Minnesota 55455, USA} \affiliation{NSF Institute on Accelerated AI Algorithms for Data-Driven Discovery (A3D3),
MIT, Cambridge, MA 02139 and University of Minnesota, Minneapolis, MN 55455}

\author[0009-0008-9546-2035]{Andrew Toivonen}
\affiliation{School of Physics and Astronomy, University of Minnesota, Minneapolis, Minnesota 55455, USA}

\author[0000-0001-7357-0889]{Argyro Sasli}
\affiliation{School of Physics and Astronomy, University of Minnesota, Minneapolis, Minnesota 55455, USA}

\author[0000-0002-8532-9395]{Frank J. Masci}
\affiliation{IPAC, California Institute of Technology, 1200 E. California Blvd, Pasadena, CA 91125, USA}

\author[0000-0001-9152-6224]{Tracy X. Chen}
\affiliation{IPAC, California Institute of Technology, 1200 E. California Blvd, Pasadena, CA 91125, USA}

\author[0000-0002-5884-7867]{Richard Dekany}
\affiliation{Caltech Optical Observatories, California Institute of Technology, Pasadena, CA  91125}

\author[0000-0003-1227-3738]{Josiah Purdum}
\affiliation{Caltech Optical Observatories, California Institute of Technology, Pasadena, CA  91125}

\author{Antoine Le-Calloch}
\affiliation{School of Physics and Astronomy, University of Minnesota, Minneapolis, Minnesota 55455, USA}

\author[0000-0003-3533-7183]{G. C. Anupama}
\affiliation{Indian Institute of Astrophysics, 2nd Block 100 Feet Rd, Koramangala Bangalore, 560034, India}

\author[0000-0002-3927-5402]{Sudhanshu Barway}
\affiliation{Indian Institute of Astrophysics, 2nd Block 100 Feet Rd, Koramangala Bangalore, 560034, India}


\begin{abstract}
    
    We present the searches conducted with the Zwicky Transient Facility (ZTF) in response to S250206dm, a bona fide event with a false alarm rate of one in 25 years, detected by the International Gravitational Wave Network (IGWN). Although the event is significant, the nature of the compact objects involved remains unclear, with at least one likely neutron star.
    ZTF covered 68\% of the localization region, though we did not identify any likely optical counterpart. We describe the ZTF strategy, potential candidates, and the observations that helped rule out candidates, including sources circulated by other collaborations. Similar to \citet{Ahumada2024}, we perform a frequentist analysis, using \texttt{simsurvey}, as well as Bayesian analysis, using \texttt{nimbus}, to quantify the efficiency of our searches. We find that, given the nominal distance to this event of 373$\pm$104 Mpc, our efficiencies are above 10\% for KNe brighter than $-17.5$ absolute magnitude. 
    Assuming the optical counterpart known as kilonova (KN) lies within the ZTF footprint, our limits constrain the brightest end of the KN parameter space. Through dedicated radiative transfer simulations of KNe from binary neutron star (BNS) and black hole–neutron star (BHNS) mergers, we exclude parts of the BNS KN parameter space. Up to 35\% of the models with high wind ejecta mass ($M_{\rm wind} \approx 0.13$ \Msun) are ruled out when viewed face-on ($\cos\theta_{\rm obs} = 1.0$).
    Finally, we present a joint analysis using the combined coverage from ZTF and the Gravitational Wave Multimessenger Dark Energy Camera Survey (GW-MMADS). The joint observations cover 73\% of the localization region, and the combined efficiency has a stronger impact on rising and slowly fading models, allowing us to rule out 55\% of the high-mass KN models viewed face-on.
    
\end{abstract}

\section{Introduction}

The fourth observing run of the International Gravitational Wave Network (IGWN) re-started operations after a commissioning break between January and April 2024, detecting more than 99 binary black hole (BBH) merger candidates and one merger with confident presence of a neutron star (NS): S250206dm. This builds on previous successful runs, that to date sum over 102 BBH mergers and 6 mergers involving an NS \citep{GWTC3}. The most studied gravitational wave (GW) detection, GW170817, was discovered in coincidence with a short gamma-ray burst (sGRB), an afterglow, and a kilonova (KN), opening a new window into multi-messenger astronomy (MMA) \citep{AbEA2017b,AbEA2017c,GoVe2017}. The subsequent study of the KN unequivocally revealed the presence of heavy elements, produced through r-process nucleosynthesis, and the study of the afterglow has allowed for the discovery of super-luminal motion and helped constrain the geometry of the system \citep{Haggard2017, Hallinan2017, Margutti2017, Troja2017, Mooley2018170817, Pozanenko2018,Makhathini2021,Balasubramanian2022,Mooley2022,Coulter2017,  Drout2017, Evans2017, Kasen2017,KaNa17, LiGo2017, SoHo2017, Valenti2017, Utsumi2017, Arcavi2018, KaKa2019}. 

Due to the plethora of scientific studies that GW170817 has enabled, multiple collaborations have developed complex responses to IGWN triggers. Particularly in the optical and near-infrared (NIR), collaborations such as the Asteroid Terrestrial-impact Last Alert System (ATLAS),  the All-Sky Automated Survey for Supernovae (ASAS-SN), Gravitational Wave MultiMessenger DECam Survey (GW-MMADS), the Gravitational-wave Optical Transient Observer (GOTO), the Wide-Field Infrared Transient Explorer (WINTER), and the Panoramic Survey Telescope and Rapid Response System (Pan-STARRS), among others, have all performed observations of the GW regions \citep{Engrave2020,Grandma2020,Goto2020,ShPr2014,ToDe2018,ChMa2016,LiGo2017,Lundquist2019,Paek2024gecko,SoHo2017,2023FrASS..10.2887H,winter2025GW}. Despite all efforts including extensive tiling and galaxy-targeted searches, no confidently associated electromagnetic counterpart has been detected \citep{CoAh2019, Goldstein2019S190426c, S190814growth, Andreoni2019S190510g, Kasliwal2020kn, Grandma2020, Morgan2020, Vieira2020GW190814, Kilpatrick2021GW190814, Alexander2021GW190814, Wet2021GW190814, Thakur2021GW190814,Tucker2022GW190814,Rastinejad2022gwsearch,Dobie2022GW190814,cabrera24,2025arXiv250315422P}.  

The Zwicky Transient Facility (ZTF; \citealt{Bellm2019, Graham2019,Dekany2019}), mounted on the Samuel Oschin 48-inch Telescope at Palomar Observatory, is a public-private project that nominally covers the entire northern night sky in $g$, $r$, and $i$ band every two nights. The high cadence of the public survey allows ZTF to have one of the most complete records of the dynamic optical sky, and enables the discovery of transients at early stages. The large field of view (FoV) of ZTF, of 47 square degrees additionally grants ZTF with the capacity to perform rapid searches of relativistic transients, such as GRBs \citep{Ahumada2022grb, Coughlin2018} and GW events \citep{anand2020,Coughlin_S190425z, kasliwal2020,Ahumada2024} across thousands of square degrees. These searches have led to the discovery of multiple afterglows, including the shortest gamma-ray burst linked to a collapsar and an orphan afterglow detected during the IGWN third observing run (O3) \citep{Ahumada2021sgrb, perley2024pim}.

Throughout this paper, we discuss the follow-up of the high-significance event S250206dm, describing the GW event in \S2, the follow-up strategy of ZTF in \S3, the candidate vetting strategy in \S4, and we discuss the implications of our non-detections in \S5. In \S6, we show a joint analysis with observations from the Gravitational Wave Multimessenger Dark Energy Camera Survey (GW-MMADS; PI: Andreoni \& Palmese), which are presented in the companion paper \citep{decamGWlei}, and an extensive analysis of the ZTF candidates and other candidates announced through the Transient Name Server (TNS) is presented in Appendix \ref{appendixCandidates}.

\section{S250206dm}

On 2025-02-06 21:25:44 UTC IGWN detected a candidate merger of two compact objects, with a false alarm rate (FAR) of 1 in 25 years. The event classification probabilities were reported in the GCN circular as 55\% NSBH, 37\% BNS, and 8\% Terrestrial \citep{2025GCN.39178....1L}. Assuming the GW event is of astrophysical origin, the machine learning inference on the GW data \citep{embright} shows that the merger had at least one NS involved (HasNS = 100\%), a 63\% probability of having a compact object (HasMassGap) in the mass gap (3--5 M$_{\Sun}$), and 30\% of leaving remnant material to power a KN (HasRemnant) \citep{2025GCN.39178....1L}.

The initial localization by BAYESTAR \citep{Singer2016} covered 2139 sq. deg., distributed between a northern (Dec $> -30$ deg) and a southern lobe (Dec $< -30$ deg), containing respectively 73\% and 27\% of the total probability \citep{2025GCN.39175....1L}. The localization was updated several times, and the final map, produced by Bilby \citep{bilby1,bilby2} and circulated 1.5 days after the event, featured a more compact northern and southern lobe. The majority of the probability shifted to the northern lobe, which contained 78\% of the probability, while the southern lobe contained the remaining 22\% \citep{2025GCN.39231....1L}. 

\begin{figure*}
    \centering
    \includegraphics[width=0.45\textwidth]{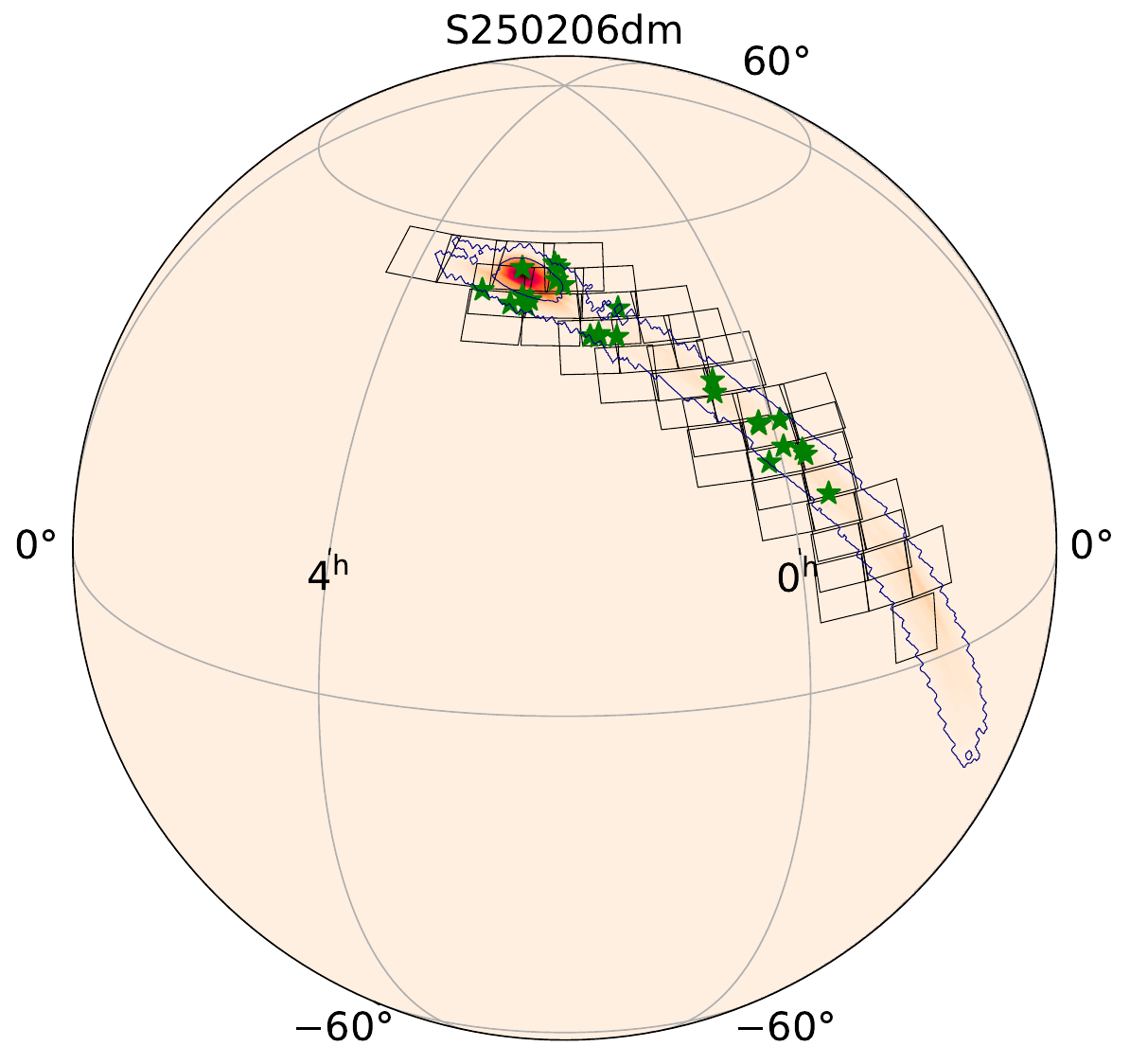}
    \includegraphics[width=0.45\textwidth]{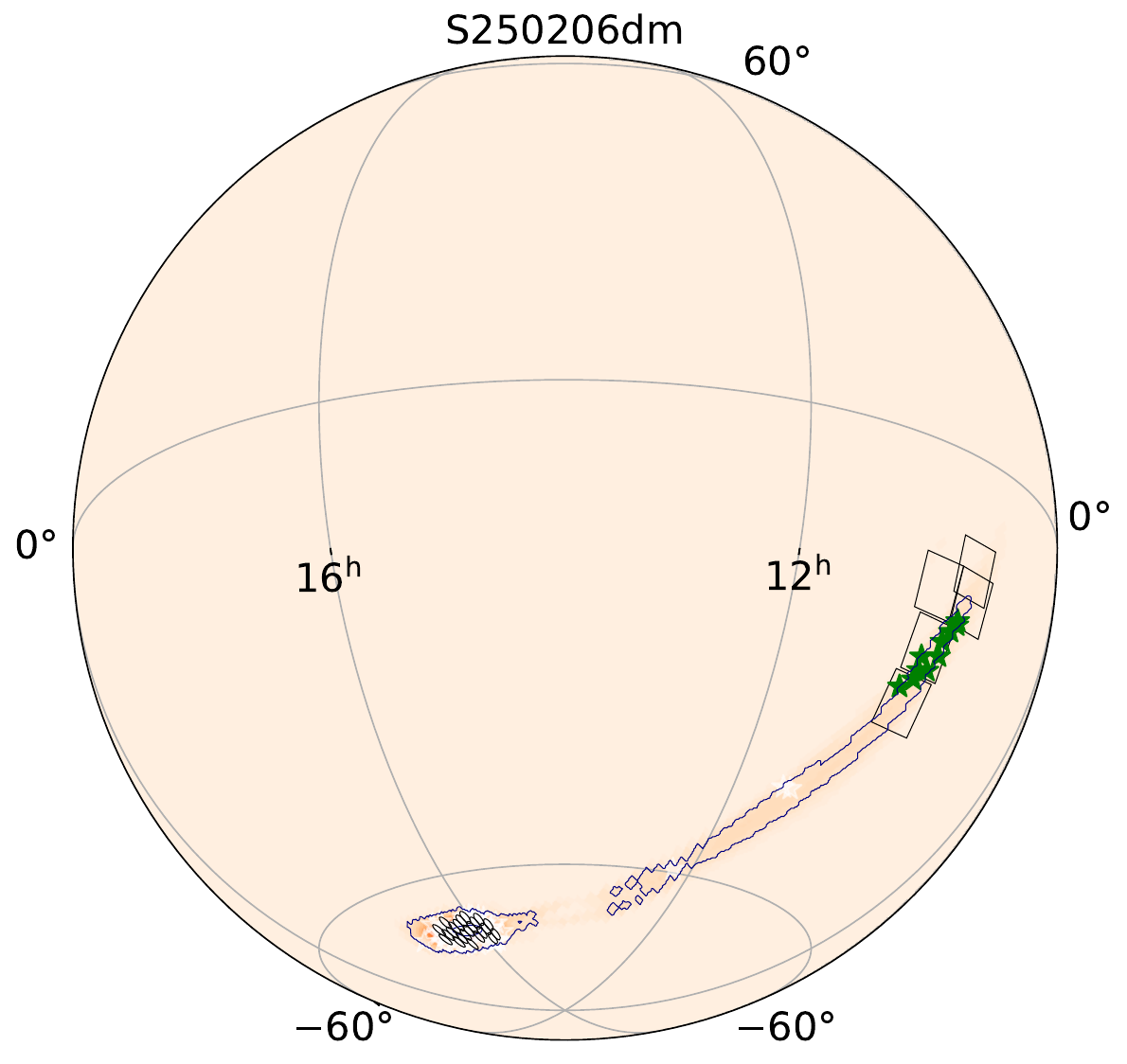}
    \caption{Localization of the high-significance event S250206dm, overlaid with the ZTF tiles (black squares), the GW-MMADS tiles (black circles), and the 90\% probability contour (navy). The green stars represent transients reported on TNS that were accessible from Palomar, while the white stars indicate candidates that were not accessible from Palomar.  }
    
    \label{fig:map_S250206dm}
\end{figure*}

\section{ZTF follow-up campaign}

\subsection{Observing plan}
ZTF observations of the GW skymap started on UT 2025-02-08 02:21 UTC -- 29 hours after the merger, due to poor weather. ZTF observations are conducted on a pre-defined grid of fields, and the GW region included ZTF fields that were setting early. In order to accommodate a larger number of fields, we decided to slice the GW localization in right ascension (RA) and feed four separate regions to the scheduler optimizer \texttt{gwemopt} \citep{Coughlin2018}. The RA slices were the following: from 22 hr\, to 0 hr, from 0 hr to 2 hr\,, from 2 hr\, to 5 hr\, and from 7 hr\, to 21 hr\,. For all these regions, we scheduled a sequence of 300 s exposures in $r$, $g$, and $r$ band for our first night, and a sequence of 300 s exposures in $g$, $r$, $i$ band for the following nights. In order to account for the chip gaps in the ZTF fields, we scheduled observations using fields in both the primary and secondary grid, as they have complementary coverage of the ZTF fields. The schedule was repeated over the course of 9 nights. The full details of the pointings are available on TreasureMap \citep{Wyatt2020tresuremap}. Within 9 days, a total of 68\% of the probability enclosed within the GW skymap was observed with ZTF at least once, while 64\% of the total probability was observed at least twice (see Fig.\ref{fig:map_S250206dm}).

\begin{figure*}[t]
    \centering
     \includegraphics[width=0.4\textwidth]{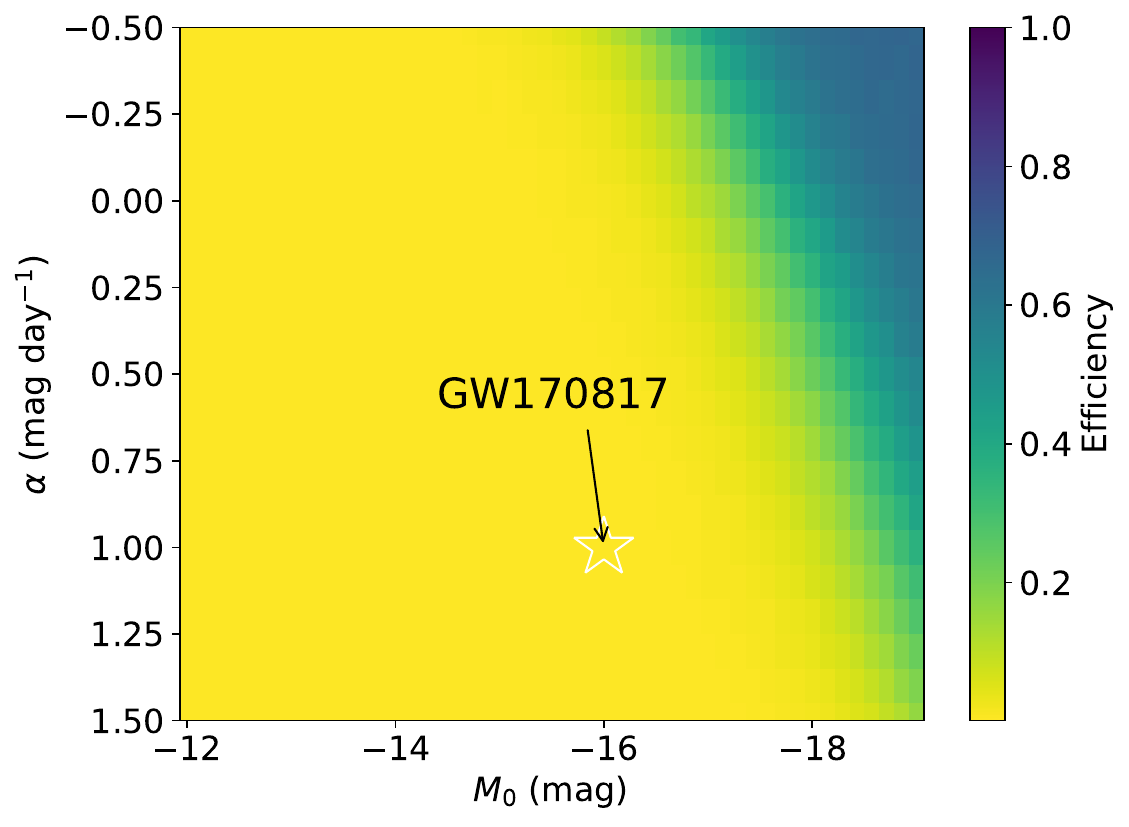}
\includegraphics[width=0.4\textwidth]{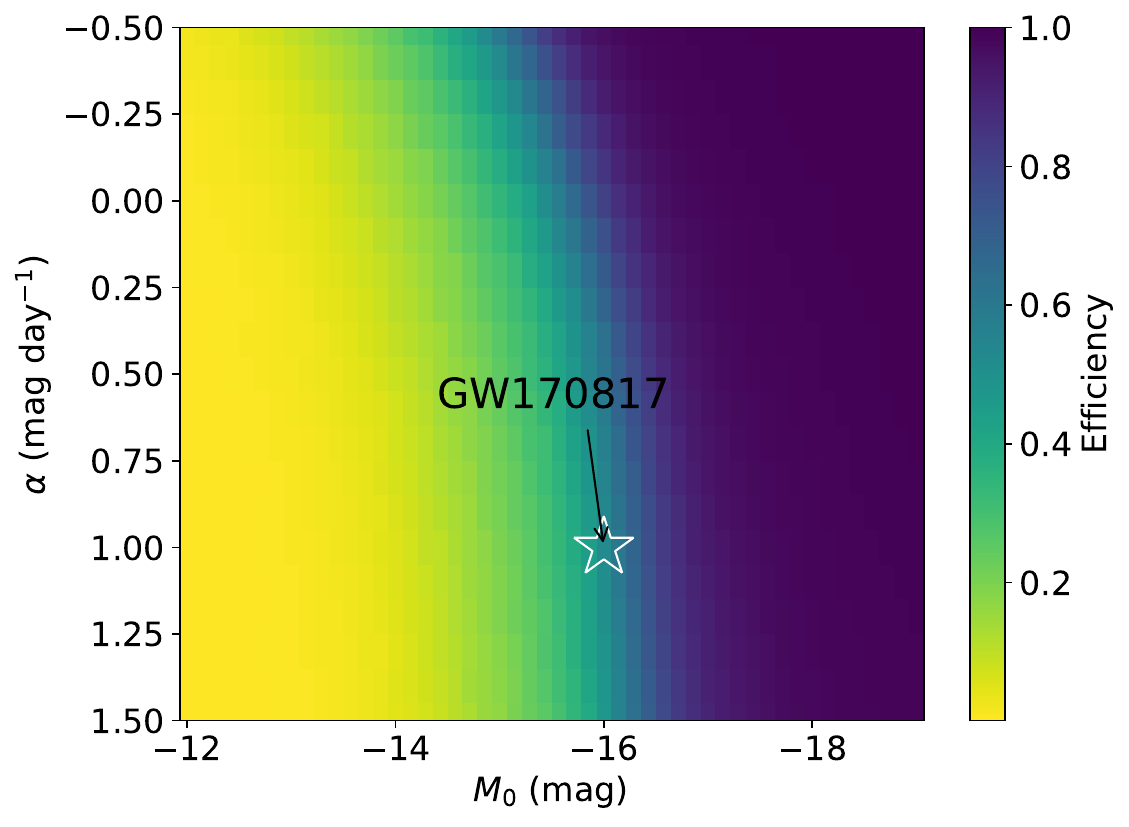}
    \caption{Kilonova single-detection efficiency with \texttt{simsurvey} for the \textit{Tophat} model evolution for (left) S250206dm and (right) for the combined set of GW triggers: S250206dm, S230518h, S230627c, S230731an, S231113bw, GW230529, GW200115, GW200105, GW190814, GW190426, and GW190425. The color bar shows the fraction of sources detected once versus the number of sources ingested in the GW volume. We mark the position of a GW170817-like KN on this plot. The ZTF efficiency of recovering a GW170817-like KN in the skymap of S250206dm is $<$1\% (left) while for the combined set of events the efficiency is 36\%.}
    \label{fig:simsurvey}
\end{figure*}

\begin{figure*}[t]
    \centering
     \includegraphics[width=0.4\textwidth]{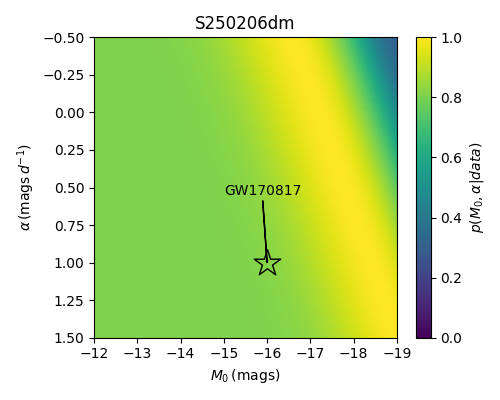}
\includegraphics[width=0.4\textwidth]{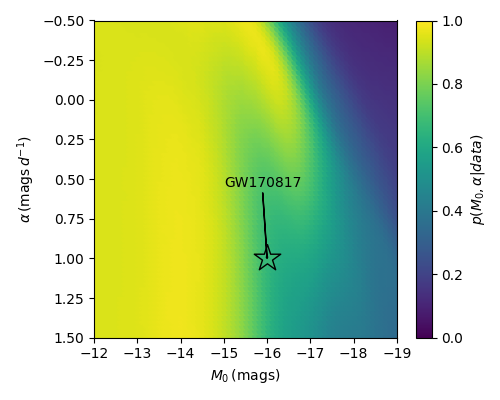}
    \caption{ The \texttt{nimbus} results of (left) S250206dm and (right) the combined set of GW triggers: S250206dm, S230518h, S230627c, S230731an, S231113bw, GW200115, GW190524, GW190426, GW190814 and GW230529 for KN model parameters assuming the \textit{Tophat} model. The x-axis shows the peak absolute magnitude $M_0$ of a model, while the y-axis shows its evolution rate $\alpha$. The color bar shows the posterior probability of each model, where yellow regions show the favored regions of parameter space given the non-detection of KNe from ZTF observations, and the bluer regions show less preferred combinations for peak $M_0$ and $\alpha$. We note that for S250602dm a GW170817-like KN is preferred at 82\% level, while for the combined analysis it falls at the 67\%.}
    \label{fig:nimbus}
\end{figure*}

\subsection{Candidate Vetting}
ZTF observed the S250206dm region for nine consecutive nights. Each night, the ZTF pipeline \citep{Masci2019}, operated at the Infrared Processing and Analysis Center (IPAC\footnote{\url{https://www.ipac.caltech.edu/}}), processes, calibrates, and performs image subtraction in near real-time. Any flux deviation exceeding 5$\sigma$ from the reference image produces an \textit{alert} \citep{Patterson2019}, which includes metadata on the transient, such as its lightcurve history, a real-bogus score \citep{Duev2019}, and other relevant information. 

We then query the ZTF stream of alerts using Kowalski \citep{Kasliwal2020kn} through Fritz \citep{vanderWalt2019,Coughlin2023skyportal}.
Our criteria for candidate selection start with filtering for transients that have a positive residual after image subtraction with respect to the reference image, and a deep learning real-bogus score \citep{Duev2019} greater than 0.3 to differentiate between real astrophysical sources and artifacts. To avoid contamination from stars, we require transients to be located more than 3 arcsec away from sources classified as probable stars by a morphology-based classifier \citep{TaMi2018} applied to sources detected by Pan-STARRS (PS1; \citealt{ChMa2016}). We request a minimum of two detections separated by at least 15 minutes to remove most moving objects and cosmic rays. To reduce the influence of artifacts from bright stars, sources must be located more than 20 arcsec from any object with a magnitude less than 15. We exclude sources that show activity before the GW event, as KNe and relativistic afterglows are only expected to occur after the merger, and finally, we require the candidate to lie within the 95\% contour of the latest and most up-to-date GW skymap.

For the ZTF sources that pass the alert filtering, we further cross-match the candidates to the Minor Planet Center to flag known asteroids, and we cross-match to WISE to reject AGNs using their WISE colors \citep{WrEi2010}. We also run forced photometry on ZTF images \citep{Masci2019} and require that there are no detections before the GW trigger.

In addition to Fritz,  we queried the \textit{Kowalski} database using the \texttt{emgwcave}\footnote{\url{https://github.com/virajkaram/emgwcave}} Python script, which retrieves candidates based on similar cuts to those mentioned earlier. \texttt{emgwcave} offers added flexibility, allowing for easy modifications to the queries. 

Next, we performed an independent search using the \texttt{nuztf}\footnote{\url{https://github.com/desy-multimessenger/nuztf}} Python package \citep{nuztf}, originally developed for the ZTF Neutrino Follow-Up Program \citep{stein_23_nu}. The \texttt{nuztf} package utilizes the \texttt{AMPEL} framework for candidate filtering \citep{ampel_19} and retrieves ZTF data with minimal latency from the \texttt{AMPEL} broker data archive \citep{ampel_19}. We applied selection criteria similar to those outlined previously, followed by automated cross-matching with various multi-wavelength catalogs to identify potential variable AGN or stars.  Additionally, \texttt{nuztf} uses ZTF observation logs from IPAC to calculate the survey coverage of a skymap, factoring in chip gaps and processing failures.

Lastly, we utilized the \texttt{ZTFReST} infrastructure \citep{Andreoni2021ztfrest} to retrieve candidates. \texttt{ZTFReST} is an open-source tool for flagging fast-fading transients based on ZTF alert photometry and forced photometry \citep{Yao2020}. 

The described selection criteria resulted in 13 candidates from the ZTF stream. All candidates were identified in the Fritz, \texttt{emgwcave} and \texttt{nuztf} searches, while only the fast-evolving ones appeared in the ZTFReST search, as the latter filters out slow-evolving transients ($\Delta m / \Delta t < 0.3$ mag/day). Candidates discovered after the first two nights of observations were circulated via GCN \citep{ahumadaGCNs250206dm}, whereas the remaining ones were reported to the Transient Name Server (TNS). The full list of ZTF candidates is shown in Table \ref{table:ztfcand}, and detailed descriptions for them can be found in Appendix \ref{appendixCandidates}.  

To determine whether a candidate is related to the GW event, we rely on further analysis to reach one of our rejection criteria. When available, we examine the spectra of the transient and derive a classification by comparing the spectra to various known transient types, as well as KN models. If the source fades beyond spectroscopic limits, we use the redshift of the host to assess whether it falls within the GW volume. Alternatively, we cross-match our sources with Gaia \citep{Gaia2018} and classify as \textit{stellar} the sources within 2 arcsec of a Gaia object with significant proper motion ($>3\sigma$). We run forced photometry over the entire history of the ZTF survey, and sources with previous detections are classified as \textit{old}. For sources that cannot be ruled out, we request further photometric follow-up and compare the photometric evolution of the sources to KN models. We classify sources as \textit{slow} if their evolution does not align with KN model predictions.

\begin{figure*}[t]
    \centering
     \includegraphics[width=0.9\textwidth]{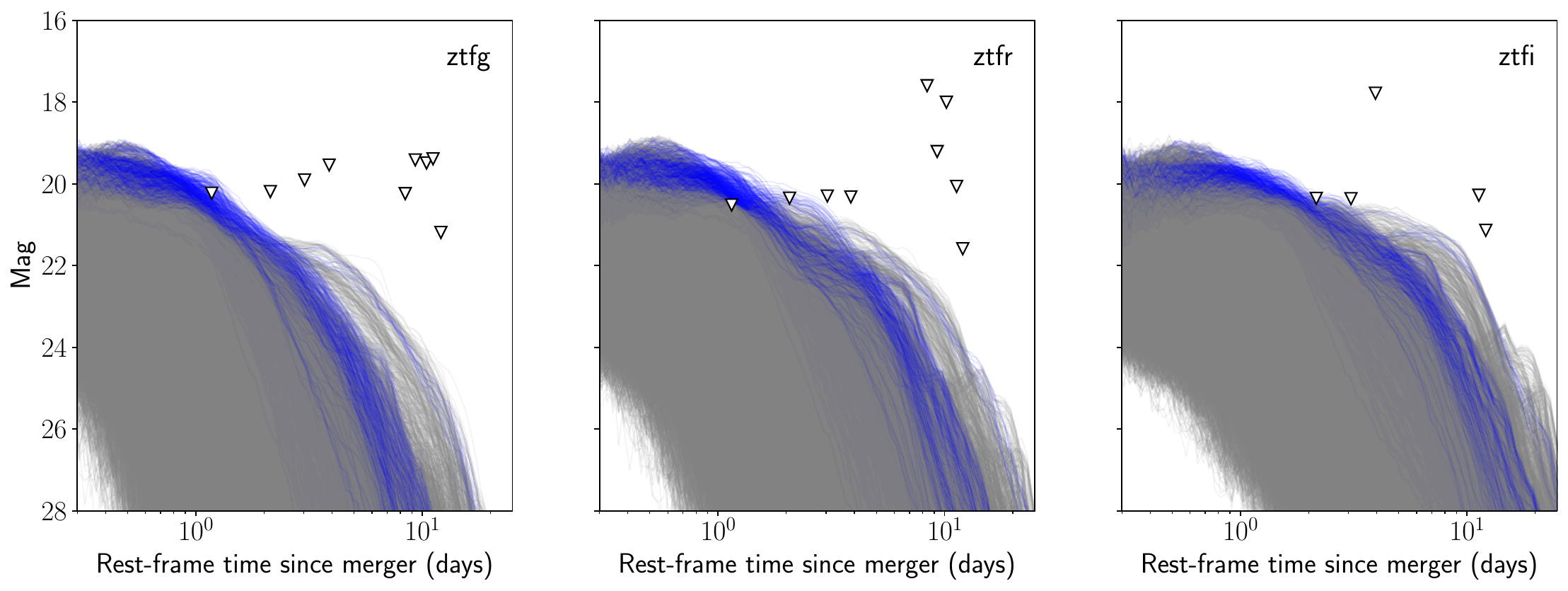}
\includegraphics[width=0.9\textwidth]{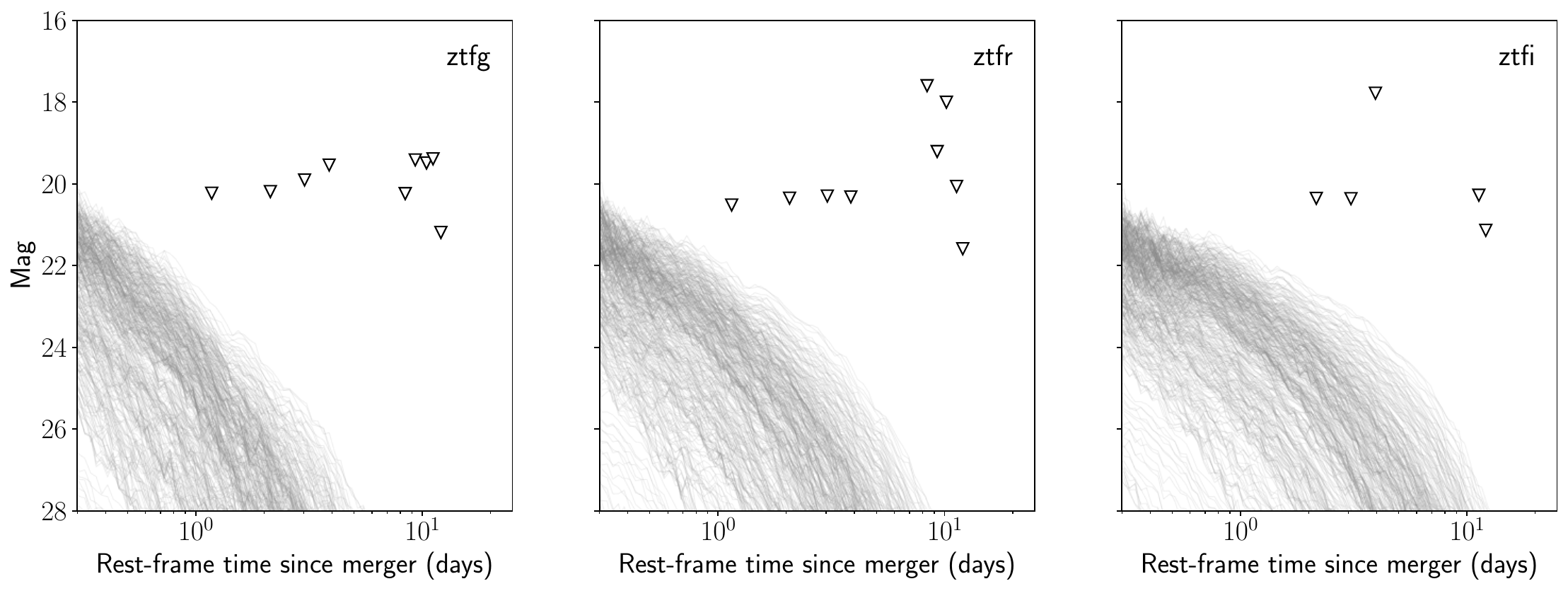}
    \caption{Light-curves of (top) BNS and (bottom) NSBH models pinned at 269 Mpc (grey), the closest 1-sigma from the GW distance distribution. We show the ZTF median limits as triangles for (left) g-band, (middle) r-band, and (right) i-band. A number of BNS models were ruled out (in blue), mostly due to the limits achieved during the first night of observations. No NSBH models were ruled out with the ZTF observations.  }
    \label{fig:KNvslimits}
\end{figure*}

All ZTF candidates were ruled out as potential counterparts to S250206dm based on the criteria described above. 

\subsection{Candidates from other facilities}

Multiple facilities conducted searches of counterparts to S250206dm. Candidates including coincident fast radio bursts (FRBs) \citep{GCNfrb}, neutrinos \citep{gcnnu}, X-ray transients \citep{GCNEP}, and a number of optical candidates were circulated through the General Coordinate Network (GCN) and TNS \citep{2025GCN.39208....1B, 2025GCN.39226....1F, 2025GCN.39215....1S, 2025GCN.39214....1B, 2025GCN.39211....1S, 2025GCN.39210....1Y, 2025GCN.39197....1A, 2025GCN.39196....1F, 2025GCN.39195....1H, 2025GCN.39193....1W, 2025GCN.39191....1H, 2025GCN.39785....1C, 2025GCN.39318....1C, 2025GCN.39285....1C, 2025GCN.39265....1L, 2025GCN.39256....1S, 2025GCN.39249....1L, 2025GCN.39234....1C, 2025GCN.39241....1P, winter2025GW}. ZTF observations covered the regions associated with the FRB, neutrino, and Einstein Probe (EP) candidates; however, no optical counterparts were identified (see Appendix \ref{appendixCoverage}). In addition to ZTF, we used facilities in the GROWTH collaboration (see Appendix \ref{appendixFollowup} and \citealt{Kasliwal2019growthmarshal}) to follow-up some of these transients in order to assess their connection to the GW event. We focused on the candidates in the northern lobe, as these are accessible from Palomar Observatory and partner facilities in the northern hemisphere. We direct the reader to \citet{decamGWlei} for an analysis on the candidates in the southern lobe of the skymap. Following the same rejection criteria discussed for ZTF candidates, we are able to rule out 15 of the 22 candidates in the northern region. 
We additionally used the photometric redshifts of the host galaxies to assess whether the associated candidates fall within the predictions of the KN model grid. We found that one candidate exhibited a luminosity inconsistent with the models. The photometric redshifts were obtained from either SDSS \citep{SDSSphotoz} or the Legacy Survey \citep{legacysurveyphotoz}, depending on availability. These redshifts were not used to rule out sources, but rather as a proxy to evaluate whether the observed brightness is consistent with model expectations.

We cannot rule out seven candidates: AT2025bcc, AT2025bey, AT2025bbp, AT2025bah, AT2025bam, AT2025bce, and AT2025baf. A summary of the follow-up and analysis of candidates detected by other facilities can be found in Table \ref{table:cad_notZTF} and a thorough description of each source in Appendix \ref{appendixCandidates}. Given the lack of confirmation of a KN through public channels, in this paper and the following analysis, we assume these candidates are not counterparts to S250206dm.

\section{ZTF observation}

In this section, we assess the efficiency of the ZTF search for an optical counterpart to S250206dm. Additionally, we present an updated analysis that includes high-significance (FAR $>$ 1 per year) events bearing an NS (either HasNS $> 0.1$, $P_{\rm BNS} >0.1$, or $P_{\rm NSBH} >0.1$) detected during the first part of the fourth observing run (O4a) of the IGWN. We also incorporate the confirmed astrophysical events from the third IGWN observing run (O3) into the analysis. To do so, we follow the methodology described in \citet{Ahumada2024}, and we apply both a Bayesian and a frequentist approach. Using the ZTF observations, we constrain the KN luminosity function based on various assumptions.

\subsection{Efficiencies}

As described in previous studies \citep{Ahumada2024, Kasliwal2020kn} we determine the efficacy of our searches in detecting a KN counterpart to S250206dm using two different approaches. With our Bayesian approach, \texttt{nimbus}, we calculate the posterior probability of a KN of a particular absolute magnitude, $M_{\rm 0}$, and decay rate, $\alpha$, in the GW skymap given our ZTF observations. This approach takes into account the probability that the GW event was astrophysical in origin. Our frequentist approach, \texttt{simsurvey}, quantifies the efficiency with which our observations within the GW skymap would detect a KN with a particular $M_{\rm 0}$,  $\alpha$, or whose lightcurve evolution is described by a KN model. These complementary analyses allow us to place constraints on the properties of a potential KN associated with S250206dm.

\begin{figure*}[t]
    \centering
     \includegraphics[width=\textwidth]{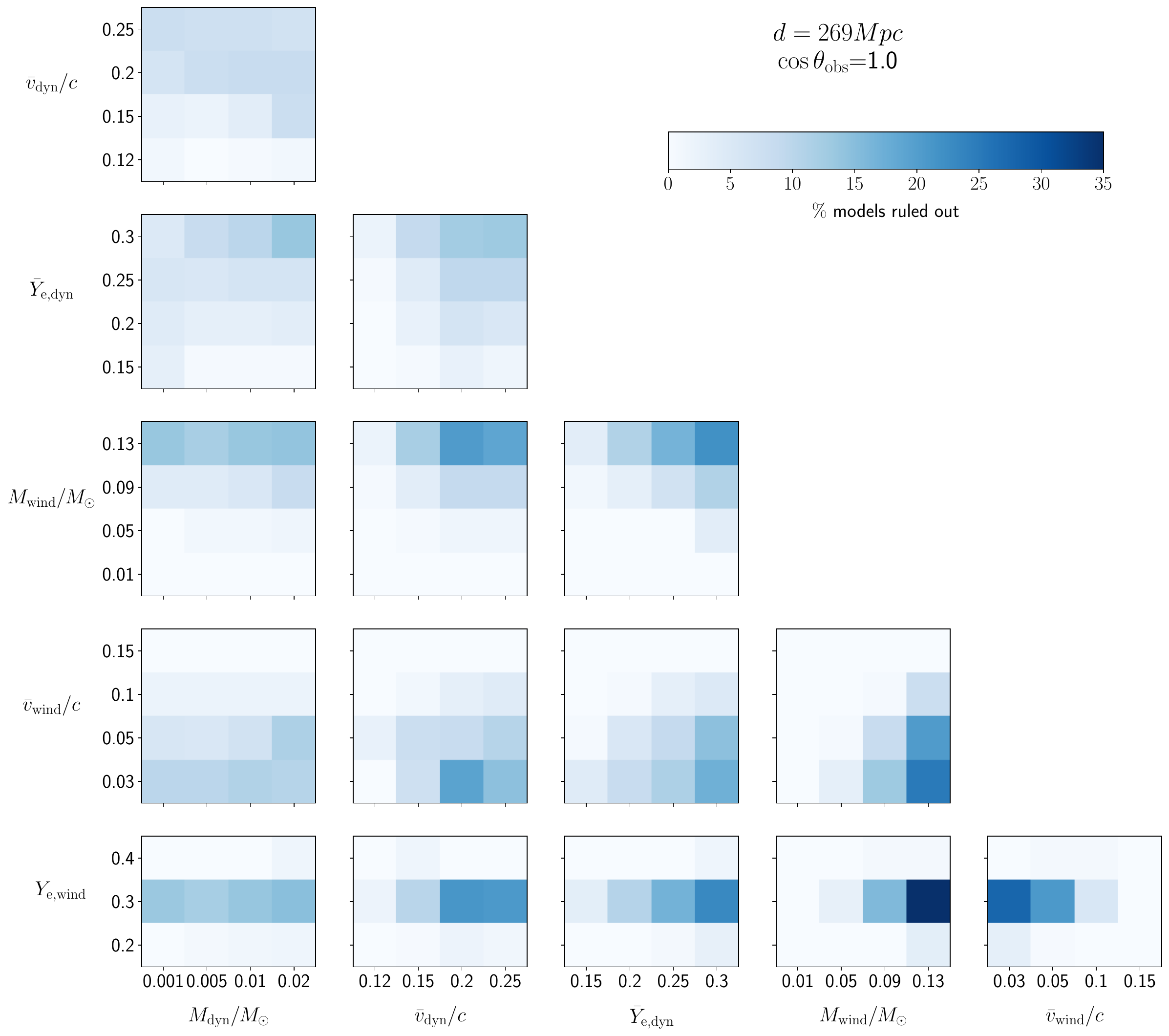}
    \caption{Corner plot showing in a colorbar the percentage of BNS models ruled out at a fixed distance of 269 Mpc and at a fixed viewing angle of 0 deg. The parameters correspond to the ejecta mass, mass-weighted averaged velocity and mass-weighted electron fraction for the dynamical and wind ejecta: $m_{\rm dyn}$, $\bar{v}_{\rm dyn}$, $\bar{Y}_{\rm e,dyn}$, $m_{\rm wind}$, $\bar{v}_{\rm wind}$ and $Y_{\rm e,wind}$  }
    \label{fig:corner}
\end{figure*}

\texttt{simsurvey} takes as inputs ZTF pointings (the area covered), limiting magnitudes, and the GW skymap, and simulates KN lightcurves as would be observed by ZTF within the skymap. The detection efficiency, defined as fraction of KNe detected amongst all simulated KNe, can be a proxy for ZTF's performance in conducting these KN searches.

Similar to \texttt{simsurvey} analyses conducted in previous studies \citep{ Kasliwal2020kn,Ahumada2024}, we calculate the efficiency with which our ZTF observations can recover KNe described by simple linear decay models (\textit{tophat}), \textit{Bulla} models from \textsc{possis} \citep{Bulla2019,Diet2020}, \textit{Kasen} models \citep{Kasen2017}, and \textit{Banerjee} models \citep{Banerjee2022}. First, adopting a model-agnostic approach, we simulate KNe with a range of $M_0$ and $\alpha$ within the skymap of S250206dm. For each combination of $M_0$ and $\alpha$, we simulate 10,000 KNe and calculate the single detection efficiency with ZTF. Our results are shown in Fig.~\ref{fig:simsurvey}. Our ZTF limits for this event are most sensitive to rising transients and fading transients brighter than $M\sim-18$\,mag. However, our detection efficiency for a GW170817-like KN is $<1$\%; thus our observations were not sensitive to KNe fainter than GW170817.

Given the median depths achieved on each night of observations for S250206dm, we calculate the detection efficiency for the brightest KN model in each grid we consider. Since our full set of \textit{tophat} models does not represent the realistic range of absolute magnitudes and decay rates expected for a KN, we choose a model with a similar peak absolute magnitude to the brightest model in the \textit{Bulla} grid ($M_0 \approx -17.6$\,mag), instead of choosing the brightest \textit{tophat} model we simulated. For the brightest \textit{Bulla} (\textit{Kasen}) [\textit{Banerjee}] model with a total $r$-process ejecta mass of 0.15 (0.10) [0.05]\,$\rm M_{\odot}$ and a peak absolute magnitude of $M_0 \approx -17.6\, (-17.0) [-16.4]$\,mag, our ZTF detection efficiency is 15 (6) [4]\%. For the corresponding \textit{tophat} model with $M_0=-17.6$\,mag and $\alpha=0.5$\,mag day$^{-1}$, we achieve 10\% efficiency with ZTF. 

In addition, we determine the joint single-detection and filtered efficiencies for GW170817 across all significant NS merger event candidates released thus far: S250206dm, S230518h, S230627c, S230731an, S231113bw, GW230529, GW200115, GW200105, GW190814, GW190426, and GW190425. Our filtering criteria for simsurvey requires two 5$\sigma$ detections with ZTF separated by 15\,min. Our joint single-detection (filtered) efficiency for a GW170817-like KN is 39 (36)\% with the \textit{Bulla} model, 38 (35)\% with the \textit{Kasen} model, 22 (20)\% with the Banerjee model, and 53 (36)\% with the \textit{tophat} model. The addition of S250206dm changes the overall joint single-detection/filtered efficiency for a GW170817-like KN by $\lesssim1$\% from \citet{Ahumada2024}, for all models considered. However, our ZTF limits for this event contribute towards constraining the bright end of the KN luminosity function.

\texttt{nimbus} \citep{Mohite_2022} is a hierarchical Bayesian Inference framework model that integrates ZTF observations across all bands, the 3D GW skymap, and the extinction values for each field. The software injects observation parameters such as time stamps, limiting magnitudes, filter information, and the observation coordinates for three days from the merger time. \texttt{nimbus} treats observations across multiple filters as a single, unified lightcurve by assuming a shared color evolution, and uses this “average-band” for our calculations. \texttt{nimbus} computes the posterior probability ($P_{nimbus}$) of a KN given the ZTF observations within the GW skymap. It then combines this posterior with the probability of the event’s localization within each field to calculate the log-posterior values for each observation. \texttt{nimbus} additionally incorporates the probability that the event is of astrophysical origin, $p_{\mathrm{astro}}$, as well as the fractional sky coverage of the event by ZTF. KN parameters for a model lightcurve can be constrained using the posterior probability derived by \texttt{nimbus}.
 
 The $p_{astro}$ value for S250206dm is $0.92$, and ZTF covered $68\%$ of the total skymap which allows for a significant posterior probability and results in constraints for the brighter KN models. Figure \ref{fig:nimbus} (left) shows the posterior probability for S250206dm. The combinations of peak magnitude and $M_0$ evolution rate $\alpha$ with a higher posterior value constitute the preferred parameter space, while those with a lower posterior value are less favored. Integrating the results for S250206dm with the posterior probabilities for events from the O4a run, Figure \ref{fig:nimbus} (right) shows the combined posterior probability for all events that have been considered significant: S250206dm, S230518h, S230627c, S230731an, S231113bw, GW200115, GW190524, GW190426, GW190814 and GW230529.

Both \texttt{nimbus}, a Bayesian method, and \texttt{simsurvey}, a frequentist approach, provide independent but complementary insights into KNe from the ZTF observations. \texttt{simsurvey} assesses the recovery efficiency of KNe with specific model parameters in the ZTF follow-up, while  \texttt{nimbus} helps identify which KN model parameters are more or less supported by the ZTF data. When comparing the results from both methods, we notice
similar overall patterns. Rising or slowly decaying bright KNe ($M \lesssim -17.5$ mag and $\alpha < 0.25$) exhibit the highest efficiencies in \texttt{simsurvey}, as our simulation recovers these KNe with efficiencies up to 60\%. The \texttt{nimbus} analysis evaluates whether a given KN model is favored by the ZTF non-detections. For S250206dm, \texttt{nimbus} disfavors a similar set of models ($P_{\texttt{nimbus}} < 0.4$ for $M \lesssim -17.5$ mag and $\alpha < 0.2$), as the ZTF data . In contrast, faint, fast-fading KNe have the lowest detection efficiencies in \texttt{simsurvey} and receive the most support from \texttt{nimbus} ($P_{\texttt{nimbus}} > 0.8$), based on the ZTF non-detections.

\subsection{Ruling out merger parameter space} Here we use our photometric upper limits to constrain the parameter space for the compact binary merger. We simulate KN spectral models using the most recent version \citep{Bulla2023} of the 3D Monte Carlo radiative transfer code \texttt{possis} \citep{Bulla2019}. Specifically, we present two new KN grids for BNS and NSBH mergers that are inspired by \cite{Anand2023} and \cite{Mathias2024}, respectively, and use revised nuclear heating rates from \cite{Rosswog2024}. The two grids will be made publicly available at \href{https://bit.ly/possis$_$models}{https://bit.ly/possis$\_$models}.

The BNS merger ejecta are described by two distinct axially-symmetric components: a first component ejected on dynamical timescales (dynamical ejecta) and a second component ejected after the merger from a disk accreted around the merger remnant (post-merger disk-wind ejecta), see, e.g., \cite{Nakar2020} for a review. Following \cite{Bulla2023}, a dependence on the polar angle $\theta$ is taken for both the density and the electron fraction in the dynamical ejecta ($\rho\propto\sin^2\theta$ and $Y_{\rm e}\propto\cos^2\theta$), while spherical symmetry and uniform composition (i.e., fixed $Y_{\rm e}$) are assumed for the wind ejecta \citep{Perego2017,Radice2018,Setzer2023}. The grid is controlled by six free ejecta parameters: dynamical mass $M_{\rm dyn}=(0.001,0.005,0.01,0.02)\,M_{\odot}$, dynamical mass-weighted averaged velocity $\bar{v}_{\rm dyn}=(0.12,0.15,0.2,0.25)\,$c, dynamical mass-weighted averaged electron fraction $\bar{Y}_{\rm e,dyn}=(0.15,0.20,0.25,0.30)$, wind mass $M_{\rm wind}=(0.01,0.05,0.09,0.13)\,M_{\odot}$, wind mass-weighted averaged velocity {$\bar{v}_{\rm wind}= (0.03,0.05,0.10,0.15)\,$c}, and wind electron fraction $Y_{\rm e,wind}=(0.20,0.30,0.40)$. The number of models is 3072, which leads to a total number of KNe of 33\,792 when accounting for the 11 different viewing angles $\theta_{\rm obs}$. These angles, which are distinct from the polar angle $\theta$ used to describe the KN geometry, are equally spaced in $\cos \theta_{\rm obs}$ from a face-on (polar) to a edge-on (equatorial) view of the system.

Similarly, the NSBH merger ejecta are described by dynamical and disk-wind ejecta components. However, we follow \cite{Kawaguchi2020} and adopt a stronger angular dependence focusing the dynamical ejecta around the merger plane as expected from NSBH systems, $\rho\propto 1/\{1+\exp[-20\,(\theta-1.2)]\}$. In addition, the compositions are the same for all the models in the grid and set to $Y_{\rm e}=0.1$ in the dynamical ejecta and $0.2<Y_{\rm e}<0.3$ in the wind ejecta \citep{Kawaguchi2020,Mathias2024}. The NSBH KN grid is constructed using binary properties and different equations of state (EoS) as free parameters. In particular, the neutron star mass $M_{\rm NS}=(1.2,1.4,1.6,1.8)\,M_{\odot}$, black hole mass $M_{\rm BH}=(4.0,6.0,8.0)\,M_{\odot}$ and black hole spin $\chi_{\rm BH}=(0.0,0.3,0.6)$ are varied within the grid, while the DD2 \citep{Hempel2010}, the AP3 \citep{Akmal1998} and the SFHo+H$\Delta$ \citep{Drago2014} are chosen as possible EoSs. Ejecta masses and velocities are computed for each model using fitting formulae as described in \cite{Mathias2024}, with 37 combinations of the free parameters that lead to the ejection of some material and therefore produce a KN (see their table 2). When accounting for the 11 different viewing angles, a total of 407 KNe are produced in this grid. 

Figure~\ref{fig:KNvslimits} shows comparison between the ZTF upper limits and KN models from the adopted BNS (top row) and NSBH (bottom row) grids in $g$ (left), $r$ (middle) and $i$ (right) filters. The distance is optimistically assumed to be at the closest $1\sigma$ end of the distribution provided by LVK, i.e. 269\,Mpc.  Even at this distance, no NSBH model can be ruled out as the resulting KNe are fainter than the upper limits in all filters. In contrast, some KN lightcurves from the BNS grid are brighter than the ZTF limits and are thus ruled out assuming the KN site was imaged during these observations. As shown in this figure, the most constraining data are those from the first observation $\sim1.2$~days after the LVK trigger and, particularly, from the $r$ filter at $\sim20.5$~mag. 

Figure~\ref{fig:corner} shows what regions of the ejecta parameter space are disfavored by our limits, assuming a face-on view of the system. Although no combination of the six ejecta parameters can be completely ruled out, we find that some combinations are clearly disfavored. For instance, high values for the wind ejecta mass and electron fraction are disfavored as they lead to bright KNe. The ZTF upper limits allow to rule out 35\% of these models. We note that models with different viewing angles than face-on, are less constrained in general (see Appendix \ref{appendixModels}).

\section{Joint ZTF and DECam observations}

The region of S250206dm was targeted by numerous instruments, both in the northern and southern hemispheres. In the south, the GW-MMADS survey was able to cover close to 9.3\% of the localization region with the Dark Energy Camera (DECam) \citep{decamGWlei}. In this section we combine the observation of ZTF and DECam to explore the efficiencies and determine how these instruments complement the GW searches. Together, these instruments covered 73.3\% of the GW error region. 

\begin{figure}[t]
    \centering
     \includegraphics[width=0.45\textwidth]{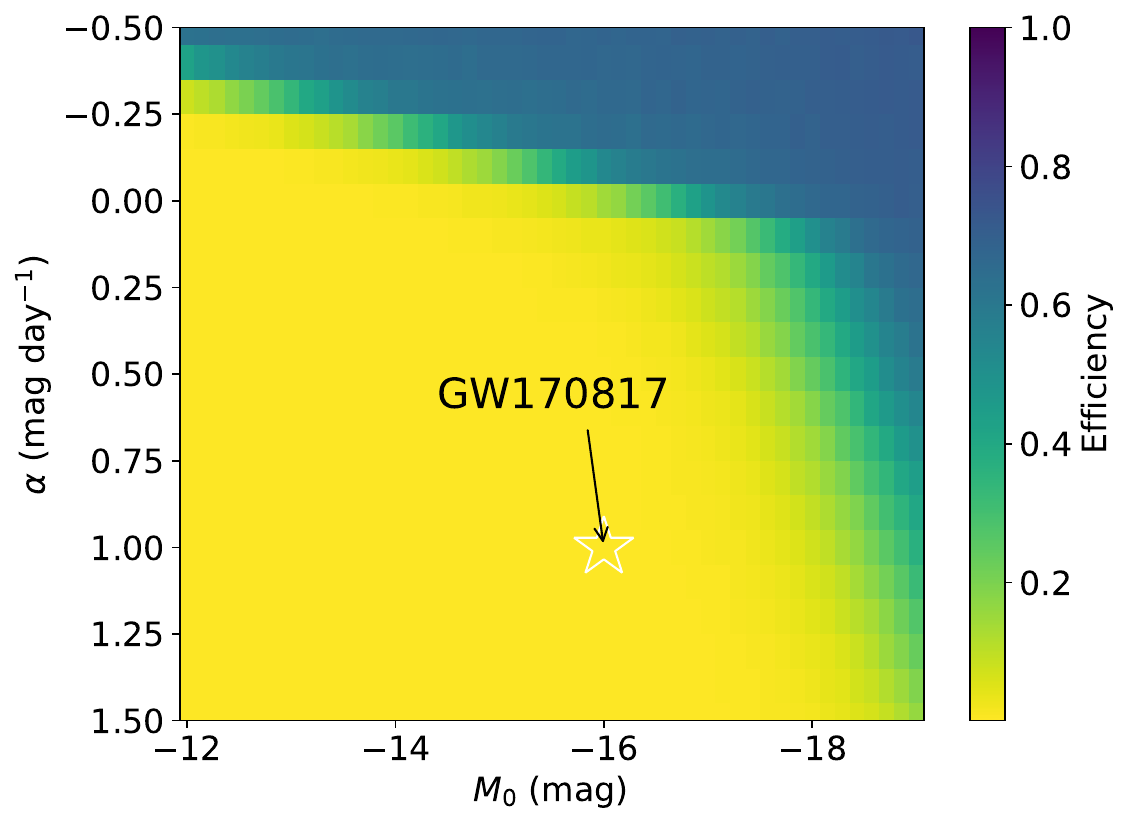}

    \caption{\texttt{simsurvey} recovery efficiency for KNe injected in the S250206dm skymap using a grid of peak absolute magnitude, $M_0$, and evolution rate, $\alpha$, using  ZTF and DECam observations. One of the main differences is that  compared to the ZTF only analysis, the joint observations are more efficient at recovering fainter KN that have a slowly rising behavior ($\alpha < 0$).     }
    \label{fig:ztf+decam_simsurvey}
\end{figure}

\begin{figure*}[t]
    \centering
     \includegraphics[width=0.55\textwidth]{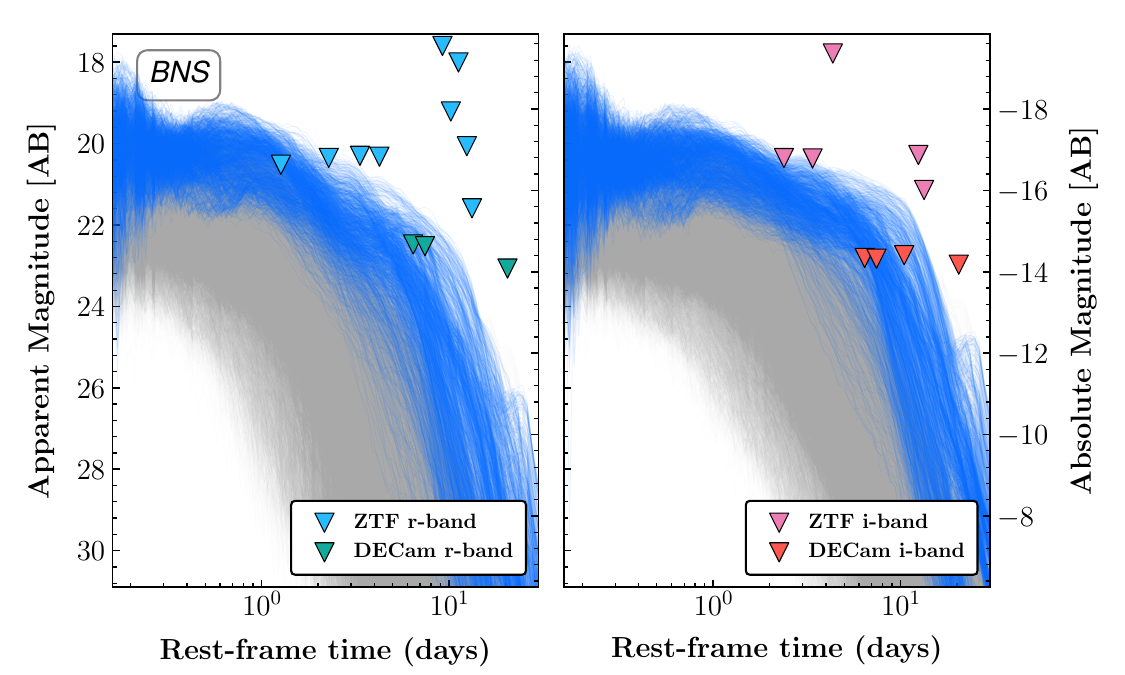}
\includegraphics[width=0.4\textwidth]{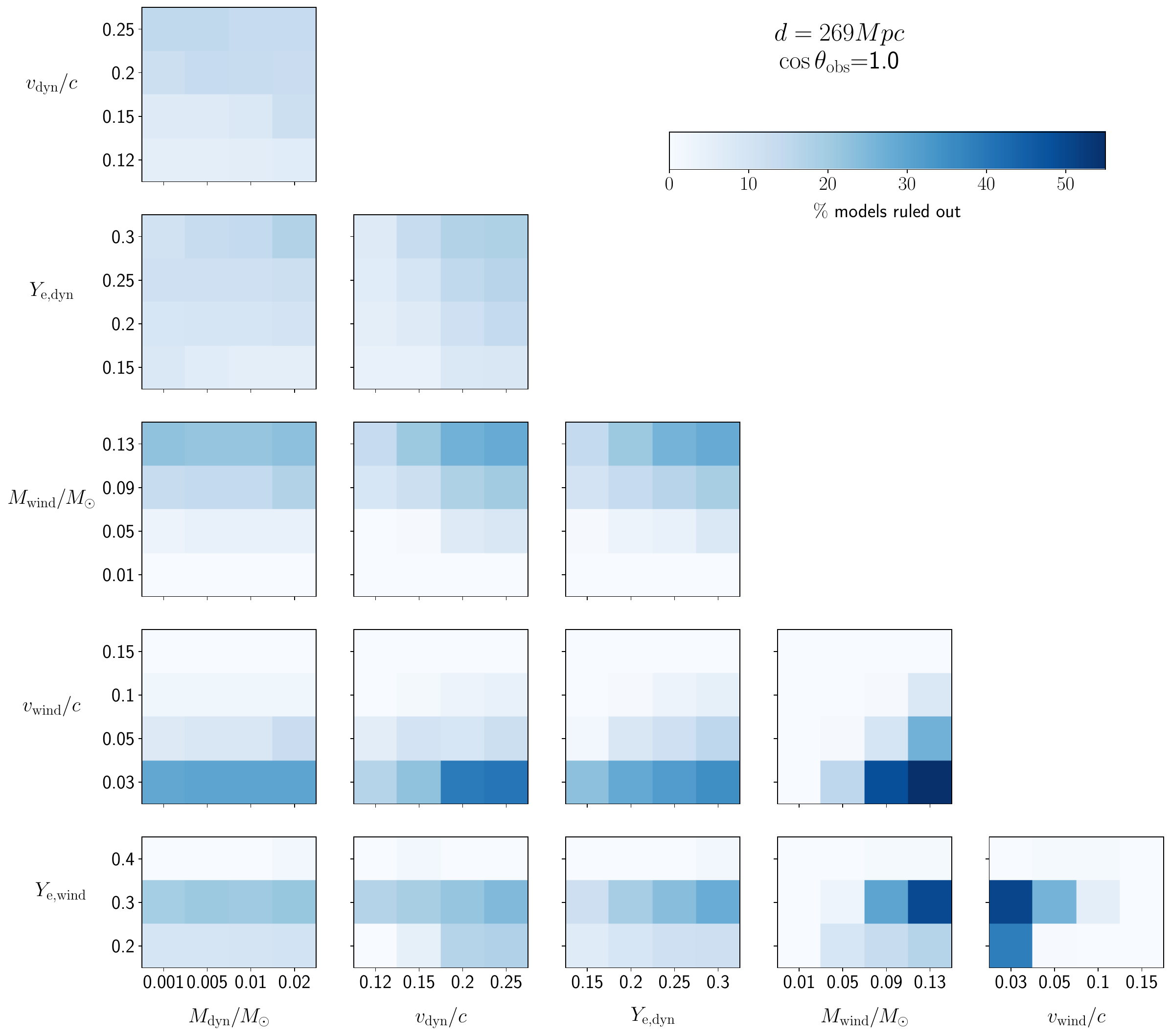}
    \caption{(left) Light-curves of BNS  models pinned at 269 Mpc, the closest 1-sigma from the GW distance distribution. We show the ZTF and DECam median limits as triangles for (left) $r$ band and (center) $i$ band. BNS models were ruled out by both ZTF and DECam at different stages in the lightcuve. (right) Similar to Fig. \ref{fig:corner}, we display a corner plot showing in a colorbar the percentage of BNS models ruled out at a fixed distance of 269 Mpc and at a fixed viewing angle of 0 deg. The maximum number of models ruled out goes up to 55\% using the DECam and ZTF observations combined. }
    \label{fig:ztf+decam}
\end{figure*}

As described in \citet{decamGWlei}, DECam started observations of the GW skymap six days after the GW event and ran for several days, reaching on average 23 mag in the $r$ band (see Fig.~\ref{fig:ztf+decam}).To assess the joint efficiency of using ZTF and DECam synoptic searches, we use \texttt{simsurvey}. Similarly to the ZTF only case, we simulate KNe in the GW skymap and feed \texttt{simsurvey} the observing logs of ZTF and DECam. The KNe are simulated independent of the observing logs and only their recovery rate depends on the executed observations. The DECam observations, although later in time, reach depths comparable to the KN models. These, combined with the ZTF shallower limits from the first few nights, result in a higher combined efficiency at ruling out KNe ($< -15$ mag) that rise ($\alpha <$ 0 mag/day) and bright KNe ($< -16.5$ mag) fading slowly ($\alpha  <$ 0.3 mag/day). For these cases, the combined efficiency is close to 60\% (see Fig.~\ref{fig:ztf+decam_simsurvey}). The efficiency drops to levels similar to the ZTF-only analysis for KNe in the 170817-like parameter space, i.e. $M_0 \approx -16 $ mag and $\alpha = 1$ mag day$^{-1}$.

The \texttt{simsurvey} approach accounts for the fact that the two instruments cover different areas of the sky and reach different depths. In addition to this analysis, we present constraints in the KN model parameter space using the ZTF and DECam limits in tandem. Although this approach assumes joint coverage, which is not the case for S250206dm, we note that for future (and past) events, large-field-of-view instruments are expected to overlap in coverage. In such cases, joint analysis will be possible.
In Fig. \ref{fig:ztf+decam}, we show the $r$- and $i$-band limits of both ZTF and DECam for S250206dm. Under this assumption, the models ruled out by the joint set of upper limits reach 55\% for face-on models with $M_{\rm wind} \approx 0.13\,$ \Msun and  $v_{\rm wind} \approx 0.03\ c$.

\section{Conclusion}\label{sec:conclusion}

We used ZTF to conduct an optical search for the electromagnetic counterpart of the compact binary merger S250206dm. The GW event had a FAR of 1 in 25 years,  a 35\% probability of being a BNS, and a 55\% probability of being an NSBH if astrophysical. In either case, and if astrophysical in origin, the probability of having an NS involved is of 100\%, whereas the probability of having a remnant that could power EM emission is 30\%. 

Due to poor weather during the first night, ZTF only began to observe the region 29 hrs after the GW event. We tiled 68\% of the region, and observed more than 470 sq deg of the latest error region\footnote{\url{https://gracedb.ligo.org/api/superevents/S250206dm/files/Bilby.offline1.multiorder.fits}}. ZTF returned to the region daily for nine days, using a mix of deep, targeted observations and shallower, serendipitous ones from its nominal public survey. We used Fritz, \texttt{emgwcave}, \texttt{nuztf}, and ZTFReST to filter the ZTF stream of alert and found 13 compelling candidates. All these ZTF candidates were ruled out as either spectroscopically or photometrically inconsistent with a KN, located outside of the GW volume, or associated with stellar sources. We analyzed candidates circulated through TNS, and with our follow-up we were able to rule out all but seven candidates. 
Assuming no KN was found, we derived the efficiency of the ZTF searches. Using a Bayesian approach, \texttt{nimbus}, we find that our searches can rule out the presence of KNe with $M_0  <-17.5$ for S250206dm ($P_{nimbus} < 0.4$), while our frequentist approach, \texttt{simsurvey}, shows that ZTF is more efficient at finding KNe in the brighter end ($\sim$60\% efficiency for KN with $M_0 <-17.5$). When analyzing all the ZTF GW follow-up combined, the addition of S250602dm only improves the efficiency of retrieving KNe on the brighter side of the parameter space. These results slightly improve the efficiencies reported in \citet{Ahumada2024}.  

Assuming the KN is in the ZTF footprint, we compared our survey upper limits to KN models generated using the latest version of the 3D Monte Carlo radiative transfer code \texttt{possis}. We present two new model grids: one for BNS mergers and another for NSBH mergers, incorporating updated ejecta properties and nuclear heating rates. The BNS grid includes 3072 models spanning a range of six ejecta parameters and 11 viewing angles, while the NSBH grid comprises 407 models built from 37 parameter combinations based on different binary properties and equations of state. By comparing these models to ZTF upper limits in the $g$, $r$, and $i$ bands, we find that none of the NSBH models are bright enough to be ruled out, whereas some BNS models exceed the observed limits and are therefore disfavored, especially those with higher wind mass and electron fraction. The most constraining observation occurred 1.2 days post-merger in the $r$ band, reaching $\sim$20.5 mag. Overall, while no single combination of BNS parameters can be entirely excluded, our results disfavour certain regions of parameter space, particularly for face-on viewing angles.

Finally, we performed a joint analysis of the ZTF and GW-MMADS DECam observations from \citep{decamGWlei} of S250206dm to assess their combined efficiency in detecting KNe. ZTF began imaging earlier than DECam, although it provided shallower limits compared to DECam ($\sim$23 mag in the $r$ band). Using the \texttt{simsurvey} framework, we simulated KNe across the GW skymap and evaluated recovery efficiencies based on the actual observing logs from both instruments. The joint dataset improves efficiency in detecting faint, slowly rising KNe (with $\alpha < 0$ mag/day) and bright, slowly fading KNe (with $\alpha > 0.3$ mag/day), reaching up to 60\% recovery in those regimes, significantly better than ZTF alone. Although DECam and ZTF covered different regions for this event, we also explored constraints assuming overlap in coverage, finding that up to 55\% of BNS models could be ruled out at 269 Mpc, particularly those with high wind mass and low wind velocity. These results highlight the value of combining wide-field optical datasets for future joint GW-KN searches, while also emphasizing that early observations since the GW event could help strongly constrain the KN parameter space.

\section{Acknowledgements}

M.M.K., S.A. and T.A. acknowledge generous support from the David and Lucile Packard Foundation.
M. B. acknowledges the Department of Physics and Earth Science of the University of Ferrara for the financial support through the FIRD 2024 grant
M.~W.~C, A.T., A.~S. and T.B. acknowledges support from the National Science Foundation with grant number PHY-2010970.
A.T. acknowledges support from the National Science Foundation with grant number PHY-2308862 and PHY-2117997
A. P. and L. H. acknowledges support by National Science Foundation Grant No. 2308193.
I. A. acknowledges support from the National Science Foundation Award AST 2505775 and NASA grant 24-ADAP24-0159 
A. Singh acknowledges support from the Knut and Alice Wallenberg Foundation through the ``Gravity Meets Light" project.
C.M.C. acknowledges support from UKRI with grant numbers ST/X005933/1 and ST/W001934/1
A.~S. acknowledges support from the National Science Foundation with grant number PHY-2010970.
M. B. is supported by a Student Grant from the Wübben Stiftung Wissenschaft.
G.C.A. thanks the Indian National Science Academy for support under the INSA Senior Scientist Programme.
A.T. acknowledges support from the National Science Foundation with grant number PHY-2308862 and PHY-2117997

Based on observations obtained with the Samuel Oschin Telescope 48-inch and the 60-inch Telescope at the Palomar Observatory as part of the Zwicky Transient Facility project. ZTF is supported by the National Science Foundation under Grants No. AST-1440341, AST-2034437, and currently Award \#2407588. ZTF receives additional funding from the ZTF partnership. Current members include Caltech, USA; Caltech/IPAC, USA; University of Maryland, USA; University of California, Berkeley, USA; University of Wisconsin at Milwaukee, USA; Cornell University, USA; Drexel University, USA; University of North Carolina at Chapel Hill, USA; Institute of Science and Technology, Austria; National Central University, Taiwan, and OKC, University of Stockholm, Sweden. Operations are conducted by Caltech's Optical Observatory (COO), Caltech/IPAC, and the University of Washington at Seattle, USA.

SED Machine is based upon work supported by the National Science Foundation under Grant No. 1106171. 

The ZTF forced-photometry service was funded under the Heising-Simons Foundation grant \#12540303 (PI: Graham).

The Gordon and Betty Moore Foundation, through both the Data-Driven Investigator Program and a dedicated grant, which provided critical funding for SkyPortal.

We acknowledge the support from the National Science Foundation GROWTH PIRE grant No. 1545949. 

This work used Expanse at the San Diego Supercomputer Cluster through allocation AST200029 -- ``Towards a complete catalog of variable sources to support efficient searches for compact binary mergers and their products'' from the Advanced Cyberinfrastructure Coordination Ecosystem: Services \& Support (ACCESS) program, which is supported by National Science Foundation grants \#2138259, \#2138286, \#2138307, \#2137603, and \#2138296.
This research has made use of the NASA/IPAC Extragalactic Database (NED), which is funded by the National Aeronautics and Space Administration and operated by the California Institute of Technology.

The Liverpool Telescope is operated on the island of La Palma by Liverpool John Moores University in the Spanish Observatorio del Roque de los Muchachos of the Instituto de Astrofisica de Canarias with financial support from the UK Science and Technology Facilities Council.

This work relied on the use of HTCondor via the IGWN Computing Grid hosted at the LIGO Caltech computing clusters.

Some of the data presented herein were obtained at Keck Observatory, which is a private 501(c)3 non-profit organization operated as a scientific partnership among the California Institute of Technology, the University of California, and the National Aeronautics and Space Administration. The Observatory was made possible by the generous financial support of the W. M. Keck Foundation. The authors wish to recognize and acknowledge the very significant cultural role and reverence that the summit of Maunakea has always had within the Native Hawaiian community. We are most fortunate to have the opportunity to conduct observations from this mountain.

The GROWTH India Telescope (GIT) is a 70-cm telescope with a 0.7-degree field of view, set up by the Indian Institute of Astrophysics (IIA) and the Indian Institute of Technology Bombay (IITB) with funding from DST-SERB and IUSSTF. It is located at the Indian Astronomical Observatory (Hanle), operated by IIA. We acknowledge funding by the IITB alumni batch of 1994, which partially supports the operations of the telescope. Telescope technical details are available at https://sites.google.com/view/growthindia/

This paper contains data obtained at the Wendelstein Observatory of the Ludwig-Maximilians University Munich. We thank Christoph Ries and Michael Schmidt for obtaining the observations. Funded in part by the Deutsche Forschungsgemeinschaft (DFG, German Research Foundation) under Germany’s Excellence Strategy – EXC-2094 – 390783311.

\bibliography{myreferences}

\appendix

\section{ZTF coverage of multi-wavelength and multi-messenger candidates} \label{appendixCoverage}

\begin{figure*}[h]
    \centering
    
    \includegraphics[width=0.45\textwidth]{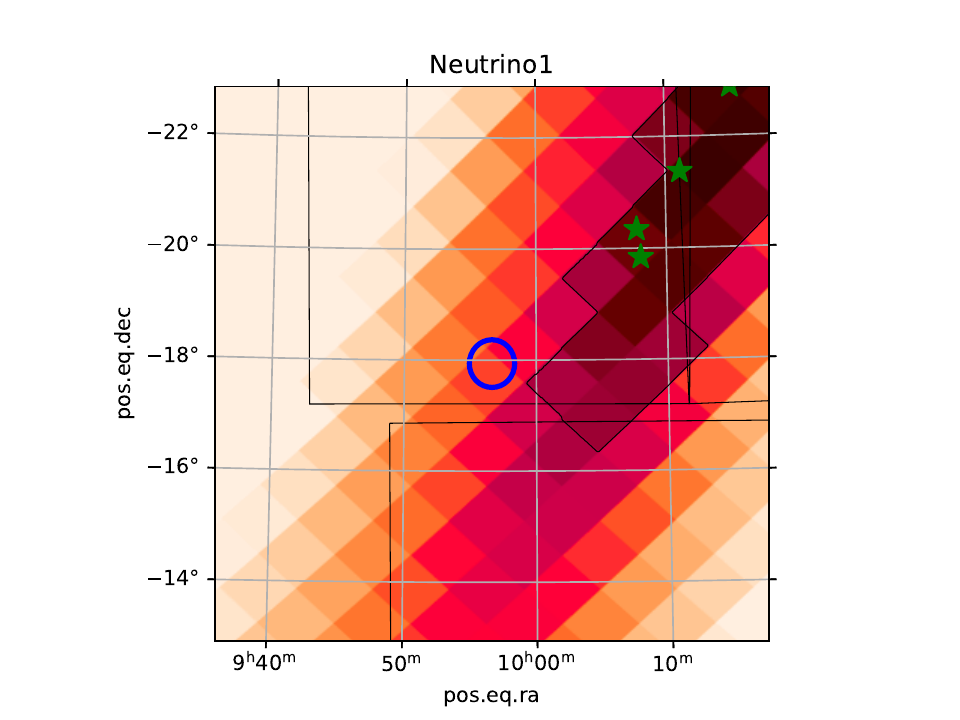}
    \includegraphics[width=0.45\textwidth]{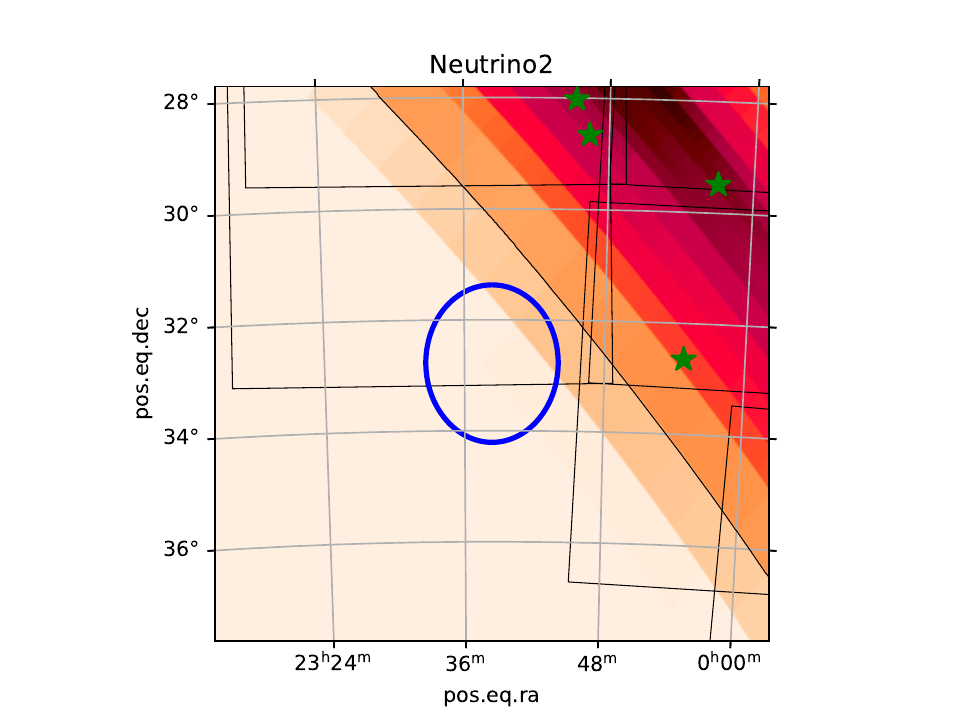}
    \includegraphics[width=0.45\textwidth]{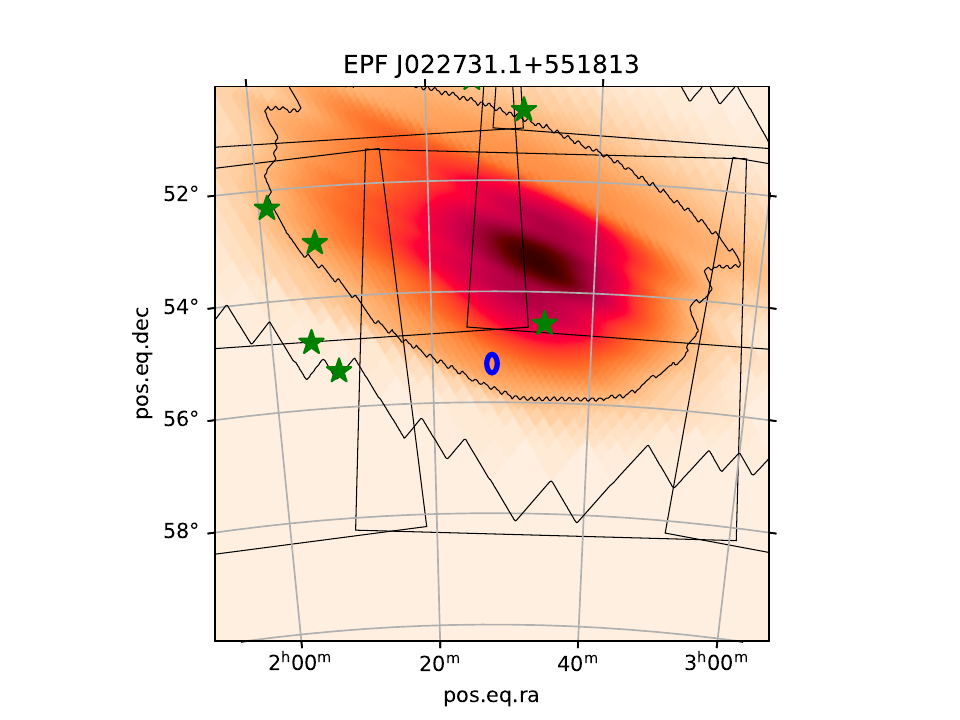}
    \includegraphics[width=0.45\textwidth]{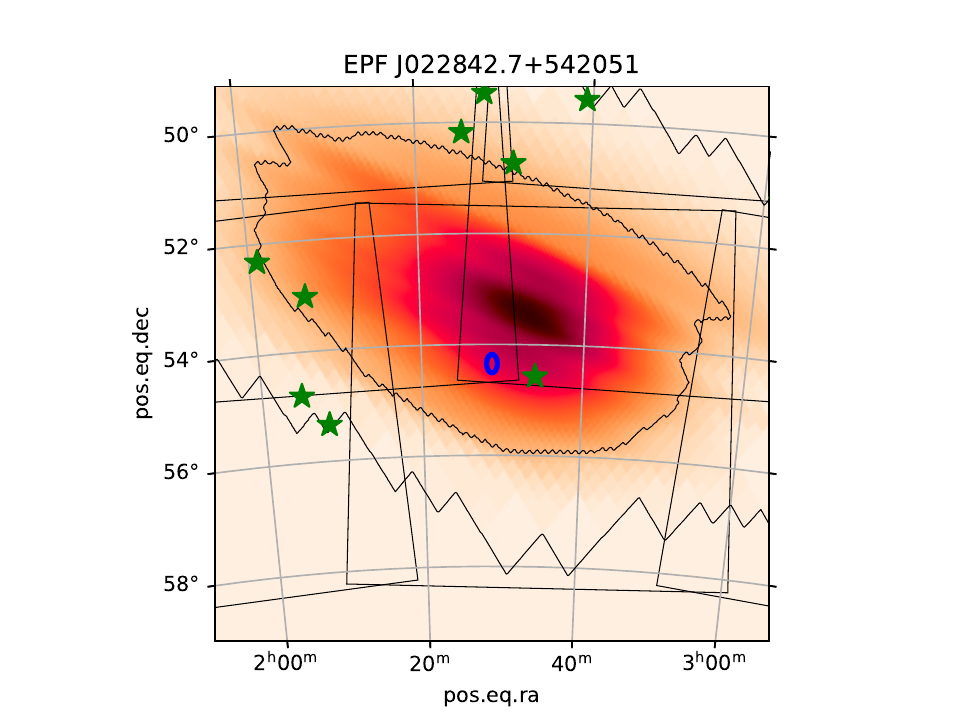}
    \includegraphics[width=0.45\textwidth]{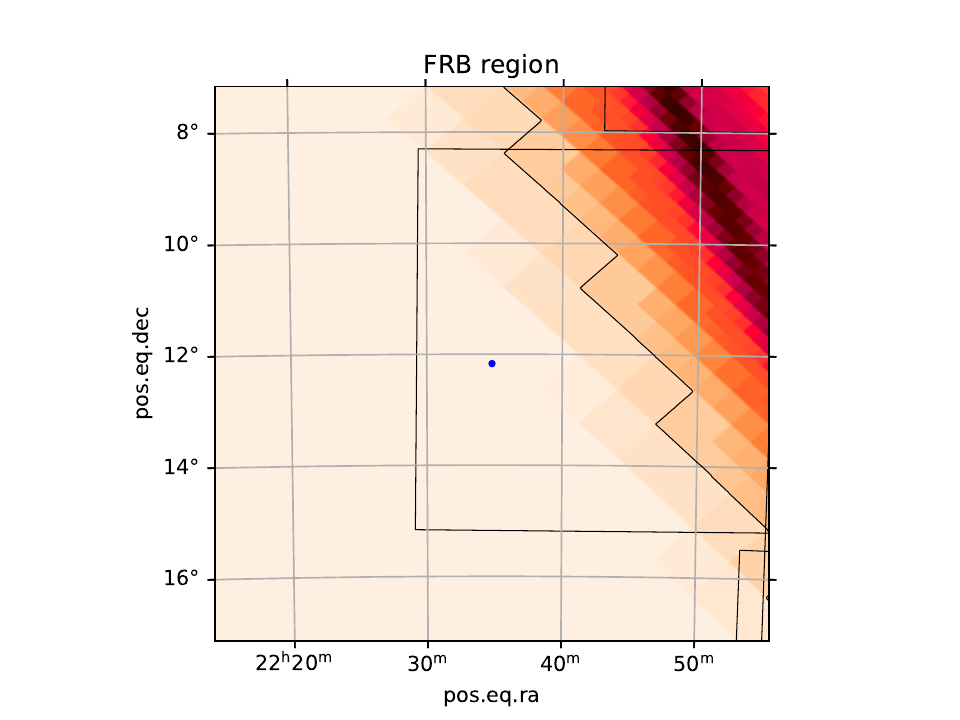}
    \caption{Localization of the high-significance event S250206dm, overlaid with the ZTF tiles and the 90\% probability contour. The blue region shows the center of different multi-wavelength and multi-messenger transients detected coincident in time with S250206dm. The green stars represent the transients circulated on TNS, accessible from Palomar, while the white stars are candidates inaccessible from Palomar.  }
    \label{fig:map_otherMMA}
\end{figure*}

\section{Follow-up details} \label{appendixFollowup}

\subsection{Photometric Follow-up} \label{subsec:imaging}

\paragraph{Palomar 60-inch} We acquired photometric data utilizing the Spectral Energy Distribution Machine (SEDM; \citealt{BlNe2018,Rigault2019,Kim2022SEDM}) mounted on the Palomar 60-inch telescope. The SEDM is a low resolution (R $\sim$ 100) integral field unit spectrometer with a multi- band ($ugri$) Rainbow Camera (RC). The follow-up request process is automated and can be initiated through \texttt{Fritz}. Standard requests typically involved 180 s exposures in the $g$-, $r$-, and $i$-bands, however, it can be customized and for some transients we used 300 s exposures. The data undergoes reduction using a Python-based pipeline, which applies standard reduction techniques and incorporates a customized version of FPipe (Fremling Automated Pipeline; \citealt{FrSo2016}) for image subtraction.

    \paragraph{GROWTH-India Telescope}
We utilized the 0.7-meter robotic GROWTH-India Telescope (GIT) \citep{Kumar2022git}, located in Hanle, Ladakh. It is equipped with a 4k back-illuminated camera that results in a 0.82 deg$^2$ field of view. Data reduction is performed in real-time using the automated GIT pipeline. Photometric zero points were determined using the PanSTARRS catalogue, and PSF photometry was conducted with PSFEx \citep{Bertin2010}. In cases where sources exhibited a significant host background, we performed image subtraction using \texttt{pyzogy} \citep{Guevel2017Pyzogy}, based on the ZOGY algorithm \citep{Zackay_2016}.

\paragraph{Liverpool Telescope} The images acquired with the Liverpool Telescope (LT) were taken using the IO:O \citep{steele2004liverpool} camera equipped with the Sloan \textit{griz} filterset. These images underwent reduction through an automated pipeline, including bias subtraction, trimming of overscan regions, and flat fielding. Image subtraction occurred after aligning with a PanSTARRS template, and the final data resulted from the analysis of the subtracted image.

\paragraph{Fraunhofer Telescope at Wendelstein Observatory}
We conducted follow-up observations using the Three Channel Imager (3KK; \citealt{2016SPIE.9908E..44L}) on the 2.1\,m Fraunhofer Telescope at Wendelstein Observatory (FTW; \citealt{2014SPIE.9145E..2DH}), located on Mt.~Wendelstein at the northern edge of the Alps. The 3KK imager enables simultaneous imaging in three channels. For our observations, we configured the blue, red, and near-infrared channels with the $r$, $i$, and $J$ bands, respectively. We applied standard data reduction techniques to derive magnitudes for the follow-up. For details on the 3KK data reduction see \citealt{Busmann2025} and \citealt{Gossl2002}.

\subsection{Spectroscopic Follow-up} \label{subsec:spectroscopy}

\paragraph{Keck I} 
We obtained spectra with LRIS on the Keck I telescope, using the 600/4000 grism on the blue side and the 600/7500 grating on the red side. This setup provided wavelength coverage from 3139-5642 \AA\, in the blue and 6236–9516 \AA\, in the red. Both arms were exposed for 600 seconds. The data were reduced using LPipe \citep{Perley2019}, with BD+28 serving as the flux calibrator. To ensure accurate relative flux calibration between the red and blue sides, we scaled the spectra by matching synthetic photometry to observed transient colors.

\paragraph{Palomar 200-inch} We observed ZTF candidates using the Palomar 200-inch Next Generation Palomar Spectrograph (NGPS). The setup configuration used a 1.5 arcsec slitmask, a D55 dichroic, a blue grating of 600/4000, and a red grating of 316/7500. We applied a custom PyRAF DBSP reduction pipeline \citep{BeSe2016} to process and reduce our data.

\paragraph{Nordic Optical Telescope} We obtained spectra of potential candidates and their host galaxies using the Alhambra Faint Object Spectrograph and Camera (ALFOSC) mounted on the Nordic Optical Telescope. We used Grism \#4 with a 1 arcsec slit. The spectra were reduced in a standard manner using a custom fork of \texttt{PypeIt} 
\citep{pypeit:zenodo,pypeit:joss_arXiv,pypeit:joss_pub}

\section{Candidate analysis} \label{appendixCandidates}

In this section, we describe the candidates found in the 95\% region of S250206dm, as well as the reasons to rule them out.

\subsection{Candidates found with ZTF}

\paragraph{AT2025bcq} Originally found as ZTF25aafiwsg, this transient was discovered at $g = 20.38$ mag, 2.3 days after the GW alert. The source is consistent with a star in the Gaia DR3 \citep{GaiaDR3} within 2 arcsec. We ruled it out as it is a stellar source.

\paragraph{AT2025bda} Discovered as ZTF25aaffynz, this transient was at $g = 20.88$ mag and rose to $g = 19.5$ mag in 11 days. We ruled it out due to its slow evolution. We note the potential host galaxy has a photometric redshift of $z_{ph}=0.789$ based on the Legacy Survey measurements. 

\paragraph{AT2025bay} Discovered as ZTF25aaffyzc, at $r = 20.51$ mag 1.2 days after the trigger, we conducted prompt spectroscopic observations with Keck I/LRIS to classify it as a Type Ia SN at $z_{spec}=0.19$ \cite{GCN2025LRIS}. Additionally, the source showed a slow evolution in the $r$ band, inconsistent with a KN.

\paragraph{AT2025brm}
Originally identified as ZTF25aafnncn, this source was discovered at $g = 20.41$ mag and 4.2 days after the merger. The source is next to a stellar source and forced photometry in the ZTF data shows a 4.5 $\sigma$ detection 30 days before the merger, which we interpret as previous activity. We rule it out as an old, unrelated source. 

\paragraph{AT2025ben}
ZTF25aafjwbr was discovered at $r = 19.44$ mag and 17 hrs after the merger. It is located in the nucleus of the elliptical galaxy WISEA J234517.01+280121.7. The source slowly rose to a peak of $r = 18.8$ mag in 3 days, thus we ruled it out based on the slow photometric evolution. 

\paragraph{AT2025bro} Detected as ZTF25aaffxsx at $r = 20.08$ mag, this source is in the outskirts of a galaxy with photometric redshift of $z_{phot} = 0.146$. This puts the transient potentially outside of the GW volume of interest. Additionally, photometric monitoring of the source showed no evolution in the first 3 days after discovery in the $r$ band, thus we ruled out this source based on its slow evolution. 

\paragraph{AT2025brn} First discovered as ZTF25aafnnbw, this transient is hostless and the first detection was 4 days after the GW event and the candidate passed our filters as there was a forced photometry detection on day 2 at $r = 19.2$ mag. We rule out this source due to its slow evolution.  

\paragraph{AT2025bcx} Detected with ZTF as ZTF25aafisft at $r = 20.9$ mag 1.2 days after the GW event, it remained active without evolving. We rule it out due to its slow evolution. 

\paragraph{AT2025bcw} Discovered as ZTF25aafgakh at $r = 20.67$ mag, this source is 1 arcsec from a red, point source. Forced photometry from ZTF revealed activity during the 50 days prior to the GW event, and the continuous monitoring with ZTF and LT did not show any evolution. Thus we reject this source as both old and not evolving. 

\paragraph{AT2025brp} Originally as ZTF25aafnmng, it was detected 1.2 days after the GW event at $r = 20.5$ mag. The source is in the outskirts of an elliptical galaxy with LS photometric redshift of $z_{phot} = 0.61$, potentially placing it outside of the GW volume. The continuous monitoring with ZTF showed no evolution in 12 days. Thus we reject it due to its lack of evolution.

\paragraph{AT2025brl} Initially discovered 4 days after the GW event, ZTF25aafnndi showed 2 previous detections (forced photometry) 2 days after the GW event. After further inspection, the source is classified as stellar activity due to its proximity (1.9 arcsec) to a star with $g = 15.7$ mag. 

\paragraph{AT2025bcr} Initially detected as ZTF25aafiske, this source is located between a galaxy and a point source. Monitoring with the LT showed no evolution over a period of four days, thus ruling it out.

\paragraph{AT2025cdh} Originally ZTF25aagfolh, this source has shown a photometric evolution similar to  a supernova, rising from $g = 21$ mag to $g = 19$ mag in 15 days and showing a slow decay. We therefore ruled it out due to its slow evolution.

\subsection{Follow-up of candidates from other facilities}

\paragraph{AT2025bmq} Originally detected by us, this source is associated with WISEA J023404.21+543420.9, at a redshift of 0.08. Forced photometry of ZTF revealed an active source with bursts between 45 and 10 days before the GW trigger. Our Wendelstein/FTW data showed a $r=20.03$ mag source 11 days after the merger. Thus we ruled out this candidate as old with respect to the GW trigger, and due to its slow evolution. 
\paragraph{AT2025bew} Discovered by Pan-STARRS at $i = 20.86$ mag, this source showed no significant evolution in our Wendelstein and LT images, as it was detected at $i = 21.6$ mag after 10 days. We rule out this candidate due to its slow evolution rate of 0.07 mag/day.
\paragraph{AT2025bev} This source was discovered by Pan-STARRS, and also detected in the ZTF stream. The monitoring with ZTF, Wendelstein and LT showed a slowly rising source that peaked at $r = 20.7$ mag 5 days after the GW event, and slowly decayed at a 0.03 mag/day rate. We ruled out this source based on its slow evolution. 
\paragraph{AT2025bbn} Originally discovered by Pan-STARRS, the source is coincident with a point source we classified as a star using Gaia data. Additionally, ZTF forced photometry shows previous activity starting 30 days prior to the GW event.
\paragraph{AT2025bbm} Similarly, this source was announced by Pan-STARRS, and has a match to a stellar source in the Gaia database. 
\paragraph{AT2025bbo} Detected by Pan-STARRS, this nuclear source associated with WISEA J013717.29+454331.9, with a photometric redshift of 0.062. The monitoring with LT showed no evolution 4 days after the Pan-STARRS detection. We acquired a spectrum with NIRES, which showed only emission lines at a common redshift of 0.07 \citep{AT2025bboNIRES}, the redshift of the host. This candidate was later retracted as an artifact \citep{at25bboretraction}. 
\paragraph{AT2025bex} Discovered by Pan-STARRS, this source is coincident with a point source with $B = 21.98$ mag. ZTF forced photometry shows a $g$-band detection 17 days before the GW event. Additionally, the monitoring with Wendelstein and LT showed a rising source 10 days after the GW event, inconsistent with a KN. 
\paragraph{AT2025bbt} This source, originally detected by Pan-STARRS, has a ZTF forced photometry detection 45 days prior to the GW event. 
\paragraph{SN2025bpv} This source was originally detected by GOTO. Our NOT spectra showed SN-like features, and fit a SN Ia template at a redshift of 0.0688. This source appears as ZTF25aagacqk in the ZTF data stream. 
\paragraph{AT2025baw} This source first discovered by Pan-STARRS is coincident with a galaxy with LS photometric redshift of 0.2495. It has multiple forced photometry detections in the ZTF data stream. We ruled out this source as old.  
\paragraph{AT2025bai} This source was first detected by Pan-STARRS, near the nucleus of a spiral galaxy with LS photometric redshift of 0.0991. The source had multiple ZTF detections before the GW event. 
\paragraph{SN2025bag} This source was first detected by Pan-STARRS, and it was additionally detected by ZTF as ZTF25aafhecq. The source has been classified as a SN Ia, and has a lightcurve extending for over 60 days.
\paragraph{AT2025azm} This source was first detected by SAGUARO, in the nucleus of a galaxy with LS photometric redshift of 0.07. This source has ZTF forced photometry detections that indicate this source was active 150 days before the GW event.
\paragraph{AT2025azn} This source was detected originally by SAGUARO, associated with WISEA J023917.49+493420.9, an elliptical galaxy at a photometric redshift of 0.0804. The NGPS spectrum of this galaxy shows lines at a common redshift of 0.35, placing the candidate outside of the GW volume. 
\paragraph{AT2025bcc} This source was first detected by Pan-STARRS as a hostless transient. Our observations with Wendelstein and LT do not show any sources down to a 5$\sigma$ of $r > 23.5$ mag, $i > 23.1$ mag, and $J> 21.7$ AB, 4 days after the GW event. We cannot rule out this transient under any of our rejection criteria. 
\paragraph{AT2025bey} This candidate was detected by Pan-STARRS in association with an elliptical galaxy. Our LT monitoring shows $r > 21.82$ mag, 4 days after the GW event. We cannot rule out this source under any of our rejection criteria. 
\paragraph{AT2025baf} This source was detected by Pan-STARRS in association with a galaxy with a photometric redshift from LS of 0.6429, putting it outside the GW volume. We disfavor a GW association, though we note this conclusion is based solely on photometric redshift. 
\paragraph{AT2025bah} This source was originally detected by Pan-STARRS, and coincident with a galaxy with an LS photometric redshift of 0.1586. This would put the transient outside the GW volume. 
\paragraph{AT2025bbp} This source, discovered by Pan-STARRS, is associated to a galaxy with an LS photometric redshift of 0.1121, putting the target outside the volume of the GW event. 
\paragraph{AT2025bam} Detected by Pan-STARRS, this source is coincident with a galaxy with a LS photometric redshift of 0.5661, putting the source outside of the GW volume. 
\paragraph{AT2025bce} This transient was first detected by Pan-STARRS. Our LT observations 7 days after the merger found no transient up to a limiting magnitude of $r>19.9$ mag. Similarly our Wendelstein data showed no source down to limiting magnitudes of $r>23.6$ mag, $i>23.2$ mag, $J>21.8$ AB mag.

\section{Forecasts of Kilonova Lightcurves for ZTF Bands}\label{appendixML}

We have developed a machine learning model using bidirectional long-short-term memory (LSTM) networks to forecast KN lightcurves based on low-latency alerts from IGWN, focusing on ZTF bands.

   We use publicly available simulated observation data from the IGWN User Guide\footnote{\url{https://emfollow.docs.ligo.org/userguide/capabilities.html}}, including 17,009 binary neutron star (BNS) and 3,148 neutron star black hole (NSBH) events\footnote{\url{https://zenodo.org/records/12696721}} that exceed the IGWN detection threshold (SNR > 8) \citep{Kiendrebeogo23}. Sky maps are generated for simulated BNS and NSBH mergers using the Bayestar localization code \citep{Singer2016}, extracting parameters such as sky position, distance and the 90\% localization area. Since the simulations provide only SNR, we map SNR to FAR using a large set of BNS injections to estimate $P_{\text{Astro}}$.
   
   We calculate probabilities from the EM-bright\footnote{\url{https://git.ligo.org/emfollow/em-properties/em-bright}} classification, which estimates the likelihood that a merger involves at least one neutron star (HasNS), produces ejecta (HasRemnant), or includes a neutron star in the 3–5 solar mass range (HasMassGap).
  
   Lastly, we use the NMMA\footnote{\url{https://nuclear-multimessenger-astronomy.github.io/nmma/fitting.html}} framework, which incorporates the POSSIS model (\citealt{Bulla2019,Diet2020}), to generate lightcurves for each simulated compact binary coalescence. 
   
   We train a machine learning model to forecast KN lightcurves using features such as area (90\%), distance, longitude, latitude, HasNS, HasRemnant, HasMassGap, and $P_{\text{Astro}}$. To ensure consistency in scale and measurement units across the training dataset, we apply the RobustScaler. The data is split into a 70/30 ratio for training and testing. In the test set, we achieve a mean squared error (MSE) of 0.24 in the g-band and 0.16 in the r-band.
   
   For the 250206dm event, we collect the necessary features (FAR, area (90\%), distance, longitude, latitude, HasNS, HasRemnant, HasMassGap, and $P_{\text{Astro}}$) and use them to forecast the KN
   lightcurve with our machine learning model. This analysis was performed offline, after the manual vetting of the candidates was complete.

None of the candidates is consistent with our forecasts for the 250206dm event. The model's performance is further improved by integrating dynamical and wind ejecta as additional features, enhancing its ability to capture KN lightcurves, and significantly improving performance, reducing the MSE of 0.16 in the g-band and 0.11 in the r-band. However, even with these improvements, none of the candidates matched our predictions.

\startlongtable
\begin{deluxetable*}{ccccccccc}
\tablewidth{0.99\textwidth}
\tablecaption{Candidates from ZTF \label{table:ztfcand}}
\tablehead{
\colhead{Team} & 
\colhead{AT name} & 
\colhead{RA} & 
\colhead{DEC} & 
\colhead{$\Delta t$} & 
\colhead{Discovery mag.} & 
\colhead{C.R.} & 
\colhead{Redshift} &
\colhead{Rejection criterion}\\
\colhead{} & 
\colhead{} & 
\colhead{[deg]} &
\colhead{[deg]} &
\colhead{(days after GW)} & 
\colhead{(AB magnitude)} &
\colhead{} &
\colhead{} &
\colhead{}
}
\startdata
ZTF & AT2025bcq & 36.312650 & 50.177961 & 2.36 & g = 20.38 mag & 0.57 & -- & stellar \\
ZTF & AT2025bda & 354.338535 & 22.979038 & 1.23 & g = 19.92 mag & 0.66 & $z_{ph}=0.789$ & slow evolution \\
ZTF & AT2025bay & 7.504874 & 37.168325 & 1.24 & r = 20.51 mag & 0.69 & $z_{spec}=0.19$ & SN Ia \\
ZTF & AT2025brm & 359.267230 & 29.487675 & 4.23 & g = 20.41 mag & 0.73 & -- & old \\
ZTF & AT2025ben & 356.321031 & 28.022989 & 3.22 & r = 19.44 mag & 0.76 & $z_{ph}= 0.074$ & slow evolution\\
ZTF & AT2025bro & 356.597145 & 28.657444 & 1.21 & r = 20.08 mag & 0.77 & $z_{ph}= 0.146$ & slow evolution\\
ZTF & AT2025brn & 2.037106 & 32.845540 & 4.23 & g = 20.34 mag & 0.79 & -- & slow evolution \\
ZTF & AT2025bcx & 7.352266 & 38.690377 & 1.24 & r = 20.75 mag & 0.81 & -- & slow evolution\\
ZTF & AT2025bcw & 21.331975 & 45.271967 & 1.25 & r = 20.67 mag & 0.84 & -- & old \\
ZTF & AT2025brp & 358.691917 & 32.650467 & 1.22 & r = 20.50 mag & 0.87 & $z_{ph}= 0.61$ & slow evolution \\
ZTF & AT2025brl & 1.907439 & 27.964044 & 4.23 & g = 19.96 mag & 0.93 & -- & stellar \\
ZTF & AT2025bcr & 20.491312 & 48.970425 & 2.28 & r = 20.59 mag & 0.93 & -- & slow evolution \\
ZTF & AT2025cdh & 156.656672 & -25.498773 & 11.51 & g = 19.16 mag & 0.95 & -- & slow evolution \\
\enddata
\end{deluxetable*}

\startlongtable
\begin{deluxetable*}{ccccccccc}
\tablewidth{0.99\textwidth}\tablecaption{Candidates circulated on TNS accessible from Palomar}\label{table:cad_notZTF}
\tablehead{
\colhead{Team} & 
\colhead{AT name} & 
\colhead{RA} & 
\colhead{DEC} & 
\colhead{$\Delta t$} & 
\colhead{Discovery mag.} & 
\colhead{C.R.} & 
\colhead{Redshift} &
\colhead{Rejection criterion}\\
\colhead{} & 
\colhead{} & 
\colhead{[deg]} &
\colhead{[deg]} &
& 
\colhead{(AB magnitude)} &
\colhead{} &
\colhead{} &
\colhead{}
}
\startdata
WL-GW & AT2025bmq & 38.517583 & 54.572472 & 2.01 & i = 20.00 mag & 0.16 &  $z_{ap}= 0.08$ & old  \\  
Pan-STARRS & AT2025bew & 31.571696 & 53.010387 & 1.42 & r = 20.86 mag & 0.44 &  -- & slow evolution  \\  
Pan-STARRS & AT2025bev & 30.246246 & 52.324672 & 1.41 & r = 21.35 mag & 0.51 &  --&  slow evolution  \\  
Pan-STARRS & AT2025bbn & 37.788441 & 50.735331 & 1.37 & r = 20.19 mag & 0.52 &  --& stellar  \\  
Pan-STARRS & AT2025bbm & 36.962653 & 49.475908 & 1.37 & r = 20.03 mag & 0.72 & -- & stellar \\  
SAGUARO & AT2025azm & 2.031278 & 32.565750 & 0.31 & Clear = 20.10 mag & 0.77 & $z_{ph}=0.074$ & old \\  
Pan-STARRS & AT2025bbo & 24.321968 & 45.725504 & 1.39 & r = 20.05 mag & 0.86 & -- & artifact \\  
SAGUARO & AT2025azn & 39.822573 & 49.572528 & 0.31 & Clear = 19.97 mag & 0.87 & $z_{spec}=0.35$ & outside volume \\  
Pan-STARRS & AT2025bex & 24.462509 & 45.339600 & 3.35 & i = 20.80 mag & 0.88 & -- & rising LT \\  
Pan-STARRS & AT2025bbt & 45.469074 & 51.129734 & 1.42 & r = 18.64 mag & 0.88 & -- & old \\  
GOTO & SN2025bpv & 156.224083 & -27.081777 & 0.69 & L = 20.19 mag & 0.93 & $z_{spec}=0.068$ & SN Ia \\  
Pan-STARRS & AT2025baw & 153.431093 & -24.613802 & 0.54 & r = 20.70 mag & 0.93 & $z_{ph}=0.24$ & old \\  
Pan-STARRS & AT2025bai & 151.935085 & -20.342932 & 0.54 & r = 19.78 mag & 0.94 & $z_{ph}=0.09$ & old \\  
Pan-STARRS & SN2025bag & 154.613735 & -26.732906 & 0.52 & r = 17.51 mag & 0.94 & $z_{spec}=0.03$ & SN Ia  \\ 
\hline
\multicolumn{9}{ l }{Candidates not ruled out}   \\
\hline
Pan-STARRS & AT2025bcc & 32.029016 & 55.342481 & 1.44 & r = 19.57 mag & 0.88 & -- & undefined \\   
Pan-STARRS & AT2025bey & 25.754872 & 45.463406 & 4.33 & z = 20.13 mag & 0.89 & -- & undefined \\   
Pan-STARRS & AT2025bbp & 153.782352 & -22.881593 & 1.54 & r = 20.77 mag & 0.92 & $z_{ph}=0.11$ & likely outside volume \\  
Pan-STARRS & AT2025bah & 152.779896 & -21.374558 & 0.53 & r = 18.69 mag & 0.92 & $z_{ph}=0.15$ &  likely outside volume \\  
Pan-STARRS & AT2025bam & 152.008807 & -19.840772 & 0.55 & r = 20.69 mag & 0.94 & $z_{ph}=0.56$ & likely outside volume  \\  
Pan-STARRS & AT2025bce & 31.232391 & 54.795342 & 1.40 & r = 19.80 mag & 0.95 & -- & undefined \\  
Pan-STARRS & AT2025baf & 158.506625 & -29.746992 & 0.52 & r = 19.33 mag & 0.95 & $z_{ph}=0.64$ & likely outside the volume \\ 
\enddata
\end{deluxetable*}

\newpage
\section{Candidates not discussed in this paper}
\startlongtable
\begin{deluxetable*}{ccccccc}
\tablewidth{0.99\textwidth}

\tablecaption{Candidates not accessible from Palomar}
\tablehead{
\colhead{Team} & 
\colhead{AT name} & 
\colhead{RA} & 
\colhead{DEC} & 
\colhead{$\Delta t$} & 
\colhead{Discovery mag.} & 
\colhead{Credible Region}  
\\
\colhead{} & 
\colhead{} & 
\colhead{[deg]} &
\colhead{[deg]} &
& 
\colhead{(AB magnitude)} &
\colhead{} 
\\
}
\startdata
GW-MMADS & AT2025bnx & 243.198339 & -68.827761 & 6.43 & r = 20.53 mag & 0.45  \\ 
GW-MMADS & AT2025bnl & 242.330798 & -69.341945 & 6.38 & i = 21.34 mag & 0.45  \\ 
GW-MMADS & AT2025bnm & 245.717365 & -69.023307 & 6.39 & i = 22.10 mag & 0.46  \\ 
GW-MMADS & AT2025bno & 242.698474 & -68.470445 & 6.38 & i = 21.11 mag & 0.49  \\
GW-MMADS & AT2025bnh & 248.159200 & -68.517481 & 6.40 & i = 21.45 mag & 0.50  \\
SOAR & AT2025ber & 247.960000 & -69.528969 & 0.44 & i = 22.00 mag & 0.50  \\ 
GW-MMADS & AT2025boa & 238.313614 & -69.284913 & 6.38 & i = 21.50 mag & 0.50  \\
GW-MMADS & AT2025bni & 248.497050 & -69.302599 & 6.40 & i = 22.50 mag & 0.51  \\
GW-MMADS & AT2025bmx & 239.335215 & -68.667346 & 6.38 & i = 20.70 mag & 0.52  \\
GW-MMADS & AT2025btp & 243.285298 & -68.054332 & 6.39 & i = 21.96 mag & 0.52  \\
GW-MMADS & AT2025bnt & 248.456112 & -70.049160 & 6.39 & i = 21.36 mag & 0.52  \\
GW-MMADS & AT2025bnn & 246.087141 & -67.969303 & 6.39 & i = 21.64 mag & 0.52  \\
GW-MMADS & AT2025btg & 245.808953 & -67.950089 & 6.40 & i = 21.31 mag & 0.54  \\
GW-MMADS & AT2025bnp & 238.092561 & -70.078637 & 6.38 & i = 21.14 mag & 0.54  \\
GW-MMADS & AT2025bng & 237.311207 & -68.793026 & 6.37 & i = 21.33 mag & 0.54  \\
GW-MMADS & AT2025bmt & 241.702375 & -68.015378 & 6.38 & i = 20.95 mag & 0.54  \\
GW-MMADS & AT2025bna & 250.352179 & -69.852791 & 6.40 & i = 22.09 mag & 0.54  \\
GW-MMADS & AT2025bnz & 252.642449 & -69.271086 & 6.41 & i = 21.03 mag & 0.56  \\ 
GW-MMADS & AT2025bne & 241.735013 & -67.822171 & 6.38 & i = 21.08 mag & 0.56  \\
GW-MMADS & AT2025bnu & 252.865009 & -69.107941 & 6.41 & i = 21.40 mag & 0.56  \\
GW-MMADS & AT2025bmr & 253.039696 & -69.238089 & 6.41 & i = 20.04 mag & 0.56  \\
GW-MMADS & AT2025bns & 253.108277 & -69.071872 & 6.41 & i = 22.20 mag & 0.56  \\ 
GW-MMADS & AT2025bth & 253.015441 & -68.239550 & 6.48 & i = 20.25 mag & 0.57  \\ 
GW-MMADS & AT2025bnb & 237.554278 & -70.317931 & 6.37 & i = 20.61 mag & 0.57  \\ 
GW-MMADS & AT2025bnd & 241.490340 & -67.782256 & 6.38 & i = 22.02 mag & 0.57  \\
GW-MMADS & AT2025bmv & 241.482057 & -67.782698 & 6.39 & i = 22.20 mag & 0.57  \\
GW-MMADS & AT2025bob & 239.568537 & -67.947727 & 6.38 & i = 21.66 mag & 0.57  \\
GW-MMADS & AT2025bmy & 235.654159 & -69.613801 & 6.37 & i = 20.83 mag & 0.57  \\
GW-MMADS & AT2025bnw & 235.314005 & -69.511897 & 6.37 & i = 22.29 mag & 0.57  \\
GW-MMADS & AT2025bnk & 236.684850 & -68.624928 & 6.37 & i = 20.30 mag & 0.58  \\
GW-MMADS & AT2025bmu & 242.175587 & -67.502065 & 6.39 & i = 21.60 mag & 0.58  \\
GW-MMADS & AT2025bti & 239.814737 & -67.758628 & 6.38 & i = 21.70 mag & 0.59  \\
GW-MMADS & AT2025btr & 254.511807 & -68.447070 & 6.48 & i = 22.29 mag & 0.59  \\
GW-MMADS & AT2025bms & 247.101545 & -70.816131 & 6.40 & i = 21.37 mag & 0.59  \\
GW-MMADS & AT2025btc & 241.670680 & -70.991068 & 6.39 & i = 22.11 mag & 0.61  \\
GW-MMADS & AT2025btl & 250.676560 & -67.511204 & 6.47 & i = 22.44 mag & 0.61  \\
GW-MMADS & AT2025btm & 247.969931 & -71.211799 & 6.40 & i = 21.38 mag & 0.61  \\
GW-MMADS & AT2025bnf & 252.514939 & -70.593309 & 6.41 & i = 20.33 mag & 0.62  \\
GW-MMADS & AT2025bte & 247.408084 & -71.315427 & 6.40 & i = 20.92 mag & 0.63  \\ 
GW-MMADS & AT2025bnc & 256.260785 & -69.873825 & 6.41 & i = 22.25 mag & 0.65  \\
GW-MMADS & AT2025bnv & 235.926019 & -67.682568 & 6.38 & i = 20.89 mag & 0.65  \\
GW-MMADS & AT2025btj & 241.632510 & -66.959770 & 6.38 & i = 22.05 mag & 0.65  \\
GW-MMADS & AT2025bnr & 239.310711 & -67.192893 & 6.38 & i = 21.09 mag & 0.66  \\
GW-MMADS & AT2025btq & 237.204147 & -67.313042 & 6.38 & i = 21.96 mag & 0.66  \\
GW-MMADS & AT2025btk & 232.860157 & -68.779324 & 6.37 & i = 20.47 mag & 0.67  \\
GW-MMADS & AT2025btd & 255.847308 & -70.370222 & 6.41 & i = 20.64 mag & 0.68  \\
GW-MMADS & AT2025bnj & 239.962646 & -66.929715 & 6.38 & i = 20.87 mag & 0.68  \\
GW-MMADS & AT2025bmz & 250.110134 & -66.773856 & 6.43 & r = 19.39 mag & 0.68  \\
GOTO & AT2025bau & 253.166158 & -66.813549 & 0.86 & L = 19.59 mag & 0.73  \\
GW-MMADS & AT2025btn & 232.587844 & -70.411893 & 6.37 & i = 19.17 mag & 0.76  \\
GW-MMADS & AT2025bnq & 237.860041 & -71.709022 & 6.37 & i = 20.75 mag & 0.76  \\
GW-MMADS & AT2025bts & 251.287795 & -71.523116 & 6.41 & i = 22.58 mag & 0.77  \\
GW-MMADS & AT2025bmw & 244.552812 & -71.938206 & 6.40 & i = 20.78 mag & 0.78  \\
GW-MMADS & AT2025bto & 237.129581 & -71.826170 & 6.37 & i = 21.56 mag & 0.81  \\
GW-MMADS & AT2025bny & 238.070346 & -72.142874 & 6.37 & i = 20.62 mag & 0.85  \\
GOTO & AT2025cat & 262.154143 & -68.789674 & 10.84 & L = 18.63 mag & 0.87  \\
GW-MMADS & AT2025btf & 235.928949 & -71.902516 & 6.42 & r = 21.25 mag & 0.88  \\
GOTO & AT2025bao & 169.237933 & -45.541262 & 0.58 & L = 19.14 mag & 0.93  \\
GOTO & AT2025bar & 170.681922 & -45.574606 & 0.69 & L = 20.23 mag & 0.94  \\
ATLAS & AT2025bfg & 184.400370 & -59.493302 & 5.12 & orange = 18.53 mag & 0.95  \\
\enddata
\end{deluxetable*}

 \begin{figure*}[h]
    \centering   
    \includegraphics[width=0.3\textwidth]{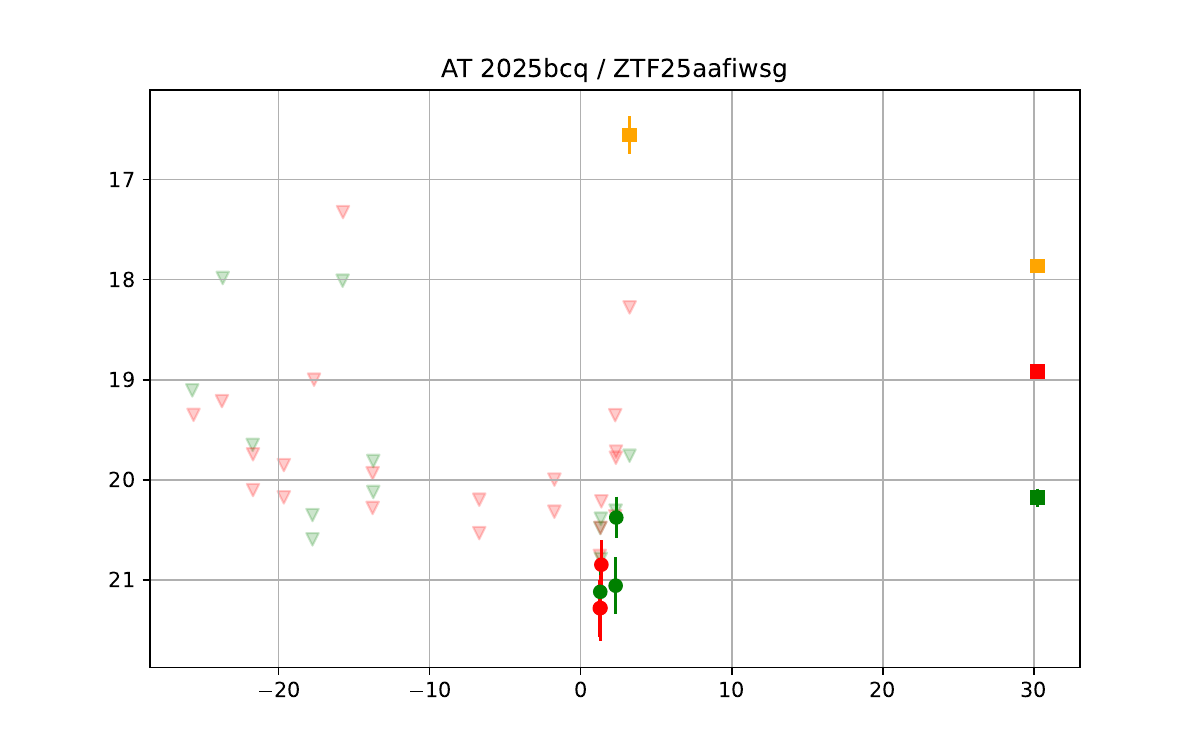}
\includegraphics[width=0.3\textwidth]{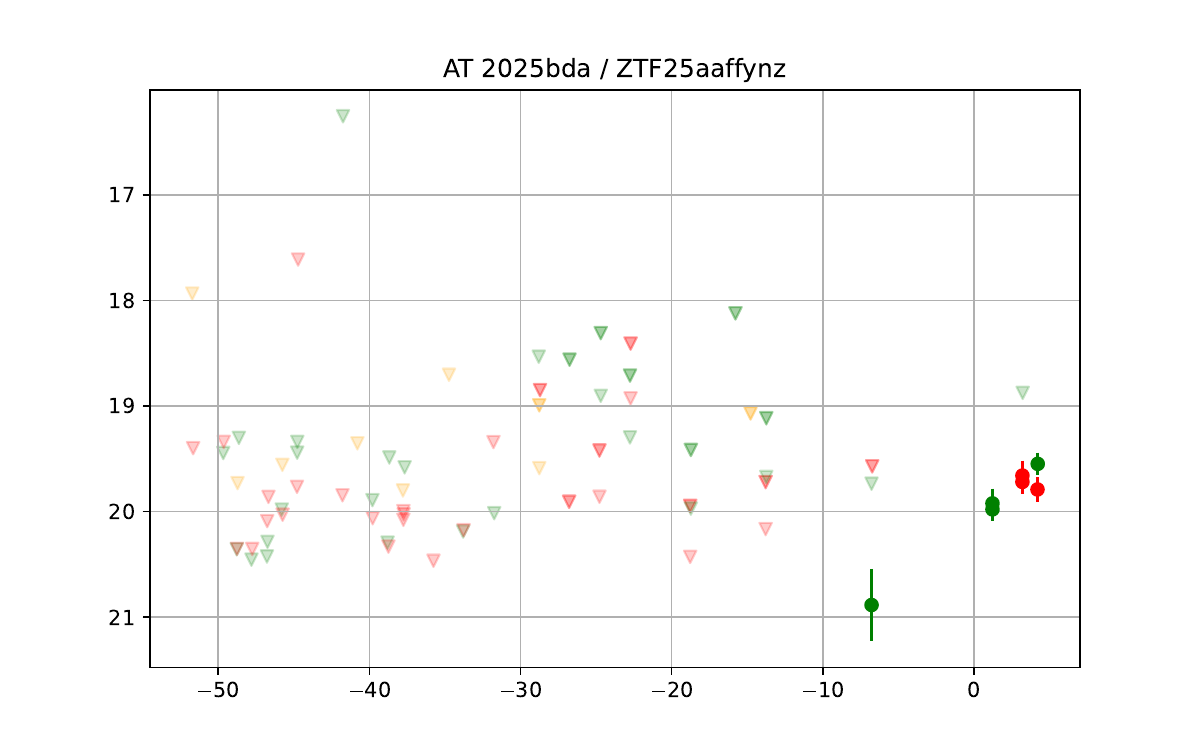}
\includegraphics[width=0.3\textwidth]{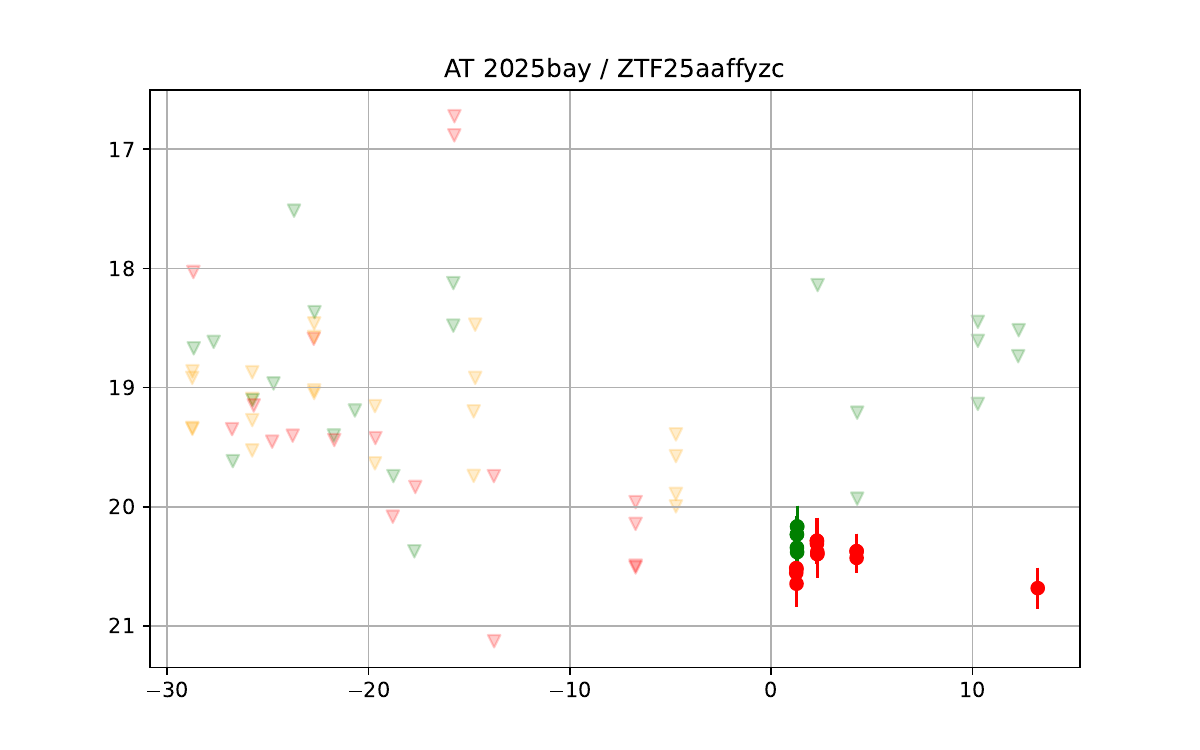}
\includegraphics[width=0.3\textwidth]{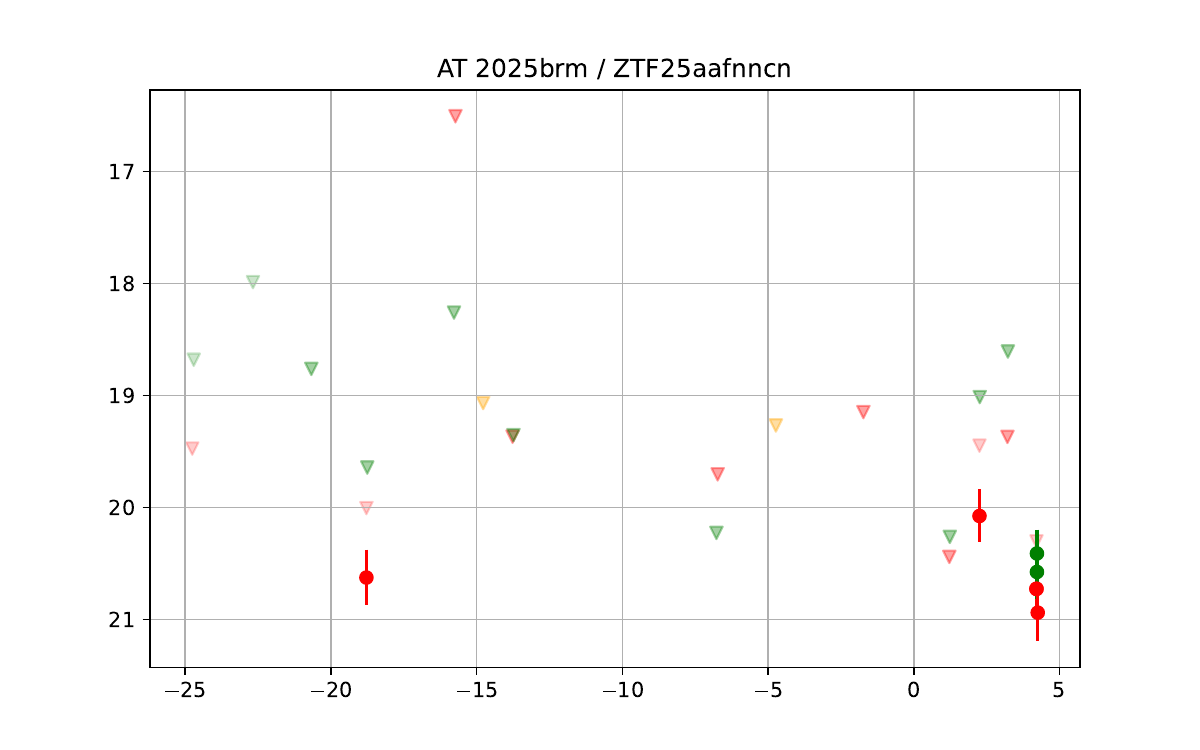}
\includegraphics[width=0.3\textwidth]{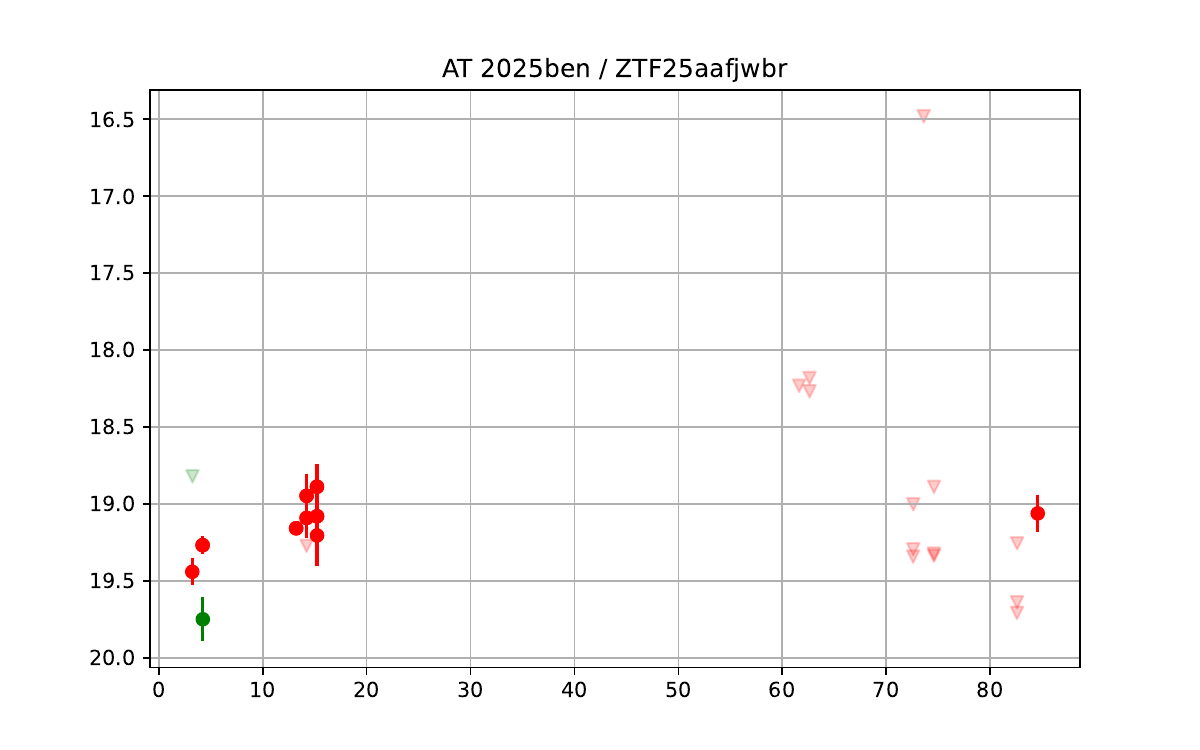}
\includegraphics[width=0.3\textwidth]{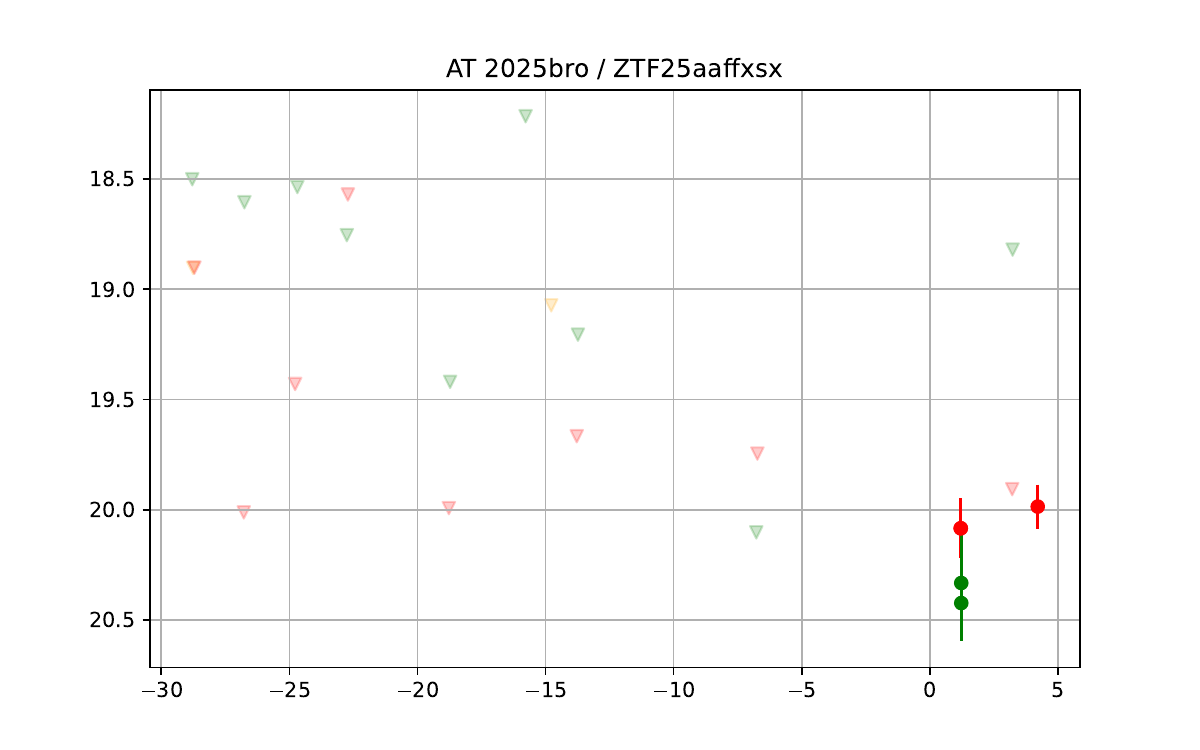}
\includegraphics[width=0.3\textwidth]{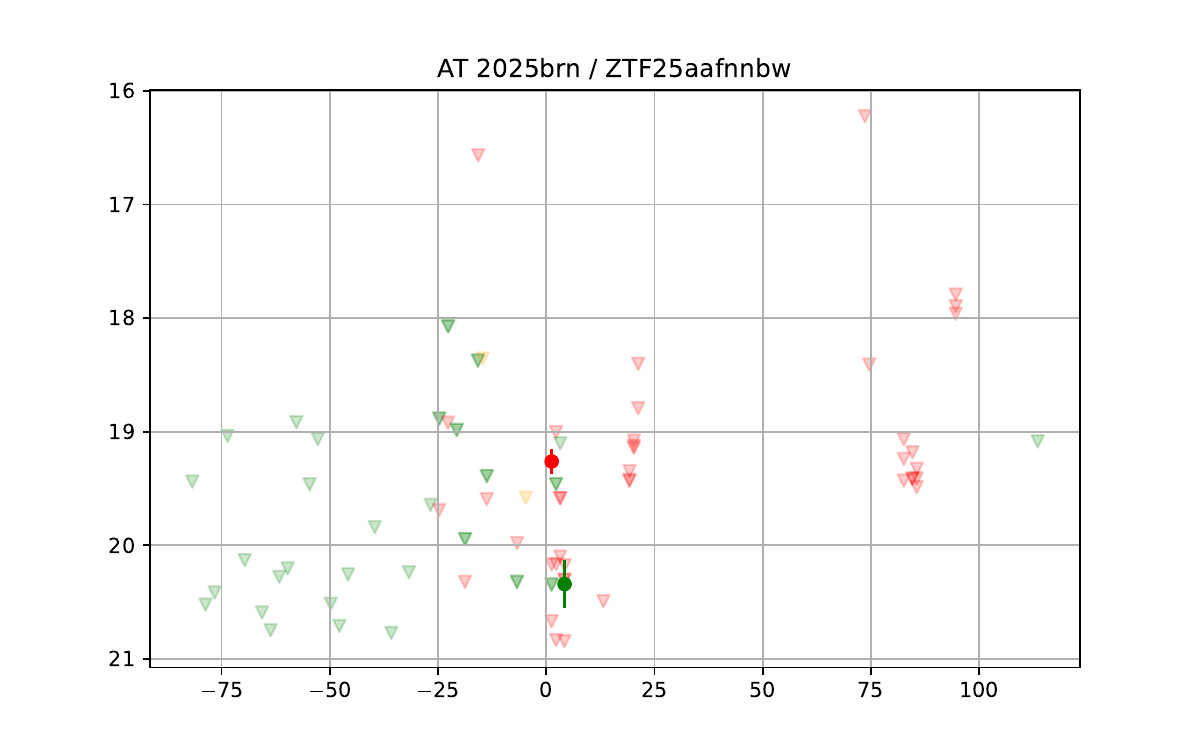}
\includegraphics[width=0.3\textwidth]{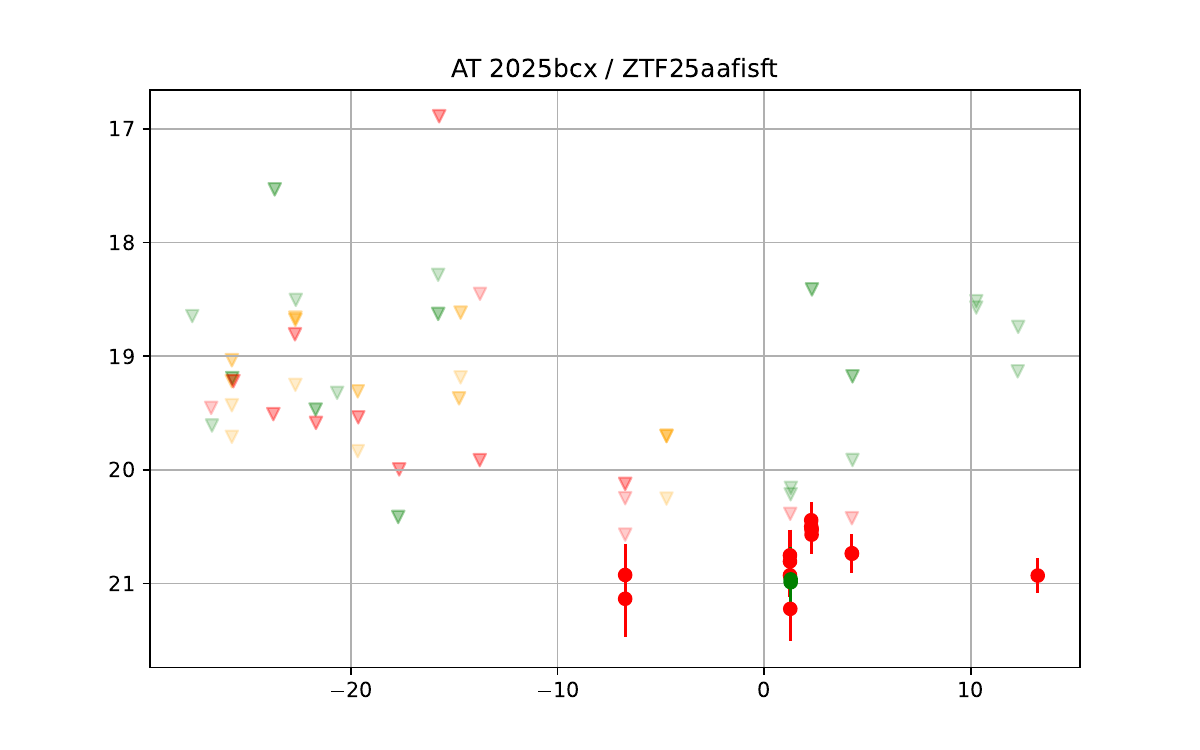}
\includegraphics[width=0.3\textwidth]{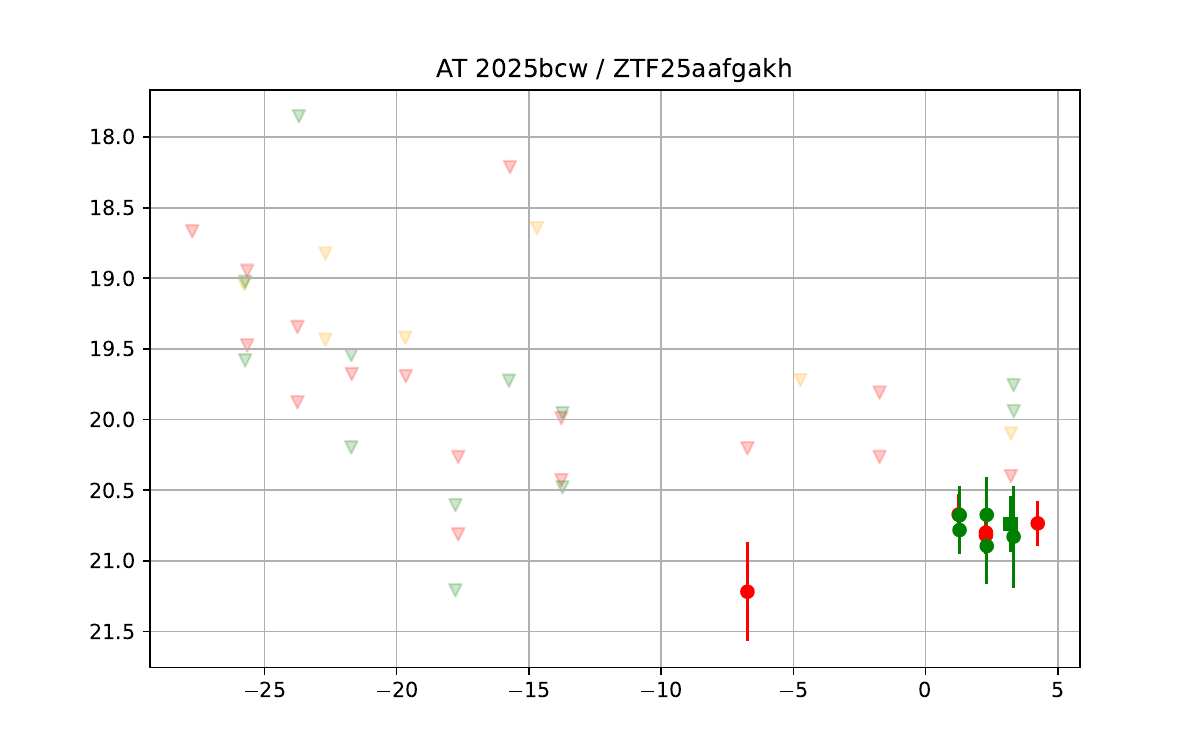}
\includegraphics[width=0.3\textwidth]{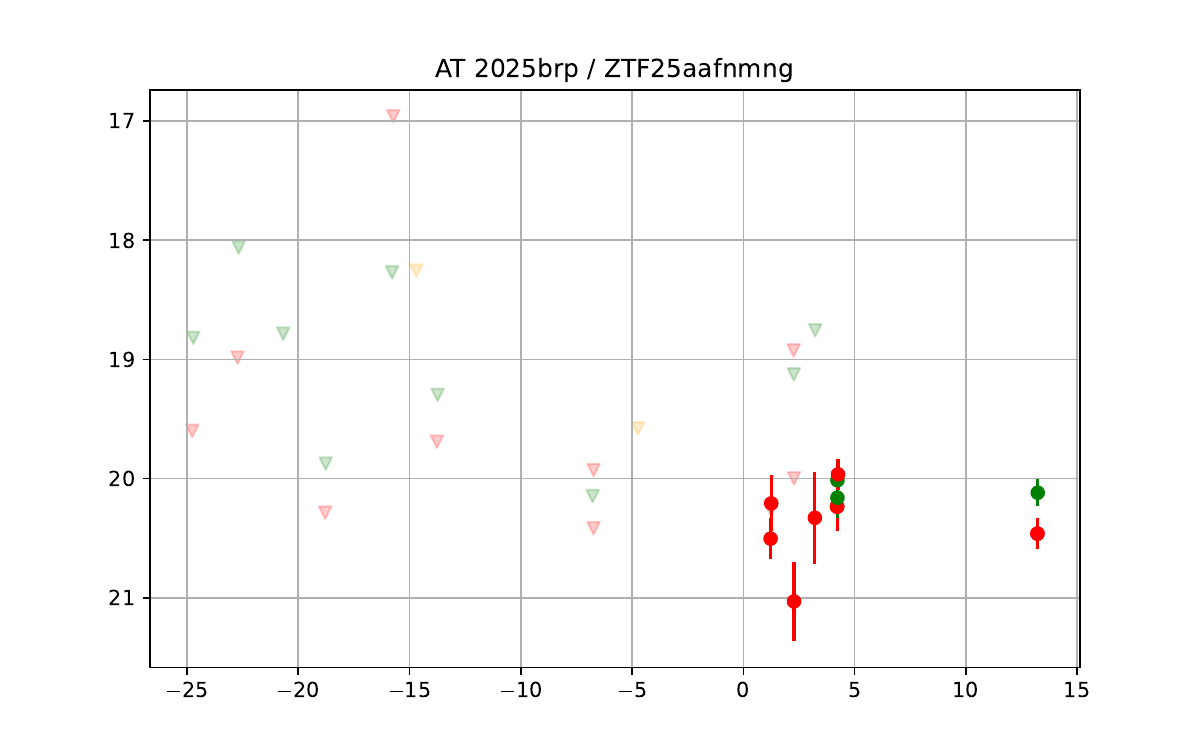}
\includegraphics[width=0.3\textwidth]{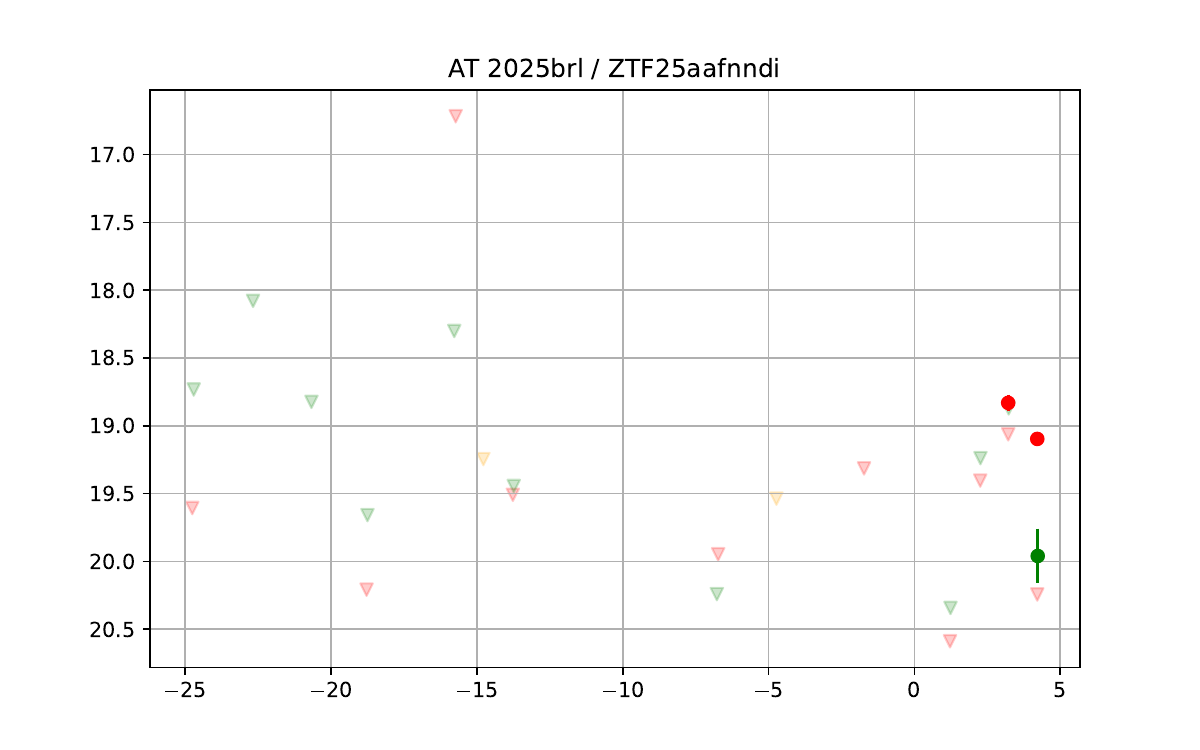}
\includegraphics[width=0.3\textwidth]{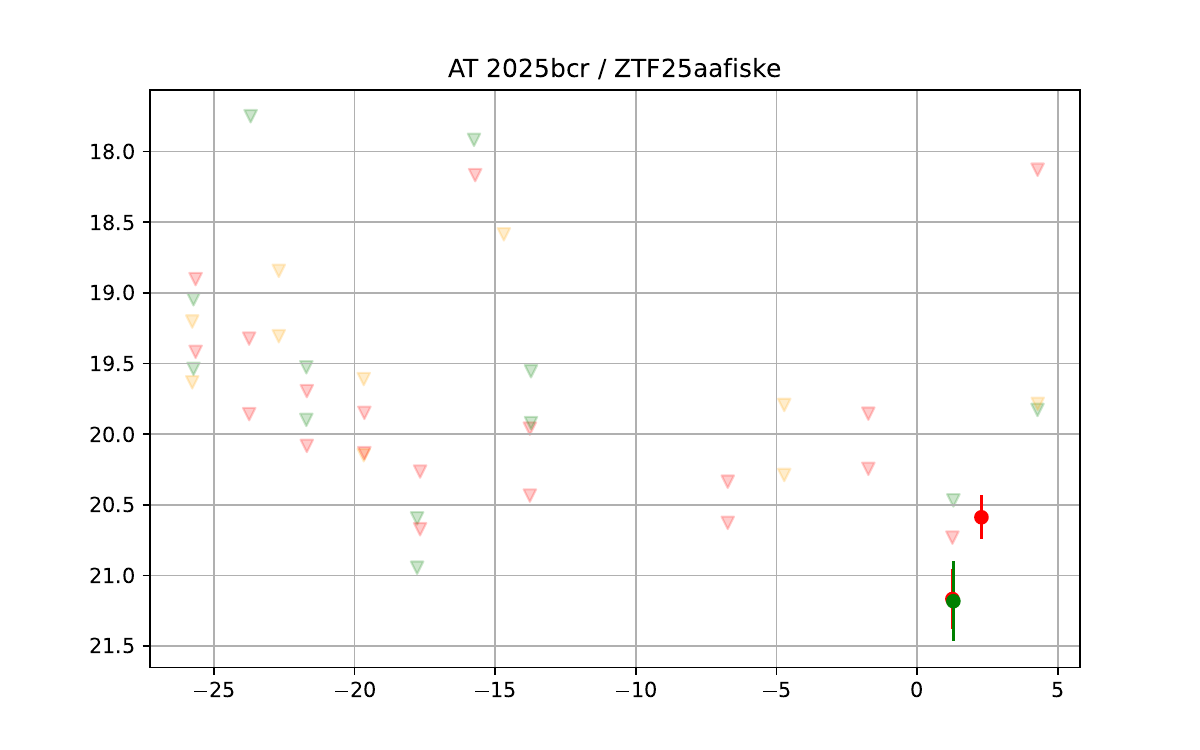}
\includegraphics[width=0.3\textwidth]{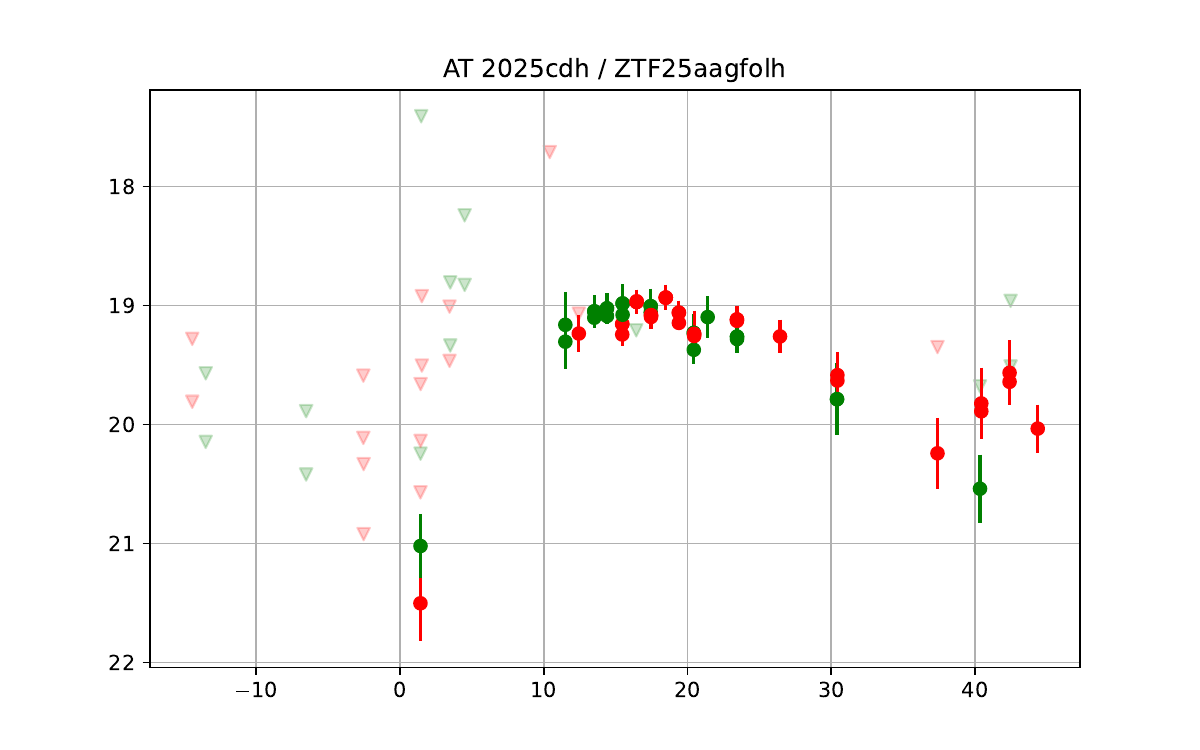}

    \caption{Light-curves for optical counterparts of S250206dm. The color (red, green, yellow, blue, cyan) represent the different filters ($g$, $r$, $i$, $L$, $c$). The different symbols (circle, square, pentagon, diamond, X, filled cross, and thin cross) represent the different facilities (ZTF, SEDM, PS1, LT, GIT, GOTO, CSS).}
    \label{fig:LC_ZTF}
\end{figure*}

\begin{figure*}[h]
    \centering
    \includegraphics[width=0.3\textwidth]{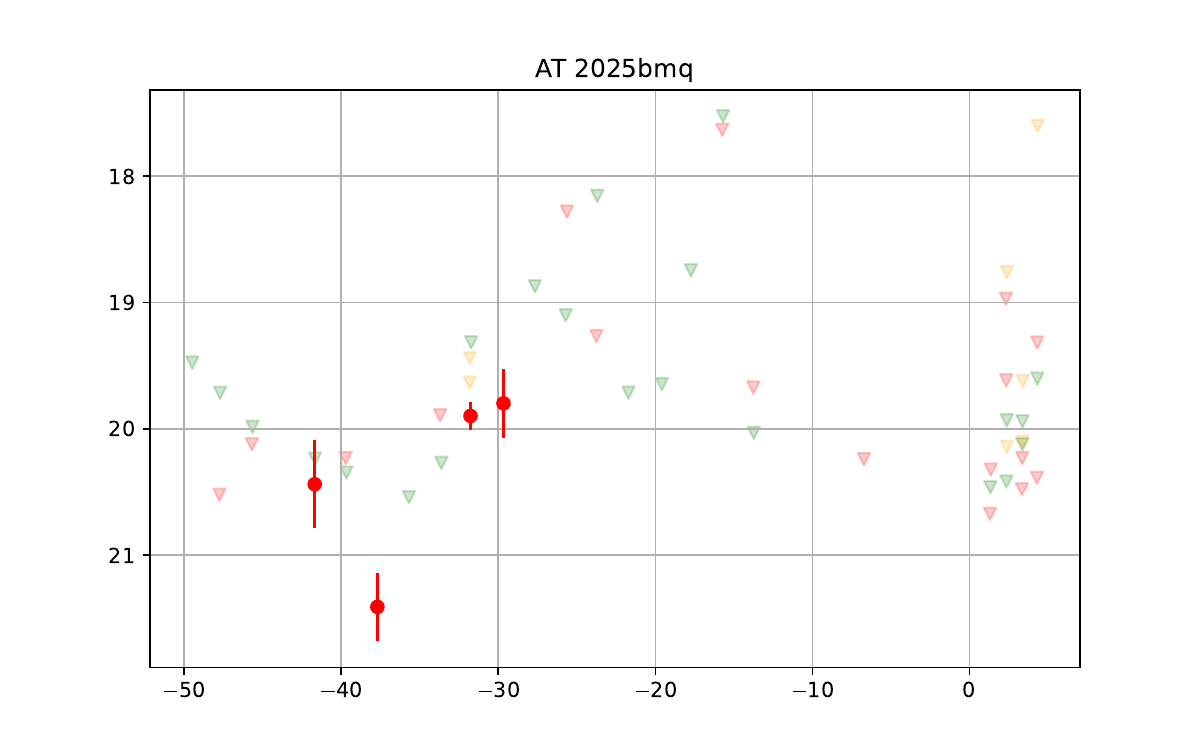}
\includegraphics[width=0.3\textwidth]{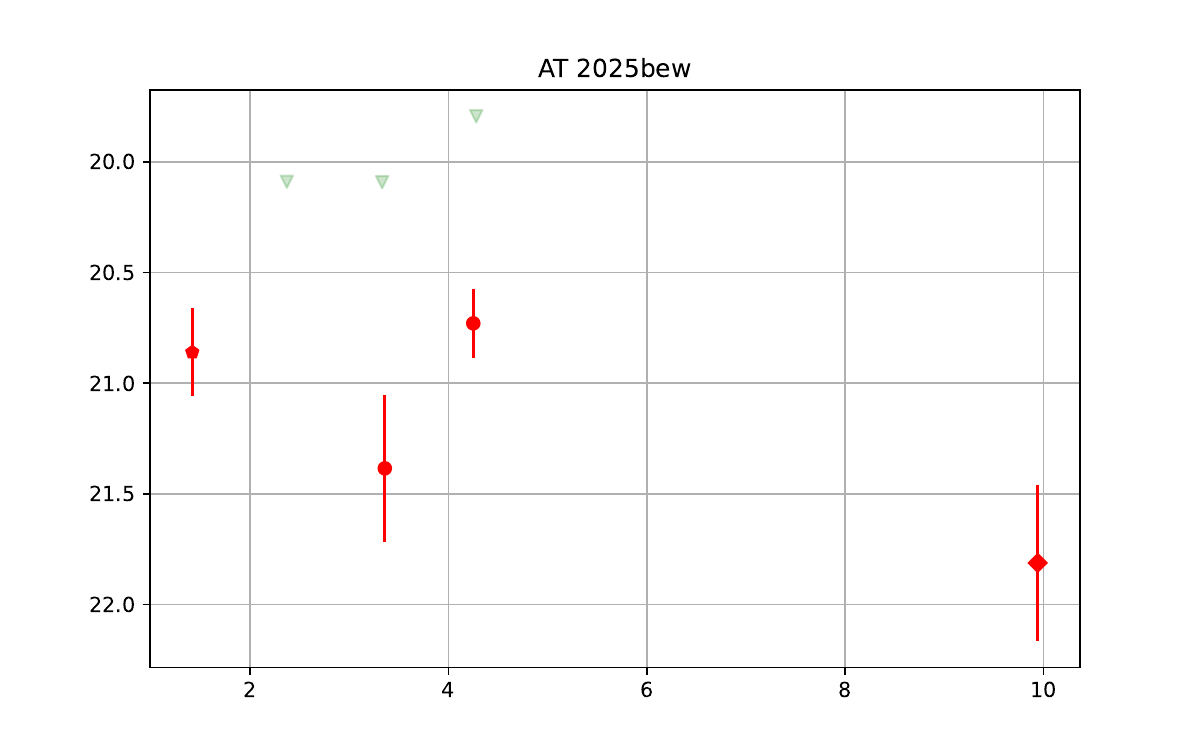}
\includegraphics[width=0.3\textwidth]{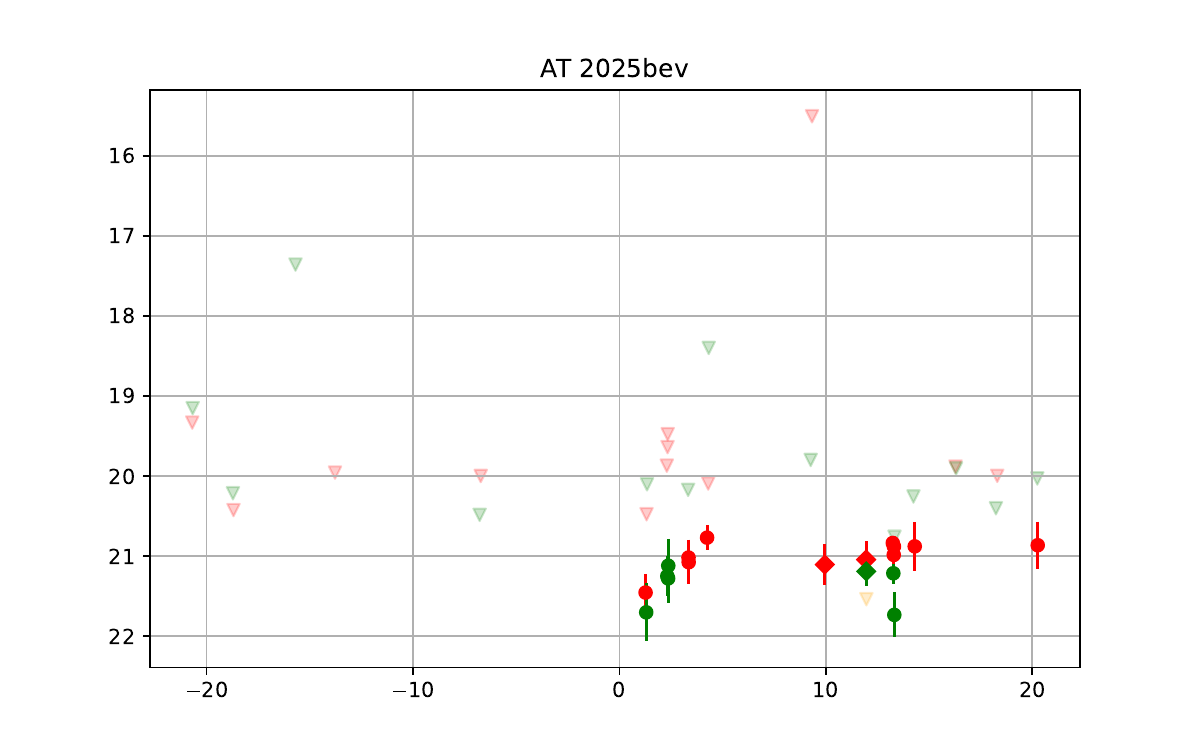}
\includegraphics[width=0.3\textwidth]{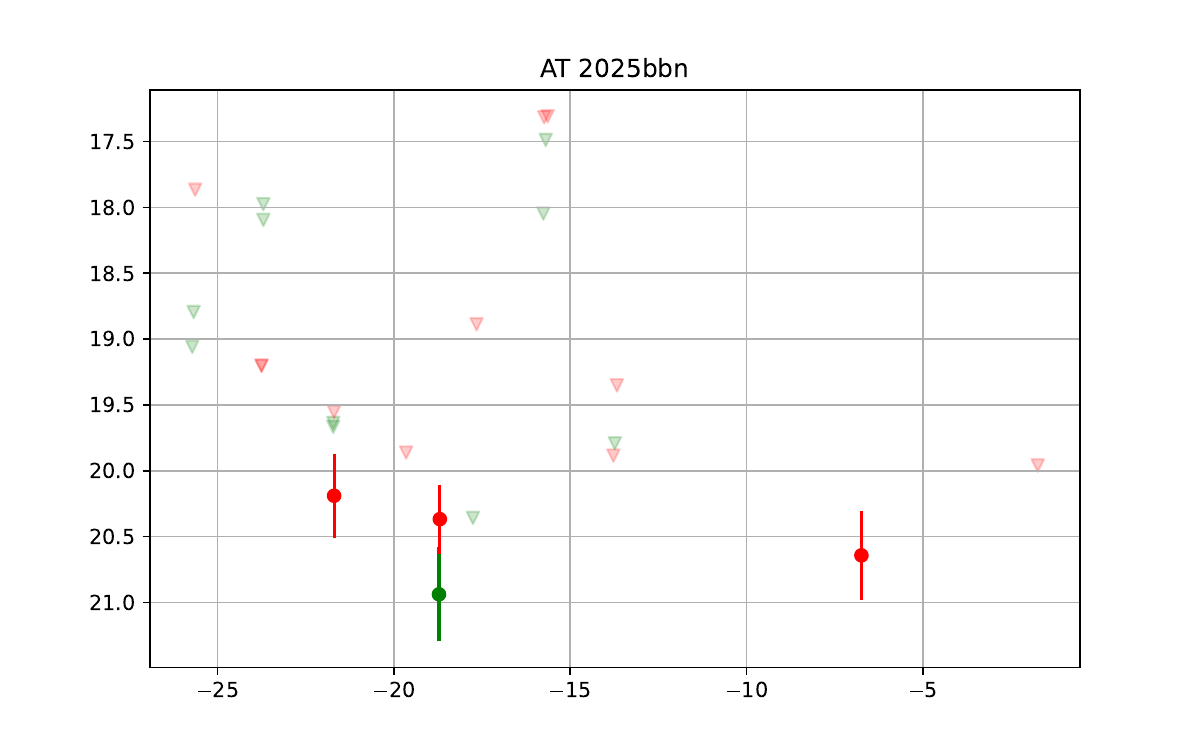}
\includegraphics[width=0.3\textwidth]{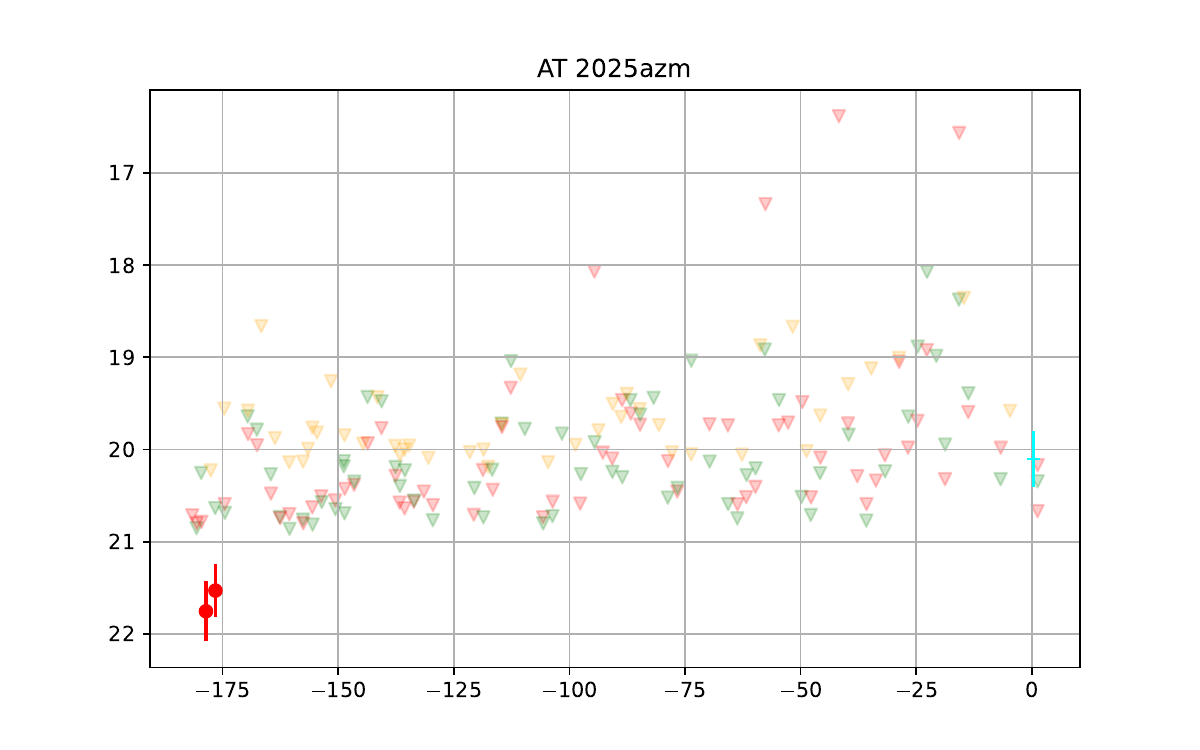}
\includegraphics[width=0.3\textwidth]{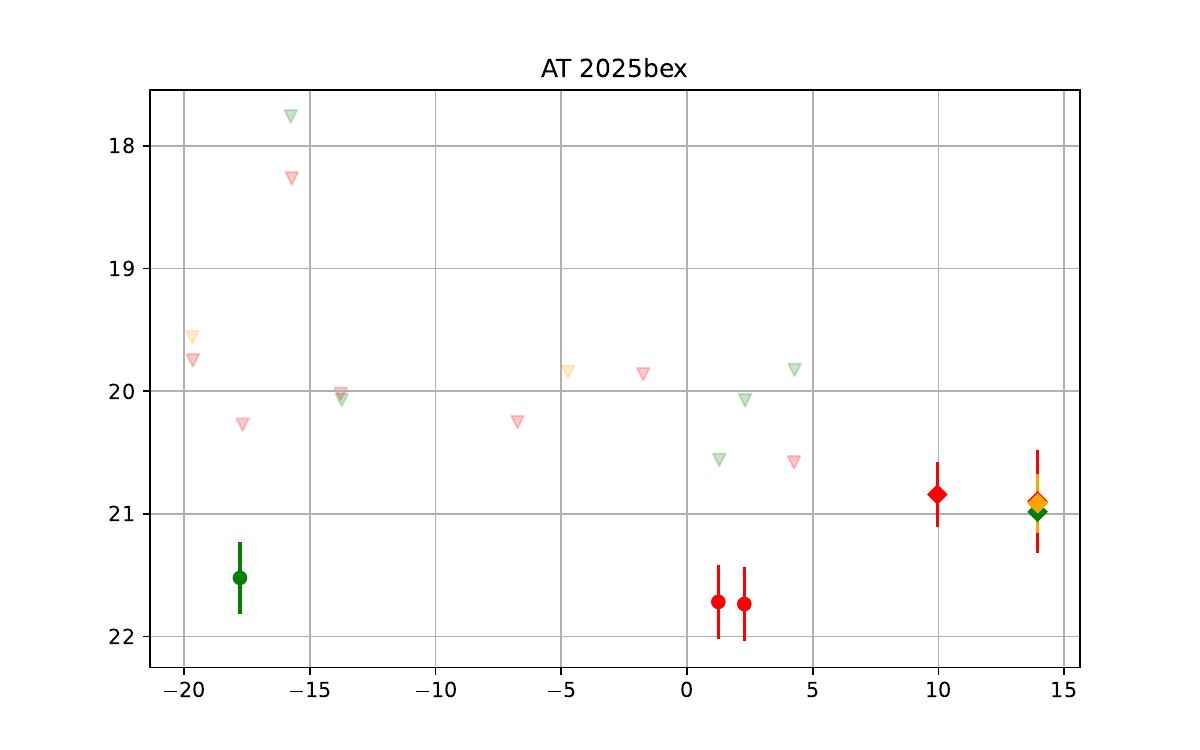}
\includegraphics[width=0.3\textwidth]{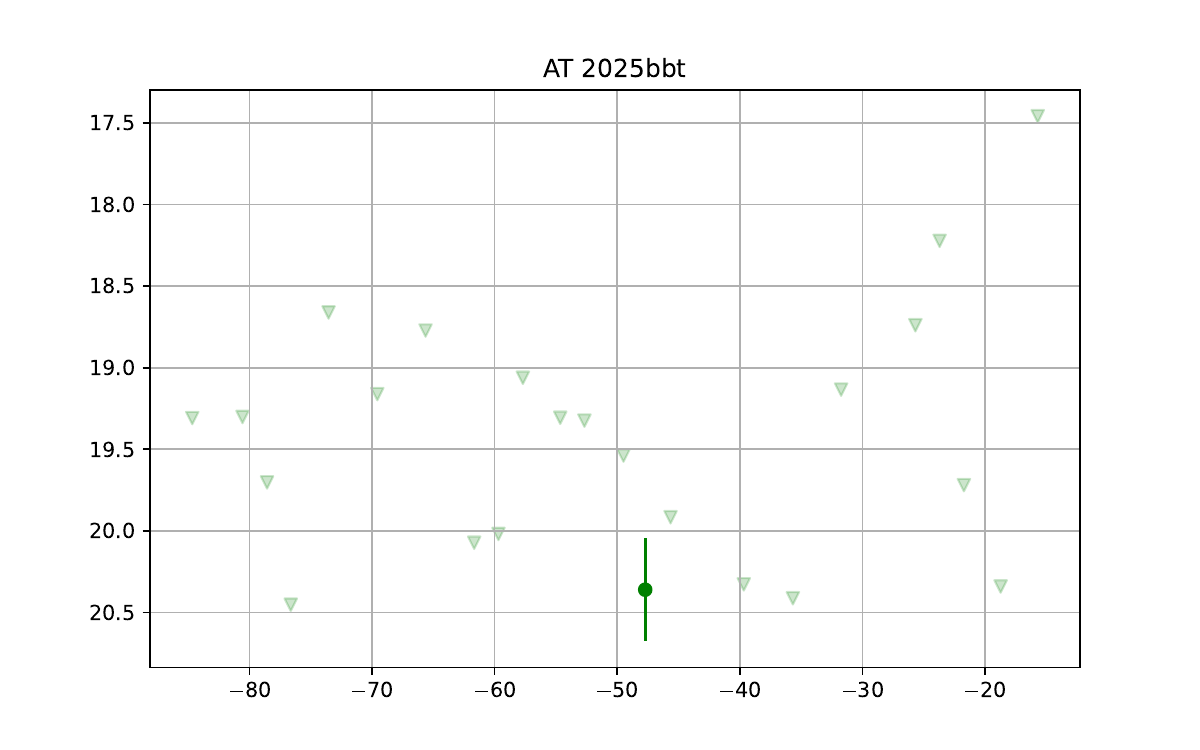}
\includegraphics[width=0.3\textwidth]{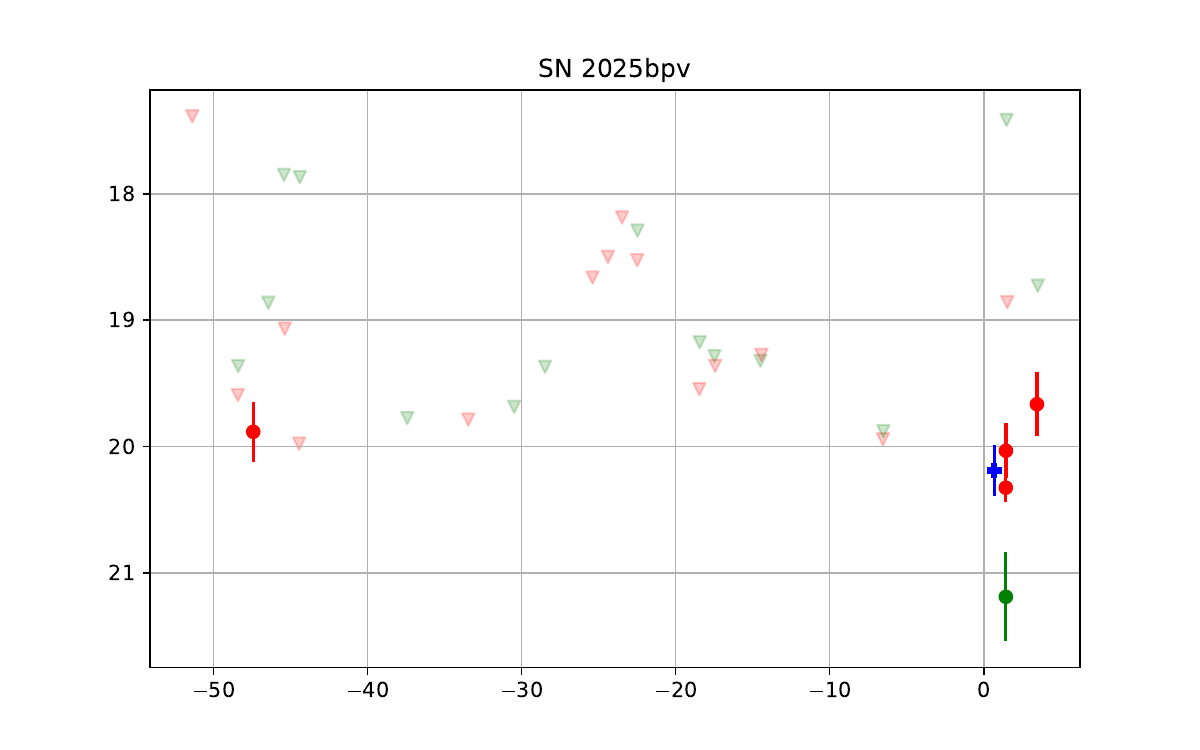}
\includegraphics[width=0.3\textwidth]{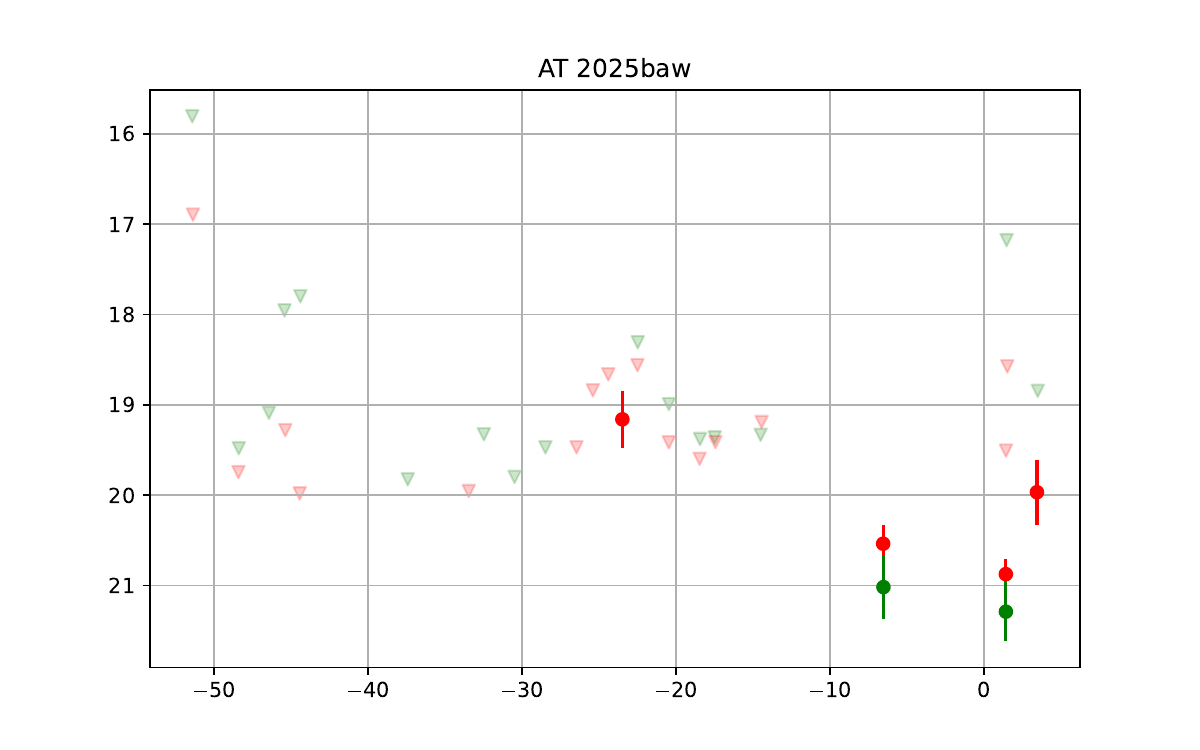}
\includegraphics[width=0.3\textwidth]{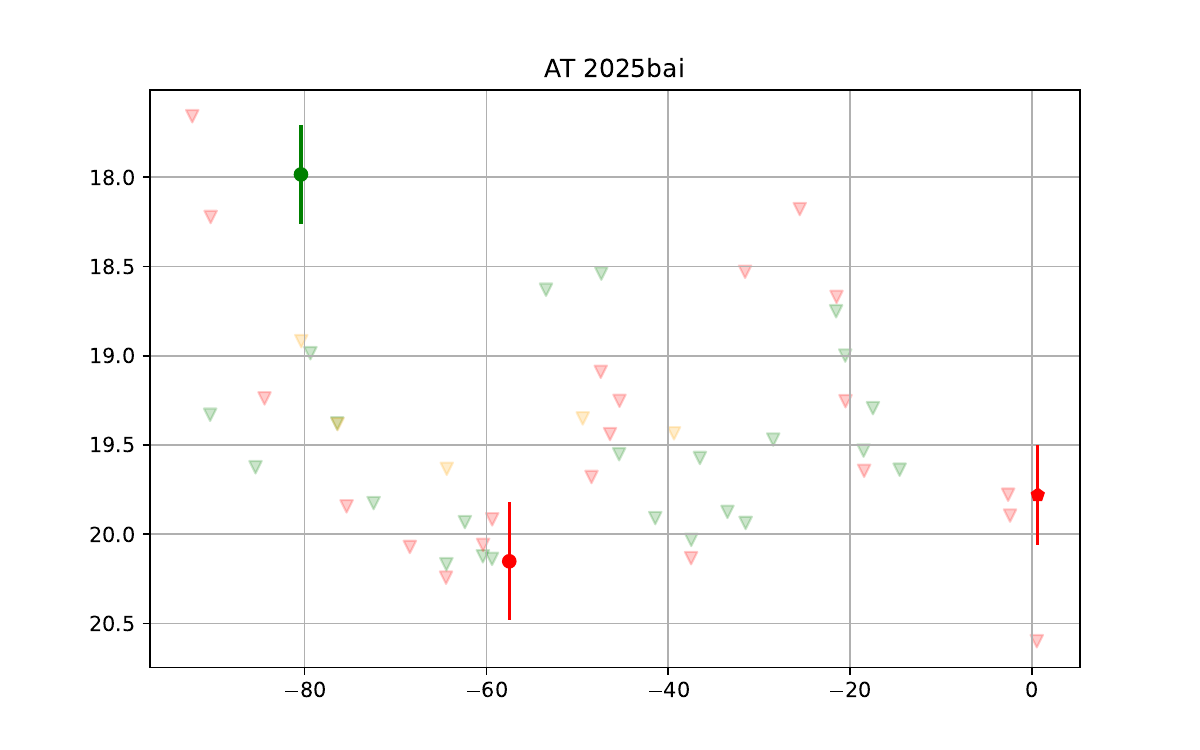}
\includegraphics[width=0.3\textwidth]{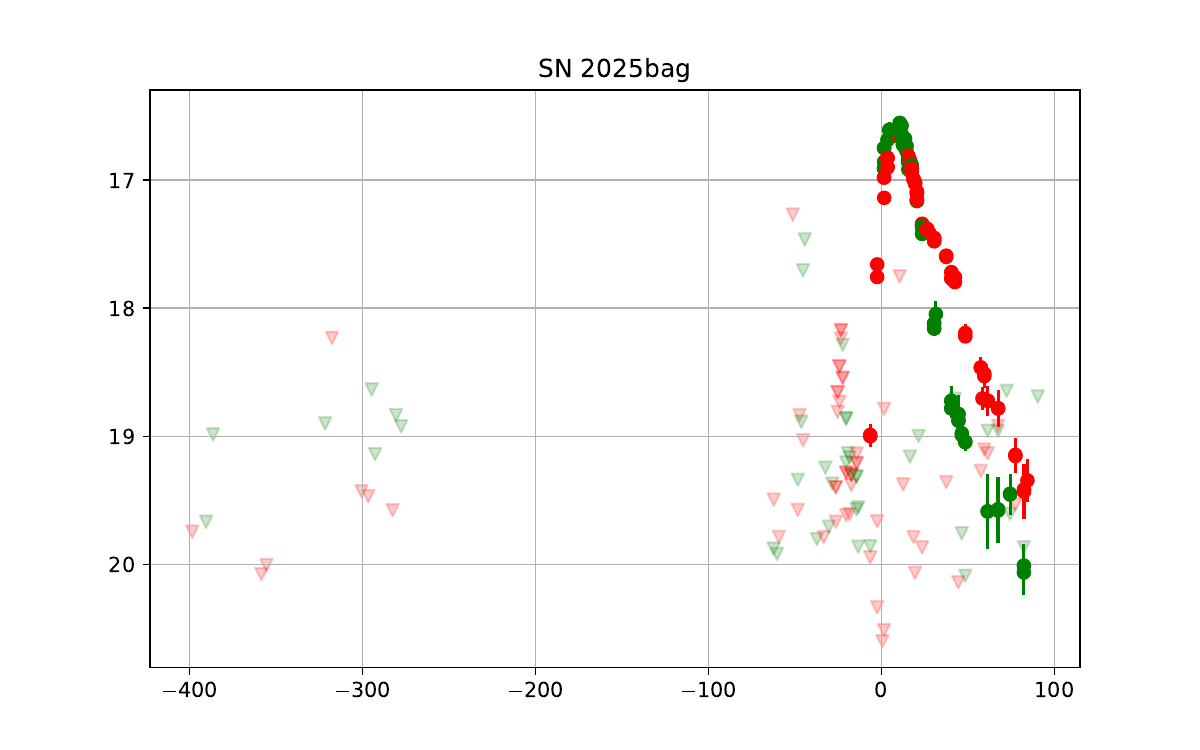}
\includegraphics[width=0.3\textwidth]{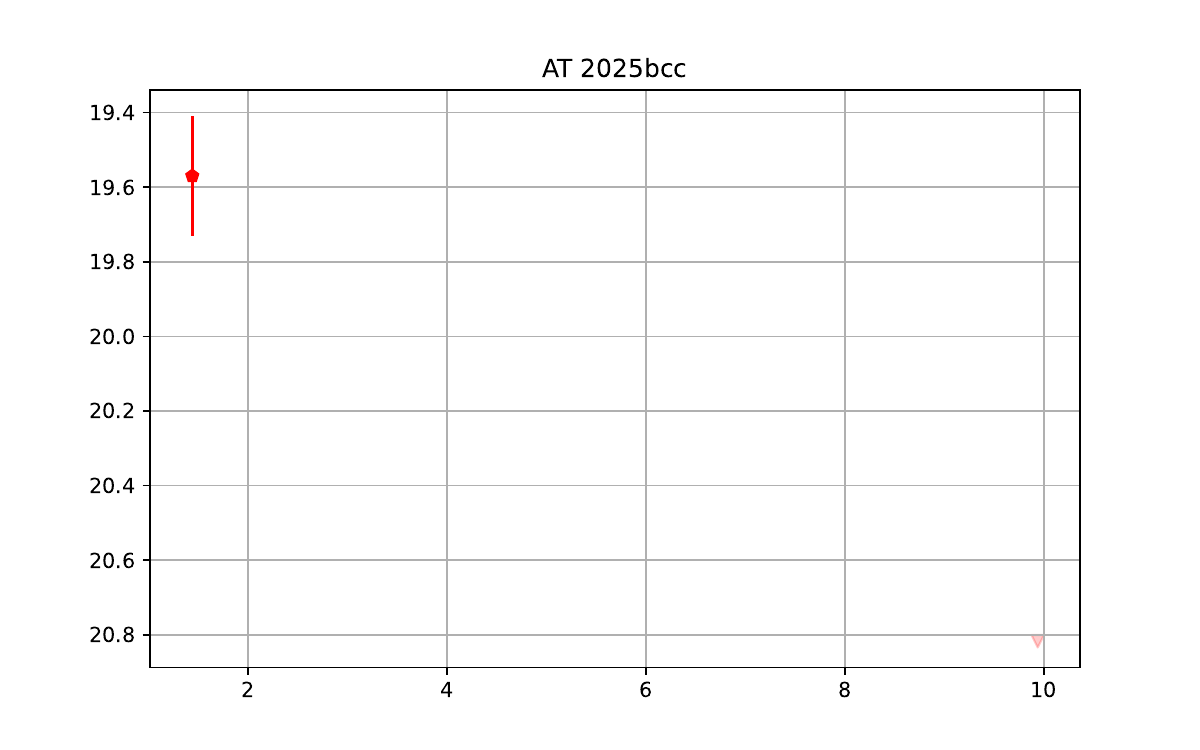}
\includegraphics[width=0.3\textwidth]{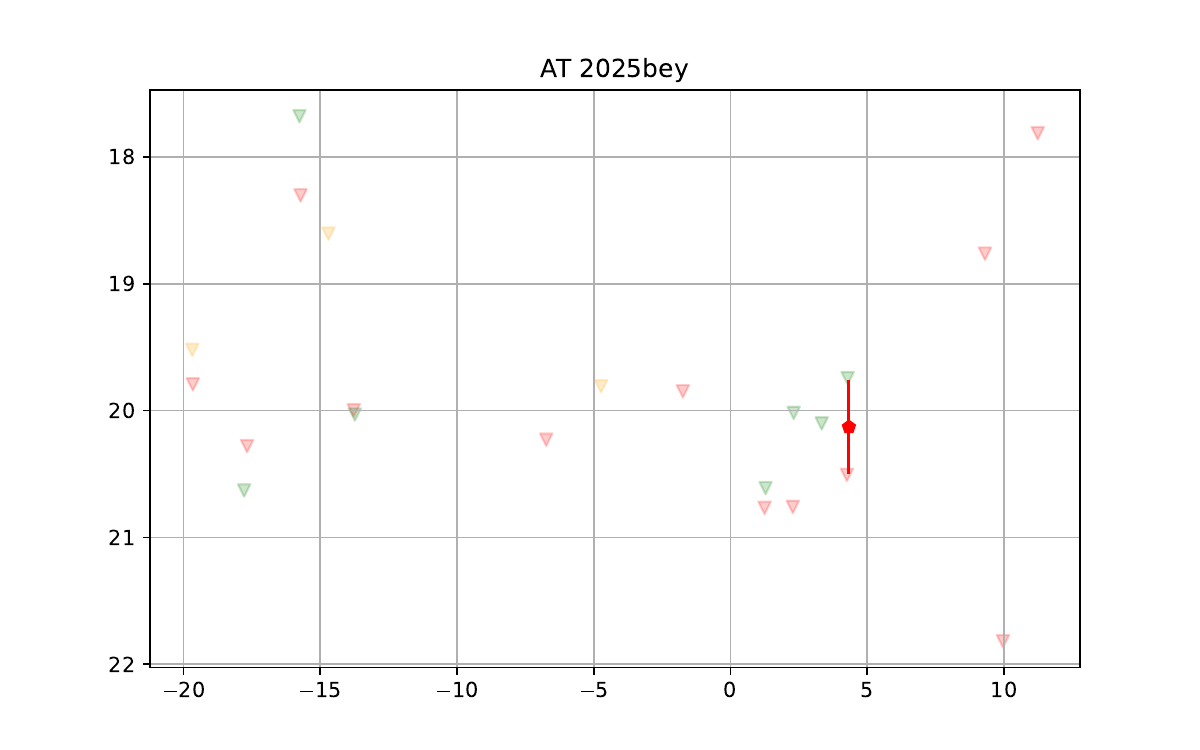}
\includegraphics[width=0.3\textwidth]{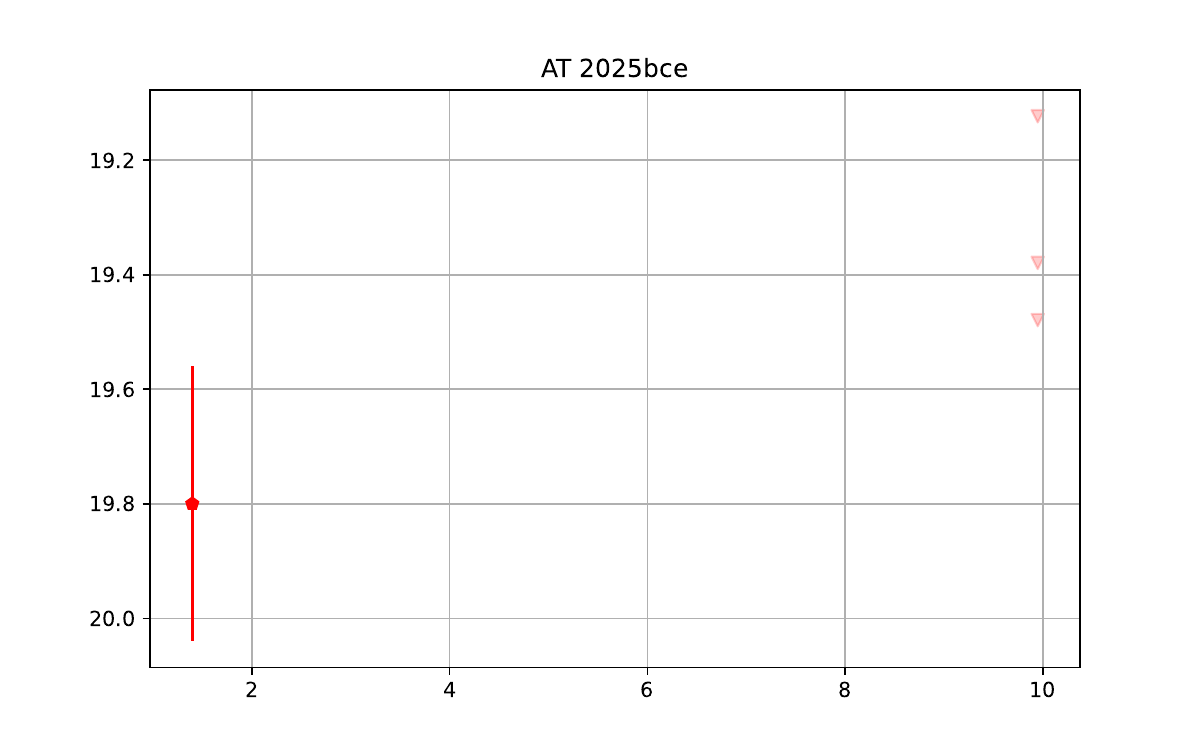}
\caption{Light-curves for optical counterparts of S250206dm. The color (red, green, yellow, blue, cyan) represent the different filters ($g$, $r$, $i$, $L$, $c$). The different symbols (circle, square, pentagon, diamond, X, filled cross, and thin cross) represent the different facilities ( ZTF, SEDM, PS1, LT, GIT, GOTO, CSS).}
    \label{fig:LC_2}
\end{figure*}

\begin{figure*}[h]
    \centering
    \includegraphics[width=\textwidth]{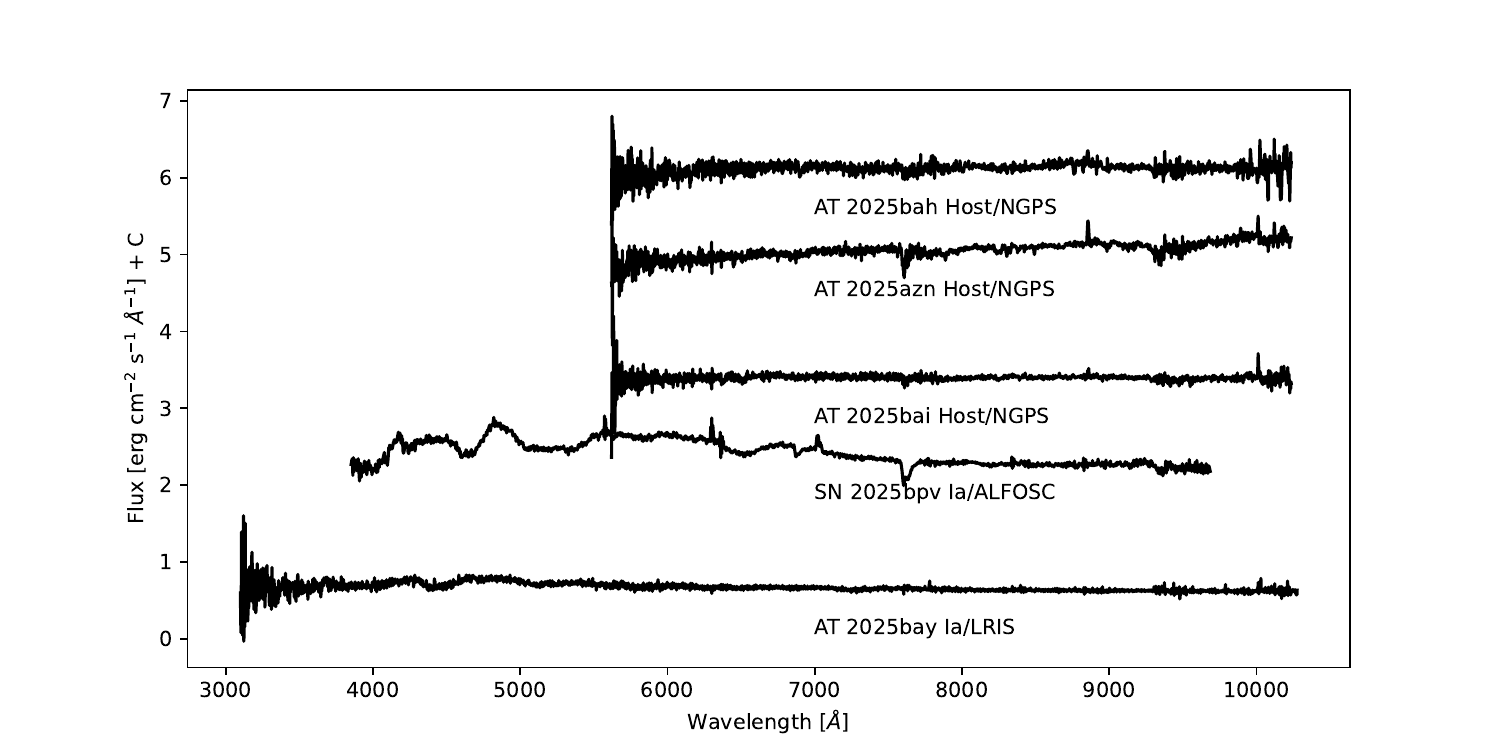}
    \includegraphics[width=\textwidth]{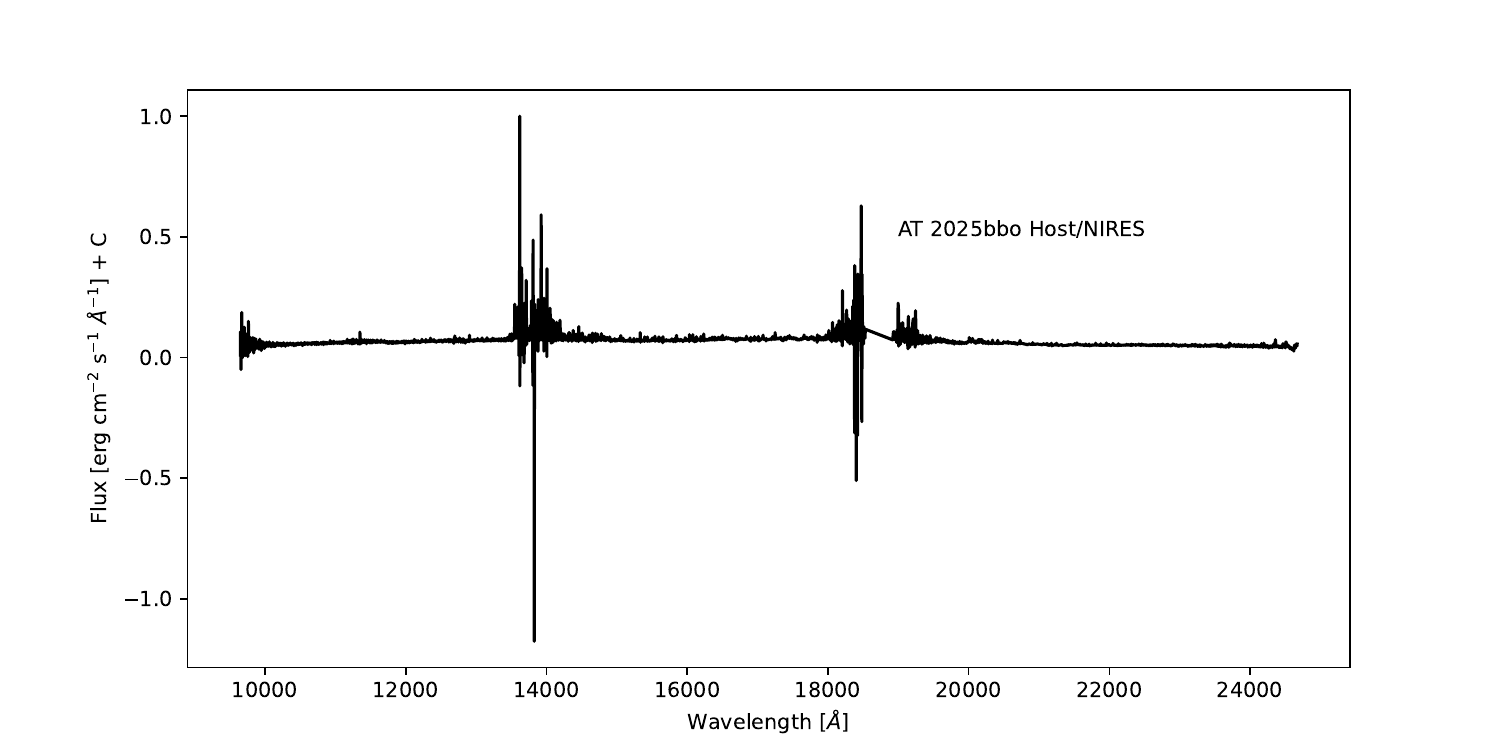}
    \caption{The optical (top) and near-infrared (bottom) spectra of the counterpart candidates of S250602dm.}
    \label{fig:spectra}
\end{figure*}

\section{Models ruled out} \label{appendixModels}

\begin{figure*}[t]
    \centering
     \includegraphics[width=0.5\textwidth]{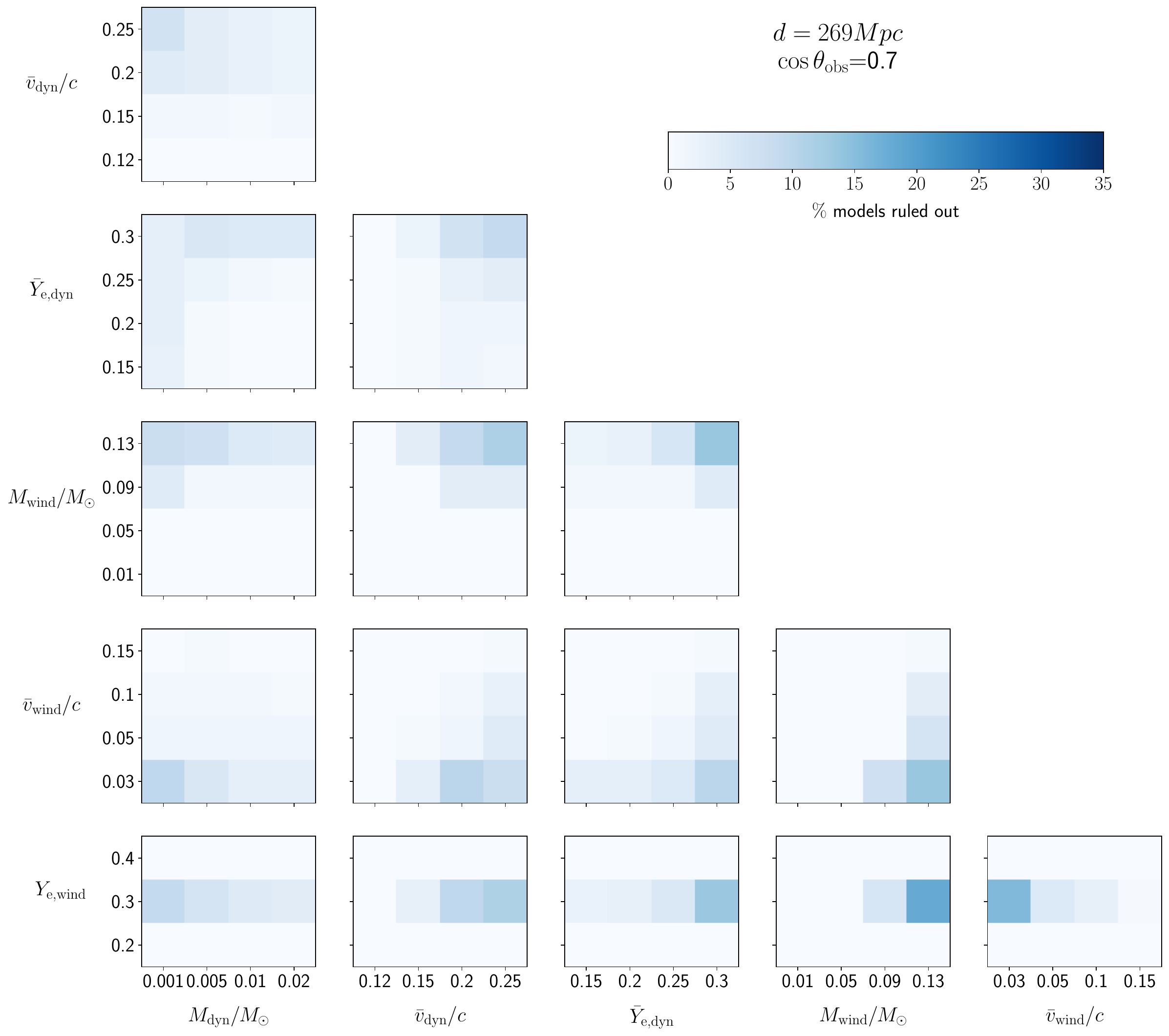}
    \caption{Corner plot showing in a colorbar the percentage of BNS models ruled out at a fixed distance of 269 Mpc and at a fixed viewing angle of 45 deg. The parameters correspond to the ejecta mass, mass-weighted averaged velocity and mass-weighted electron fraction for the dynamical and wind ejecta: $m_{\rm dyn}$, $\bar{v}_{\rm dyn}$, $\bar{Y}_{\rm e,dyn}$, $m_{\rm wind}$, $\bar{v}_{\rm wind}$ and $\bar{Y}_{\rm e,wind}$  }
    \label{fig:corner2}
\end{figure*}

\begin{figure*}[t]
    \centering
     \includegraphics[width=0.5\textwidth]{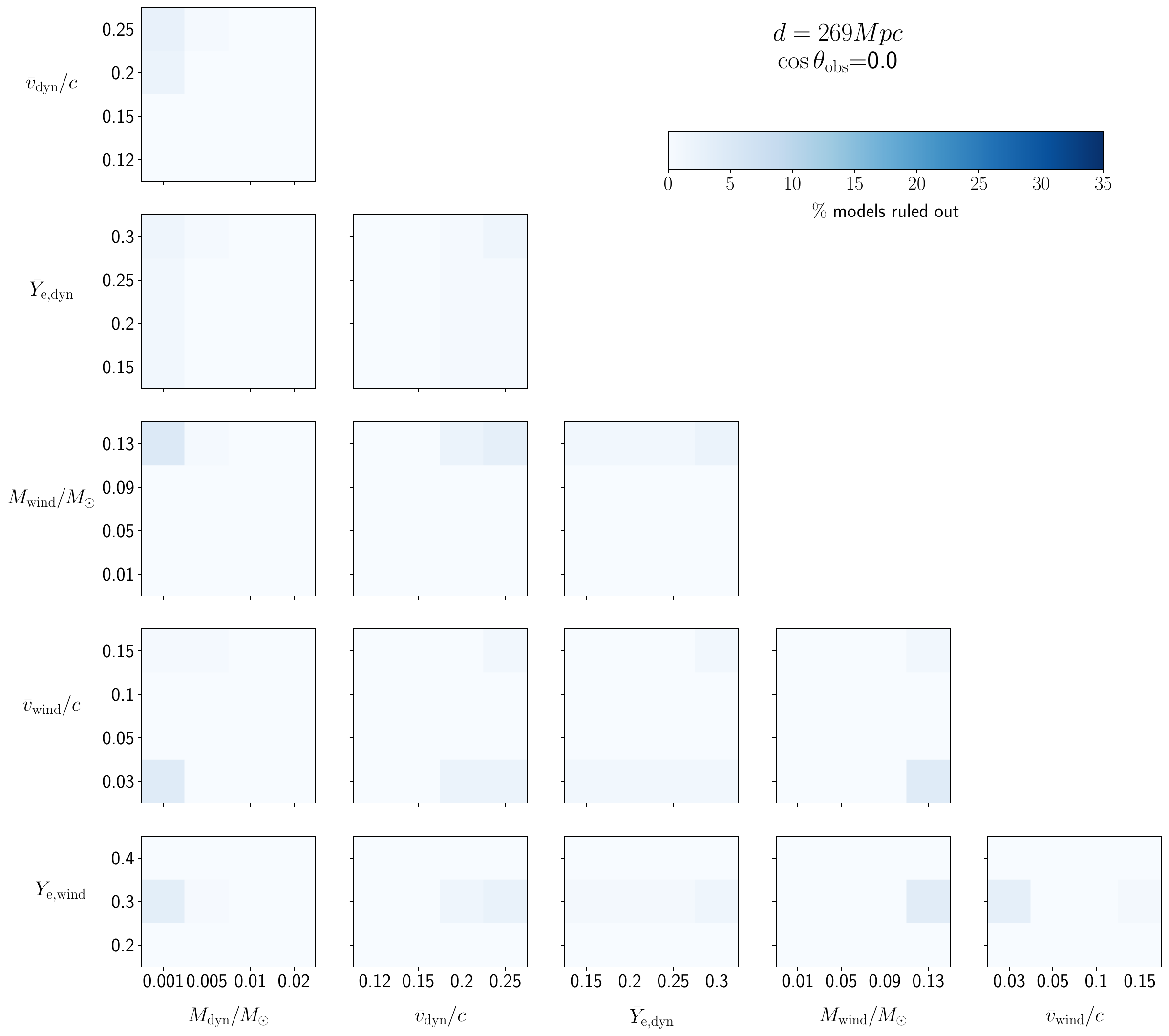}
    \caption{Corner plot showing in a colorbar the percentage of BNS models ruled out at a fixed distance of 269 Mpc and at a fixed viewing angle of 90 deg. The parameters correspond to the ejecta mass, mass-weighted averaged velocity and mass-weighted electron fraction for the dynamical and wind ejecta: $m_{\rm dyn}$, $\bar{v}_{\rm dyn}$, $\bar{Y}_{\rm e,dyn}$, $m_{\rm wind}$, $\bar{v}_{\rm wind}$ and $\bar{Y}_{\rm e,wind}$  }
    \label{fig:corner3}
\end{figure*}

\begin{figure*}[t]
    \centering
     \includegraphics[width=0.5\textwidth]{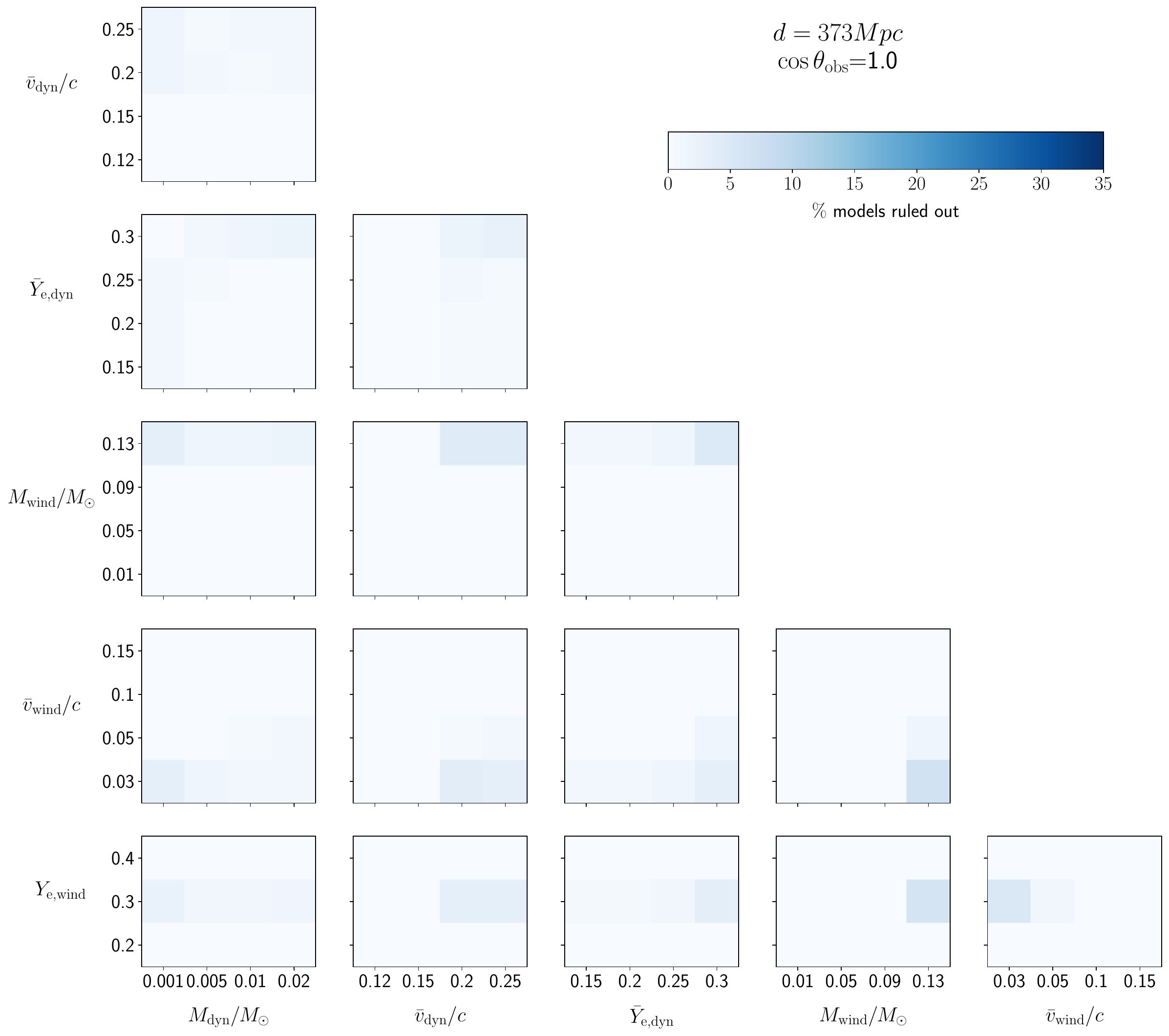}
    \caption{Corner plot showing in a colorbar the percentage of BNS models ruled out at a fixed distance of 373 Mpc and at a fixed viewing angle of 0 deg. The parameters correspond to the ejecta mass, mass-weighted averaged velocity and mass-weighted electron fraction for the dynamical and wind ejecta: $m_{\rm dyn}$, $\bar{v}_{\rm dyn}$, $\bar{Y}_{\rm e,dyn}$, $m_{\rm wind}$, $\bar{v}_{\rm wind}$ and $\bar{Y}_{\rm e,wind}$  }
    \label{fig:corner4}
\end{figure*}

\begin{figure*}[t]
    \centering
     \includegraphics[width=0.5\textwidth]{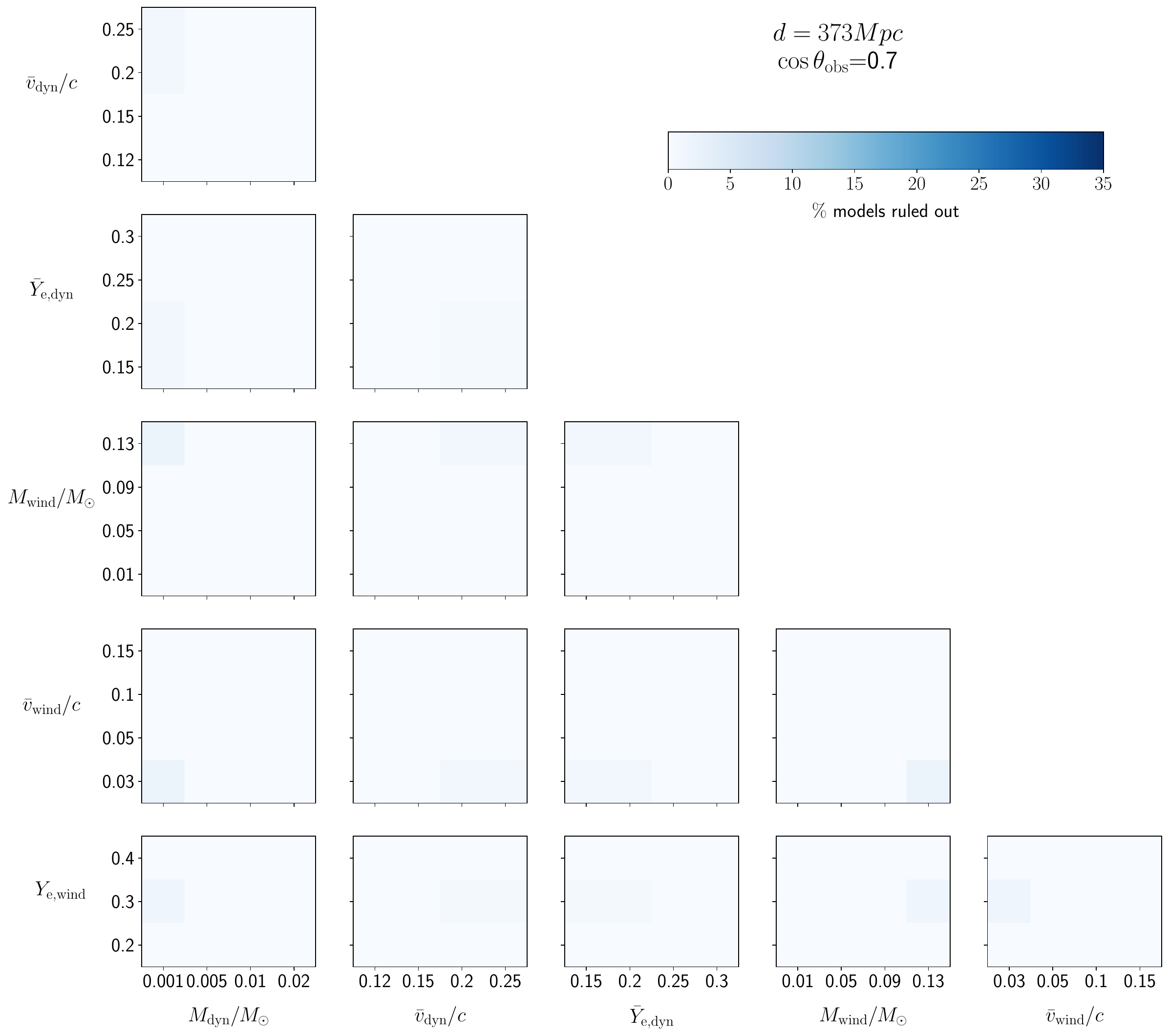}
    \caption{Corner plot showing in a colorbar the percentage of BNS models ruled out at a fixed distance of 269 Mpc and at a fixed viewing angle of 45 deg. The parameters correspond to the ejecta mass, mass-weighted averaged velocity and mass-weighted electron fraction for the dynamical and wind ejecta: $m_{\rm dyn}$, $\bar{v}_{\rm dyn}$, $\bar{Y}_{\rm e,dyn}$, $m_{\rm wind}$, $\bar{v}_{\rm wind}$ and $\bar{Y}_{\rm e,wind}$  }
    \label{fig:corner5}
\end{figure*}

\end{document}